\documentclass[reprint,twocolumn,showpacs,longbibliography]{revtex4-1}

\DeclareUnicodeCharacter{0308}{¨}

\usepackage{graphicx}
\usepackage{dcolumn}
\usepackage{bm}
\usepackage[utf8]{inputenc}
\usepackage[T1]{fontenc}
\usepackage{mathptmx}
\usepackage{etoolbox}
\usepackage{color}
\usepackage{xcolor}
\usepackage{amsmath,amssymb}

\usepackage[normalem]{ulem} 

\newcommand{\Wu}{$\frac{\text{W}}{\text{m}\cdot \text{K}}$\,}
\newcommand{\hhbar}{\mathchar'26\mkern-8mu h}
    \definecolor{myRed}{RGB}{196, 0, 0}
    \definecolor{myBlue}{RGB}{23, 56, 196}
    \definecolor{Mah}{RGB}{209, 84, 0}

\makeatletter
\def\@email#1#2{%
 \endgroup
 \patchcmd{\titleblock@produce}
  {\frontmatter@RRAPformat}
  {\frontmatter@RRAPformat{\produce@RRAP{*#1\href{mailto:#2}{#2}}}\frontmatter@RRAPformat}
  {}{}
}%
\makeatother

\begin{document}


\title[]{Optical and thermal characterization of a group-III nitride semiconductor membrane by microphotoluminescence spectroscopy and Raman thermometry}

\author{Mahmoud Elhajhasan}
\author{Wilken Seemann}
\author{Katharina Dudde}
\author{Daniel Vaske}
\author{Gordon Callsen*}
\email{gcallsen@uni-bremen.de}

\affiliation{Institut für Festk\"orperphysik, Universit\"at Bremen, Otto-Hahn-Allee 1, 28359 Bremen, Germany}

\author{Ian Rousseau}
\author{Thomas F. K. Weatherley}
\author{Jean-François Carlin}
\author{Raphaël Butt\'{e}}
\author{Nicolas Grandjean}

\affiliation{Institute of Physics, \'{E}cole Polytechnique F\'{e}d\'{e}rale de Lausanne (EPFL), CH-1015 Lausanne, Switzerland}

\author{Nakib H. Protik}

\affiliation{Institut für Physik und IRIS Adlershof, Humboldt-Universit\"at zu Berlin, 12489 Berlin, Germany}

\author{Giuseppe Romano}

\affiliation{MIT-IBM Watson AI Lab, IBM Research, Cambridge, MA, USA}

\date{\today}

\begin{abstract}
We present the simultaneous optical and thermal analysis of a freestanding photonic semiconductor membrane made from wurtzite III-nitride material. By linking micro-photoluminescence spectroscopy with Raman thermometry and other spectroscopic techniques, we demonstrate how a robust value for the thermal conductivity $\kappa$ can be obtained using only optical, non-invasive means. For this, we consider the balance of different contributions to thermal transport given by, e.g., excitons, charge carriers, and heat carrying phonons. In principal, all these contributions can be of relevance in a photonic membrane on different lengthscales. Further complication is given by the fact that this membrane is made from direct bandgap semiconductors, designed to emit light based on an In$_{x}$Ga$_{1-x}$N ($x=0.15$) quantum well embedded in GaN. Thus, III-nitride membranes similar to the one in the focus of this study, have already successfully been used for laser diode structures facing thermal limitations. To meet these intricate challenges, we designed a novel experimental setup that enables the necessary optical and thermal characterizations in parallel. After the optical characterization by micro-photoluminescence, we follow a careful step-by-step approach to quantify the thermal properties of our photonic membrane. Therefore, we perform micro-Raman thermometry, either based on a heating laser that also acts as a probe laser (1-laser Raman thermometry), or based on two lasers, providing the heating and the temperature probe separately (2-laser Raman thermometry). For the latter technique, we can obtain temperature maps over several tens of micrometers with a spatial resolution less than $1\,\mu\text{m}$. As a result, the temperature probe volume using the 2-laser Raman thermometry technique can be increased by a factor exceeding $100$ compared to the conventional 1-laser Raman thermometry technique, which impacts the derivation of the thermal conductivity $\kappa$. Only based on our largest temperature probe volume we derive $\kappa\,=\,95^{+11}_{-7}\,$\Wu for the  \textit{c}-plane of our $\approx\,250\text{-nm}$-thick photonic membrane at around room temperature, which compares well to our \textit{ab initio} calculations, applied to a simplified structure, yielding $\kappa\,=\,136\,$ \Wu. Based on these calculations, we explain the particular relevance of the temperature probe volume, as quasi-ballistic transport of heat-carrying phonons with high relevance for $\kappa$ occurs on length scales beyond the penetration depths of the heating laser and even its focus spot radius. The 1-laser Raman thermometry technique, therefore, fails to determine realistic $\kappa$ values, while with the 2-laser Raman thermometry technique one can probe temperatures in sufficiently large volumes. The present work represents a significant step towards non-invasive, highly spatially resolved, and still quantitative thermometry performed on a photonic membrane made of a direct bandgap semiconductor, which is of particular relevance for photonic applications.

\end{abstract}

\maketitle

The optical properties of semiconductors are extensively studied, and are therefore becoming increasingly well understood for the full size variation from bulk crystals down to nanostructures. Measuring fundamental optical phenomena like absorption, reflectivity, light scattering, and photoluminescence (PL) on various length- and timescales has nowadays become a widespread capability in many laboratories. The nanostructures commonly in scope encompass the entire dimensionality range from quantum dots (0D), over nanowires (1D), to quantum wells (2D). Consequently, a wide range of prototypes and devices was realized, scaling from, e.g., light detectors \cite{fan_analysis_2020}, over sensors \cite{khan_gallium_2020}, to light-emitting- \cite{kneissl_emergence_2019}, and laser-diodes \cite{kneissl_ultraviolet_2007}. For instance, laser-diodes often utilize nanostructures like quantum wells (QWs) as a built-in light source \cite{nakamura_ingan-based_1996,kneissl_ultraviolet_2003}. However, the high level of sophistication reached for the analysis of the photonic properties of these structures \cite{nippert_temperature-dependent_2016,sheen_highly_2022,nippert_auger_2018} is often contrasted by a limited understanding of the interfering thermal phenomena \cite{jagsch_quantum_2018}. Commonly, either the photonic or the thermal properties of micro- and nanostructures are studied, while the interrelation of both phenomena is rarely reported \cite{gesemann_thermal_2010,wen_thermal_2022}, prohibiting any future mutual optimization. So far, a link between photonic and phononic properties has been established in the field of nano-optomechanics, focusing mostly on low-frequency phonons (MHz - GHz range) \cite{aspelmeyer_cavity_2014}. However, thermal transport represents an even more intricate phenomenon that can be mediated by a weighted balance of, e.g., charge carriers \cite{cahill_thermometry_2002,quan_impact_2021}, excitons \cite{wu_bilayer_2014}, and most prominently high-frequency, thermal phonons up to the THz range \cite{sood_heat_2018}.

Before any link between optical and thermal material properties can be studied, one first has to develop a robust experimental technique that enables such studies. Clearly, any interlinked understanding of optical and thermal material properties could potentially pave the way towards mutual optimizations. Established thermal characterization techniques like, e.g., 3$\omega$-measurements \cite{cahill_thermal_1990} are not straightforwardly compatible with photonic micro- and nanostructures for size reasons and the required fabrication of metal contacts. A similar observation seems valid for time-domain thermal reflectance (TDTR) measurements \cite{cahill_thermometry_2002,cahill_nanoscale_2014} as another established thermal characterization approach, which most frequently relies on metal transducers. Purely thermally induced reflectivity changes in semiconductors are generally considered small over wide energy ranges, if not located in close energetic vicinity to electronic and excitonic resonances \cite{matatagui_thermoreflectance_1968,liu_thermal_2019}. Thus, in practice metal transducers with larger thermoreflectance coefficients in the wavelength range of interest are commonly required for TDTR measurements \cite{wang_thermoreflectance_2010,wilson_thermoreflectance_2012}. The processing of such metal transducers can potentially deteriorate photonic structures. Therefore, a fully non-invasive thermal characterization technique is required, which in the best case can readily be combined with standard optical analyses like absorption and reflectivity measurements, as well as (time-resolved) PL spectroscopy. In a best case scenario, this thermal characterization technique should also feature diffraction limited spatial resolution similar to, e.g., micro-TDTR \cite{maire_heat_2017}, together with a large applicable temperature range and sufficient temperature resolution. Certainly, for many applications any optical and thermal characterization at room temperature and beyond would be sufficient. Nevertheless, most interesting thermal transport physics connected to, e.g., the occurrence of various thermal phonon transport regimes, would also motivate thermometry down to cryogenic temperatures \cite{cepellotti_phonon_2015,lee_hydrodynamic_2015,anufriev_heat_2017}. In summary, a quantifying thermal characterization technique with excellent integrability, high spatial resolution, as well as a wide operating temperature range and sufficient temperature resolution is desired.

Raman thermometry appears as a promising thermal characterization technique in the light of these requirements \cite{beechem_invited_2007,beechem_invited_2015,jaramillo-fernandez_raman_2018}. Raman measurements can not only be performed in a micro-PL ($\mu$PL) setup, one can also directly map micro-Raman ($\mu$Raman) spectra with a spatial resolution close to the diffraction limit \cite{beechem_invited_2007} over a wide range of temperatures. However, these promising aspects of Raman thermometry pair with a number of challenges, whenever the Raman laser is not exclusively used as a temperature probe to map, e.g., the Joule heating of a transistor structure under operating conditions \cite{kuball_raman_2001}. As soon as the Raman laser is simultaneously used for the sample heating and the temperature probing based on Raman spectra, the situation complicates \cite{jaramillo-fernandez_raman_2018,sandell_thermoreflectance_2020}. The applied laser heating challenges the data analysis in contrast to, e.g., common 3$\omega$ or TDTR measurements that apply Joule heating to a semiconductor via metal contacts or transducers. In general, the thermal resistance of the metal/semiconductor interface represents a manageable challenge for 3$\omega$ and TDTR experiments and directly correlated thermal simulations \cite{cahill_nanoscale_2014}. In contrast for Raman thermometry, depending on the light penetration depths of the heating laser, one has to either model the laser heating by a semiconductor surface heating or a fully volumetric heating, which requires precise knowledge of the absorption coefficient and its wavelength dependence \cite{lax_temperature_1977,jaramillo-fernandez_raman_2018}. Even more severe, in stark contrast to the Joule heating via metallized semiconductor surfaces, the over bandgap optical excitation of semiconductor samples by a suitable heating laser can excite charge carriers \cite{cahill_thermometry_2002,quan_impact_2021} or even generate optical excitations like excitons \cite{wu_bilayer_2014} and exciton-polaritons \cite{ciers_polariton_2020}. As a result, not only thermal phonons are generated, but all these additional fundamental excitations, which diffuse away from a laser-induced heatspot. This situation is then further complicated by the optical decay of, e.g., excitons and exciton-polaritons during their propagation. Any measured temperature rise in a semiconductor that is directly induced by laser heating (i.e., without a metal transducer) is challenged by all these additional energy transport mechanisms. Commonly, with thermal phonons we refer to high frequency heat carrying phonons that predominantly contribute to thermal transport at all temperatures in the scope of the present work. For sake of completeness, also radiative heat transfer and related near-field phenomena \cite{biehs_thermal_2007} can be added to the list of these mechanisms. However, these mechanisms can safely be neglected for the present work due to the occurring temperature differences and distances as well as dominating contributions from the phononic type of thermal transport.

Consequently, Raman thermometry that relies on laser heating can only lead to a solid thermal characterization of a semiconductor, if its optical properties are thoroughly studied in-parallel. Not only the absorption must be known \cite{beechem_invited_2015}, but also the overall quantum efficiency of the sample is required for an estimation of how much of the absorbed light contributes to the heating of the sample. Here, the fact that common Raman thermometry applies the same laser for the laser heating and the temperature probing represents a strong limitation as the electronic \cite{weatherley_imaging_2021} or thermal excitations \cite{minnich_thermal_2011} connect to different characteristic mean free path lengths. Thus, such one-laser Raman thermometry (1LRT) is best extended to two-laser Raman thermometry (2LRT), transferring the role of the temperature probe to a second laser, which can be scanned away from the laser heat spot, over the entire structure of interest. Based on temperature maps extracted from Raman mapscans, it is consequently possible to observe temperature distributions that are caused by a weighted balance of the different mechanisms contributing to thermal transport. Thus, as the related mean free path lengths vary, 2LRT measurements have the potential to disentangle different contributions to thermal transport by varying the distance between the temperature probe and the laser-induced heatspot. Initially, Hsu \textit{et\,al.} have applied 2LRT measurements to carbon nanotubes \cite{hsu_direct_2011}, while also its applicability to silicon membranes \cite{reparaz_novel_2014} and polycrystalline nanomembranes of MoS$_2$ \cite{sledzinska_thermal_2016} was shown. However, all existing experimental 2LRT setups, as described in detail by Reparaz \textit{et\,al.} \cite{reparaz_novel_2014}, require full optical access from the sample front- and backside to overlap the heating and the probe laser focus spots on the sample. While such an approach is still feasible, e.g., for micro-structured silicon membranes \cite{graczykowski_thermal_2017}, most photonic micro- and nanostructures can only be accessed from one side as a full substrate removal is often detrimental for delicate photonic structures \cite{vico_trivino_high_2012,jagsch_quantum_2018, trivino_continuous_2015}.

In this contribution we apply 1LRT and 2LRT to a photonic membrane structure, which has widely been described in literature \cite{vico_trivino_high_2012,trivino_continuous_2015,rousseau_quantification_2017,jagsch_quantum_2018,Rousseau2018a,Rousseau2018c}. This membrane is mostly made from GaN and it incorporates an In$_{x}$Ga$_{1-x}$N ($x=0.15$) quantum well (QW) as built-in light source. Structures very similar to this photonic membrane have already formed the basis for publications reporting on, e.g., 1-dimensional nanobeam lasers \cite{trivino_continuous_2015,jagsch_quantum_2018} and 2-dimensional photonic crystals \cite{vico_trivino_high_2012} with excellent optical properties. Trivi$\tilde{\text{n}}$o \textit{et\,al.} reported on record low threshold blue lasing of InGaN/GaN nanobeam lasers \cite{trivino_continuous_2015}, which were based on the photonic membrane also in use for the present work. As such photonic structures can experimentally only be accessed from one side, a fully customized 1LRT and 2LRT setup was constructed, which heats samples and probes their temperatures from the same side by combining conventional micro-Raman mapping and a laser scanner. To highlight the functionality of this setup, all experimental results are introduced in a careful step-by-step approach. aiming to achieve quantitative Raman thermometry despite the numerous challenges posed by a photonic membrane made from direct bandgap semiconductor material. This also includes in-parallel an analysis of the PL signal, aiming to discard an impact of free carriers and excitons on our determination of the thermal conductivity $\kappa$. We show that based on 1LRT or 2LRT measurements one determines strongly different thermal conductivities, which are linked to the temperature probe volume. 1LRT measurements are not ideally suited for a quantitative determination of $\kappa$ in our membranes, whereas 2LRT measurements yield $\kappa\,=\,95^{+11}_{-7}\,$\Wu for our $\approx\,250\text{-nm-thick}$ photonic III-nitride membrane. We then compare $\kappa$ derived from \textit{ab initio} calculations to our experimental results, while analyzing the dependence of the related cumulative $\kappa$ on the phonon mean free path lengths $l^{MFP}$ provides an explanation for the different experimental results derived from 1LRT and 2LRT measurements. Consequently, only 2LRT measurements allow for a truly thermal quantitative analysis of our photonic III-nitride membrane, providing a strong motivation for any future application of this experimental technique.

The paper is structured as follows. In Sec.\,\ref{sec:Experiment}, we first introduce the photonic membrane made from III-nitride material (Sec.\,\ref{sec:Sample}) that we will optically ($\mu$PL) and thermally (Raman thermometry) characterize based on our customized optical setup (Sec.\,\ref{sec:Setup}). Herein, we describe the mode of operation of our optical setup in a four-step-process, which gradually raises the level of complication for our measurements. This rigorous step-by-step approach starts with \textbf{Step A}, the presentation of the $\mu$PL spectroscopy in Sec.\,\ref{sec:PL}. Subsequently, in Sec.\,\ref{sec:Raman} this pre-characterization of our photonic membrane is complemented by spatially resolved, non-resonant and selected resonant Raman measurements on GaN, which constitutes \textbf{Step B} in our hierarchy. Such resonant and non-resonant Raman spectroscopy represents the basis for the Raman thermometry described in Sec.\,\ref{sec:1LRTand2LRT}, which is not yet spatially resolved (\textbf{Step C}). This section is subdivided into Sec.\,\ref{sec:1LRT} and Sec.\,\ref{sec:2LRT}, which introduce the 1LRT and the 2LRT$_0$ measurements without any spatial displacement between the heat and the probe lasers. Herein, we also introduce our numerical simulations that allow us to extract $\kappa$ from our 1LRT and 2LRT without spatial displacement (2LRT$_0$) measurements. Based on these experimental and numerical foundations from Secs.\,\ref{sec:Experiment}\,-\,\ref{sec:1LRTand2LRT}, we can turn towards the last \textbf{Step D} in Sec.\,\ref{sec:Heat} that describes 2LRT measurements. Here, we show our temperature mapscans that enable the determination of $\kappa$, which is additionally modeled in Sec.\,\ref{sec:Modeling} based on state-of-the-art \textit{ab initio} calculations. As we experimentally find a particular scaling behavior in between the $\kappa$ values derived from 1LRT, 2LRT$_0$, and 2LRT measurements, we discuss this observation in Sec.\,\ref{sec:Comparison} based on our theoretical findings from Sec.\,\ref{sec:Modeling}. Finally, we present our discussion and outlook in Sec.\,\ref{sec:DiscussionOutlook}, before summarizing our results in Sec.\,\ref{sec:Summary}. 
In addition to this paper, more technical details and concepts are further described in the Supplementary Material (SM) within S-Secs.\,I\,-\,IV.

\section{\label{sec:Experiment}Experimental details \protect}

%
%
%
\begin{figure*}[]
    \includegraphics[width=14 cm]{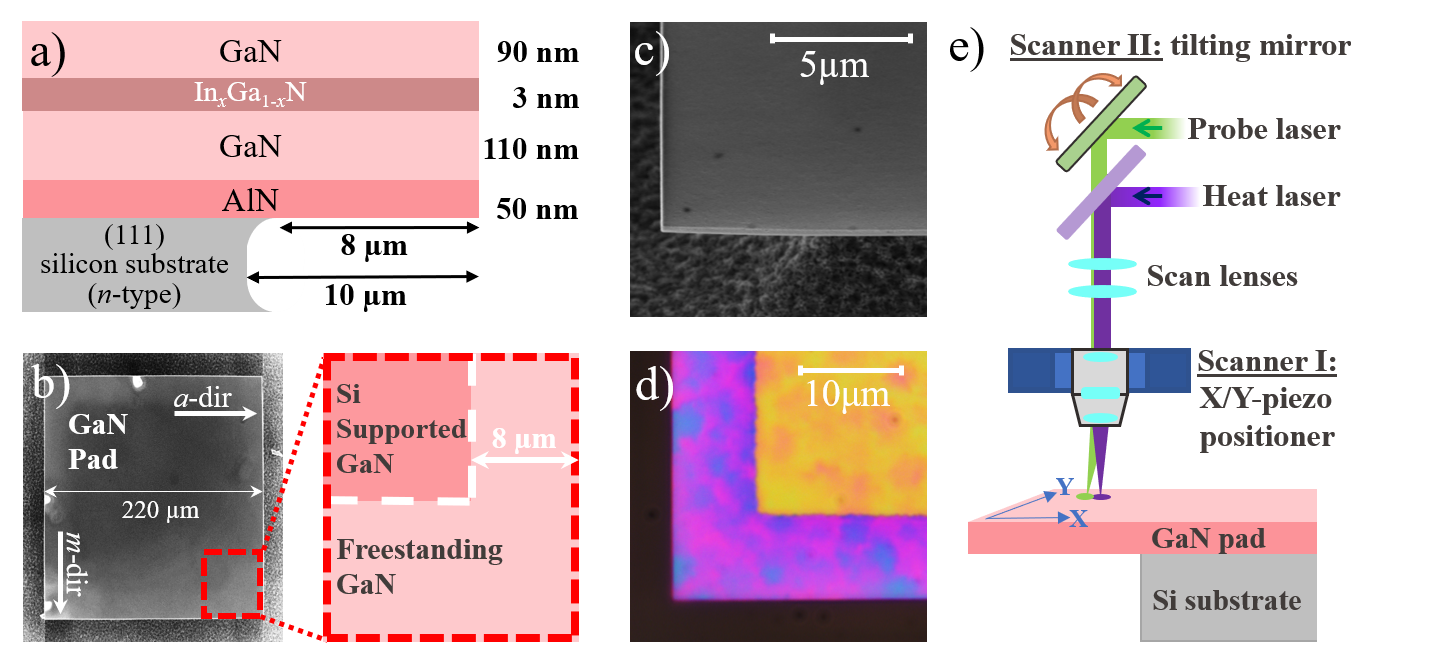}
    \caption{(a) Layer sequence of the photonic membrane made from group-III nitrides grown on \textit{n}-type silicon (111). The origin of the layer sequence is explained in the main text. (b) Top-view SEM image of the photonic membrane that mainly consists of \textit{c}-plane GaN. The \textit{a}- and \textit{m}-plane crystal directions are indicated. The dashed red rectangle along with the sketch on the right indicates the freestanding part of the photonic membrane in one of its corners, which extends up to a widths of $\approx$\,8\,$\mu$m. (c) Top-view SEM image of the photonic membrane illustrating the underetch, the roughened silicon substrate, as also the surface and the sidewalls of the photonic GaN membrane. (d) Top-view light microscope image of the same corner region of the photonic membrane. Here, the color fluctuations indicate thickness fluctuations of the photonic membrane. (e) Simplified sketch of the experimental two-laser setup that enables micro-photoluminescence spectroscopy (optical characterization) in parallel to one-laser- and two-laser-Raman spectroscopy (thermal characterization).}
    \label{fig:exp_setup}
\end{figure*}

The overall structure of the photonic membrane is introduced in Figs.\,\ref{fig:exp_setup}(a)\,-\,\ref{fig:exp_setup}(d). In addition, a simplified sketch of the experimental setup that enables $\mu\text{PL}$ spectroscopy (optical characterization) in parallel to one-laser- and two-laser-Raman spectroscopy (thermal characterization) is given in Fig.\,\ref{fig:exp_setup}(e). A more detailed drawing and explanation of the experimental setup can be found in S-Sec.\,I.

\subsection{\label{sec:Sample}The photonic membrane}

The layer stack that forms the basis for the photonic membrane is made from \textit{c}-plane, wurtzite AlN, GaN, and In$_{x}$Ga$_{1-x}$N ($x=0.15$) grown on an \textit{n}-type (111) silicon substrate as sketched in Fig.\,\ref{fig:exp_setup}(a). Such material growth on silicon (111) is key to the formation of a membrane structure, as silicon can be selectively etched. The group-III nitride material was grown by metal-organic vapor phase epitaxy (MOVPE) in a horizontal Aixtron\textsuperscript{\textregistered} 200/4 RF-S low pressure reactor and details regarding the growth of such epilayers on silicon (111) can be found in Ref.\,\cite{trivino_gan-based_2015}. First, 50\,nm of \textit{c}-plane AlN were deposited on silicon (111) to avoid melt-back etching \cite{krost_gan-based_2002}. Afterwards, 110\,nm of GaN, the 3-nm-thick In$_{x}$Ga$_{1-x}$N ($x=0.15$) QW, and finally another 90\,nm of GaN were grown. Here, the QW is deliberately placed 10\,nm above the geometric center of the surrounding GaN, taking into account the lower refractive index of AlN \cite{Rousseau2018c}. This way, the coupling of this built-in light source to the fundamental electromagnetic mode is enhanced. The overall membrane thickness of $\approx\,250\,\text{nm}$ is larger than required for single electromagnetic mode operation of this membrane, which is of relevance as soon as, e.g., a photonic crystal is processed for the blue spectral range \cite{Rousseau2018c}. However, this represents a trade-off, aiming to achieve the smoothest possible GaN surface without pits or cracks. The entire stack of III-nitride material is not intentionally doped and the GaN material exhibits a free carrier concentration of $3\,\times\,10^{16}\,\text{cm}^{-3}$. Detailed secondary ion mass spectroscopy of the photonic membrane can be found in Ref.\,\cite{Rousseau2018c}, showing that the main impurities in our photonic membrane are silicon and carbon with concentrations of $\approx\,3\times\,10^{16}\,\text{cm}^{-3}$ and $\approx\,5\times\,10^{16}\,\text{cm}^{-3}$.

The photonic motivation for smooth surfaces \cite{rousseau_quantification_2017} is also beneficial for the in-plane thermal transport that is studied in this work. To quantify light scattering losses in group-III nitride photonic crystals in the blue spectral range, Rousseau \textit{et\,al.} \cite{rousseau_quantification_2017} have conducted detailed atomic force microscopy (AFM) on the front- and backside of precisely the membrane that is studied in the present work. After all required processing steps to form the photonic membrane, they determined a GaN rout mean square (RMS) top surface roughness of 1.4\,nm, whereas the AlN backside exhibited an RMS roughness of 0.6\,nm, which was extracted from 500\,x\,500\,$\text{nm}^{2}$ AFM scans. The backside roughness of the nitrogen-polar AlN was accessed by exfoliation and subsequent flipping of a large ($\approx\,80\,\times\,80\,\mu\,\text{m}^{2}$) square of an entire membrane similar to the GaN pad shown in Fig.\,\ref{fig:exp_setup}(b). In addition to this surface roughness, a minor waviness was observed for the particular membrane in $25\,\times\,25\,\mu\text{m}^{2}$ AFM scans, which yields an RMS waviness of 4.9\,nm and $\approx$\,30 nm peak-to valley height variation over distances of 5-10\,$\mu\text{m}$. These height variations lead to the color contrast shown in Fig.\,\ref{fig:exp_setup}(d) due to thin film interference. In addition, the AFM scans on the AlN surface on the backside of the photonic membrane yielded a high density of pin pricks forming dips that correspond to dislocations with a density of $3\,-\,4\,\times\,10^{10}\,\text{cm}^{2}$, which is a typical value for AlN grown on (111) silicon \cite{eric_feltin_hetero-epitaxie_2003}. Similar AFM scans of the top GaN yield the same range for the dislocation density, which corresponds to a distance of $50\,-\,60\,\text{nm}$ between these defects. For the sake of simplicity, in the following the epilayer from Fig.\,\ref{fig:exp_setup}(a) will be described as the GaN pad (with silicon support) or membrane (without silicon support), as its main constituent GaN will dominate the thermal properties that lie in the scope of this work.

The processing that leads to the GaN pad structure illustrated in the secondary electron microscopy (SEM) image from Fig.\,\ref{fig:exp_setup}(b) is based on electron beam lithography, dry etching, and chemically selective vapor phase etching. First, hydrogen silsesquioxane is applied as negative tone resist for the electron beam lithography, which also acts as a hard mask, enabling the transfer of written patterns into the underlying epilayer via dry etching. Exactly during this pattern transfer, any potential edge roughness of the etch mask is transferred to the underlying epilayer. As no apparent edge roughness can be observed within the spatial resolution limits of the applied electron microscope, only an upper bound for the RMS sidewall roughness of $\lesssim\,5\,\text{nm}$ can be approximated. It should be noted that this sidewall roughness does not represent a general limit as even atomically flat sidewalls can be achieved by anisotropical wet etching of GaN by tetramethylammonium hydroxide \cite{palacios_wet_2000}. To this end, the corresponding sidewalls of any etched structure must be aligned to the crystallographic axes as it is the case for the present GaN pad, cf. Fig.\,\ref{fig:exp_setup}(b). The edge region of this GaN pad is then released from the silicon substrate over $\approx\,8\,\mu\text{m}$ by chemically selective vapor phase etching using XeF$_2$. The resulting structural situation is sketched in Fig.\,\ref{fig:exp_setup}(b) and further detailed by the SEM image from Fig.\,\ref{fig:exp_setup}(c). Additional details regarding the sample growth and processing can be found in Ref.\,\cite{Rousseau2018c} and the Supplemental Material of Ref.\,\cite{rousseau_quantification_2017}, reporting on high quality one-dimensional photonic crystal cavities based on the epilayer in use for the present work. For comparison, we also perform 1LRT measurements on a freestanding piece of a state-of-the-art \textit{c}-plane GaN wafer with a dislocation density $\sim\,10^{6}\,\text{cm}^{-2}$, which already formed the basis for detailed optical studies on InGaN-based light-emitting diodes \cite{haller_burying_2017}.

\subsection{\label{sec:Setup}The customized optical setup}

Figure\,\ref{fig:exp_setup}(e) introduces the spectroscopic setup that enables spatially resolved $\mu\text{PL}$ and $\mu$Raman mapscans by the independent displacement of two laser spots on the surface of the epilayer. Whereas the basic features of this setup are described in the following, its full description is given in S-Sec.\,I. First, an ultraviolet (UV) laser that provides above bandgap excitation of GaN (purple), is guided towards the sample via a suitable beamsplitter and a microscope objective. We use either a 266\,nm or a 325\,nm continuous wave (cw) laser for this purpose. Both lasers can then directly be used as excitation lasers for $\mu\text{PL}$ spectroscopy. The excitation spot of both lasers can be translated across the surface of the GaN pad over a range of $\approx\,100\,\mu\text{m}$ by actuating the closed-loop X/Y-piezo stage labeled as scanner I in Fig.\,\ref{fig:exp_setup}(e). In addition, coarser movements of the entire sample structure are enabled by a long-range X/Y/Z-piezostage (not shown) that holds the sample in a cryostat, where it is located at a base pressure of $\approx\,1\times\,10^{-6}\,\text{mbar}$. Even though the cooling option of this cryostat with its built-in X/Y/Z-piezostage is not in use for the present work, it still acts as the heat-sink for then entire sample, keeping it at an ambient temperature of $295\,\pm\,1\,\text{K}$. Upon UV excitation, the resulting $\mu\text{PL}$ signal is guided backwards through the microscope objective towards the detection system comprising a suitable monochromator and charge-coupled device (CCD) detector with a spectral resolution of $\approx\,1\,cm^{-1}$ in the visible region around 532\,nm.

Resonant Raman spectra based on one of the UV lasers (266\, nm or 325\,nm) are measured in the same manner. Here, varying the power of these lasers gives direct access to Raman thermometry based on one laser (1LRT), which originates the labeling of these lasers as "heating lasers". However, low signal strength in addition to the underlying $\mu\text{PL}$ signal arising from various optical transitions in GaN, can significantly challenge the data recording as further detailed in Sec.\,\ref{sec:RamanComparison}. In addition, non-resonant Raman mapscans of the epilayer can be recorded based on two visible (VIS) lasers (488\,nm and 532\,nm), which we describe as probe lasers. Therefore, any movement of the visible laser spot can either be achieved by scanner I or the angular tilting mirror (scanner II) as soon as the two scanning lenses are part of the beampath, cf. Fig.\,\ref{fig:exp_setup}(e). These lenses are present to ensure that the visible laser beam enters the objective in its center, and the angle between the beam and the microscope axis creates the beam displacement on the sample surface. Especially the scanning option based on scanner II is of high importance for 2LRT measurements. Such measurements can be best described by the following four measurements steps that form the basis for Secs.\,\ref{sec:PL}\,-\,\ref{sec:Heat}:

\hfill

\textbf{Step A - $\mu$PL spectra \& positioning the heating laser (Sec.\,\ref{sec:PL})}: The positioning of the heating laser based on $\mu\text{PL}$ mapscans yields a spatially and spectrally resolved intensity distribution based on scanner I [Figs. \ref{fig:PL}(a) \& \ref{fig:PL}(b)]. During such measurements the spot size of the heating laser is determined as described in S-Sec.\,I. As the spot size of the heating laser is a crucial parameter for the subsequent data analysis, we made it common practice to probe the laser spot size for every sequence of measurements. Small laser focus deviations can lead to non-tolerable laser spot size variations that scale with the numerical aperture of the microscope objective in use. Commonly, after positioning the heating laser on the desired sample location, excitation power-dependent series of $\mu\text{PL}$ spectra are recorded as a first optical and thermal characterization of the sample. For this step it is crucial that both scanning lenses illustrated in Fig.\,\ref{fig:exp_setup}(e) are already in place in preparation of step B. Measuring series of $\mu\text{PL}$ spectra also supports the determination of the maximal applicable excitation power range for the heating laser. For this, the reproducibility of the $\mu\text{PL}$ spectra is checked after cycling the excitation power of the heating laser. The positioning and focal lengths of the two scanning lenses are such that both lenses do not impact the heating laser besides minor wavefront distortions. See S-Sec.\,I in the SM for further details.

\hfill

\textbf{Step B - Raman pre-characterization \& positioning the probe laser (Sec.\,\ref{sec:Raman})}: While the position of the heating laser's focus spot is kept on the sample (fixed position for scanner I), the probe laser's focus spot is scanned across the sample by scanner II in combination with the scanning lenses, which is the normal operation mode of a laser scanning microscope \cite{stock_self-organized_2010}. For the scanning, only the probe laser beam is tilted in the entrance aperture of the microscope objective. During the scanning action of this laser, non-resonant Raman spectra are recorded for each desired point on the sample. Due to the presence of the angular tilting mirror, the detection beam path for such non-resonant Raman spectra must deviate from the detection beam path for resonant Raman and $\mu\text{PL}$ spectra described under step A, cf. S-Sec.\,I. As a result, a mapscan of non-resonant Raman spectra is recorded, which, e.g., can illustrate the distribution of strain across the sample's surface. As the heating laser is still off during this recording, we call this result an unheated Raman mapscan (Fig. \ref{fig:Raman}a).

\hfill

\textbf{Step C - 1LRT characterization \& 2LRT$_0$ (Sec.\,\ref{sec:1LRTand2LRT})}: After positioning the heating and probe lasers, conventional resonant Raman thermometry (1LRT) is performed. For such measurements, the heating laser simultaneously acts as the temperature probe laser, meaning that no spatially resolved temperature maps can be recorded. Thus, the heating laser is the source of the resonant Raman signal, which serves as a local thermometer. In this work, we show 1LRT results as a first step towards 2LRT measurements, since such measurements commonly appear in literature. Later in the paper (Sec.\,\ref{sec:Comparison}), we will show that such temperature measurements directly at the heat spot imply several complications for our photonic membrane. In addition to the 1LRT characterization and based on the non-resonant Raman scan from step B and the $\mu\text{PL}$ mapscan from step A, it is now possible to overlap the spots of the heating and probe lasers. As a result, non-resonant Raman spectra can be recorded, while the heating power is varied. We call this technique two-laser Raman thermometry without spatial displacement (2LRT$_0$). 

\hfill

\textbf{Step D - 2LRT mapscans (Sec.\,\ref{sec:Heat})}: Spatial resolution can now be introduced by controlling scanner II, which exclusively allows movements of the probe but not the heating laser spot across the sample in our setup. A well-suited UV-VIS microscope objective must be utilized, featuring the same focal length for two significantly different laser wavelengths and a sufficient planarity of the corresponding focal planes. As a result, 2LRT mapscans can be recorded for varying powers of the heating laser. Here, the unheated Raman mapscan can be used to free the heated mapscans from any Raman mode shifts and mode broadenings induced by, e.g., strain or varying defect concentrations as further explained in Sec.\,\ref{sec:Heat}. Finally, 2LRT mapscans comprising Raman mode shifts or Raman mode broadenings can be translated into two-dimensional temperature maps based on a particular temperature calibration, cf. S-Sec.\,III. These temperature maps will then serve as the basis for the extraction of key parameters like the thermal conductivity $\kappa$.

\section{\label{sec:PL}Characterization of the photonic membrane by $\mu\text{PL}$ spectroscopy \& positioning of the heating laser (step A) \protect}

%
%
%
\begin{figure*}[]
    \includegraphics[width=\linewidth]{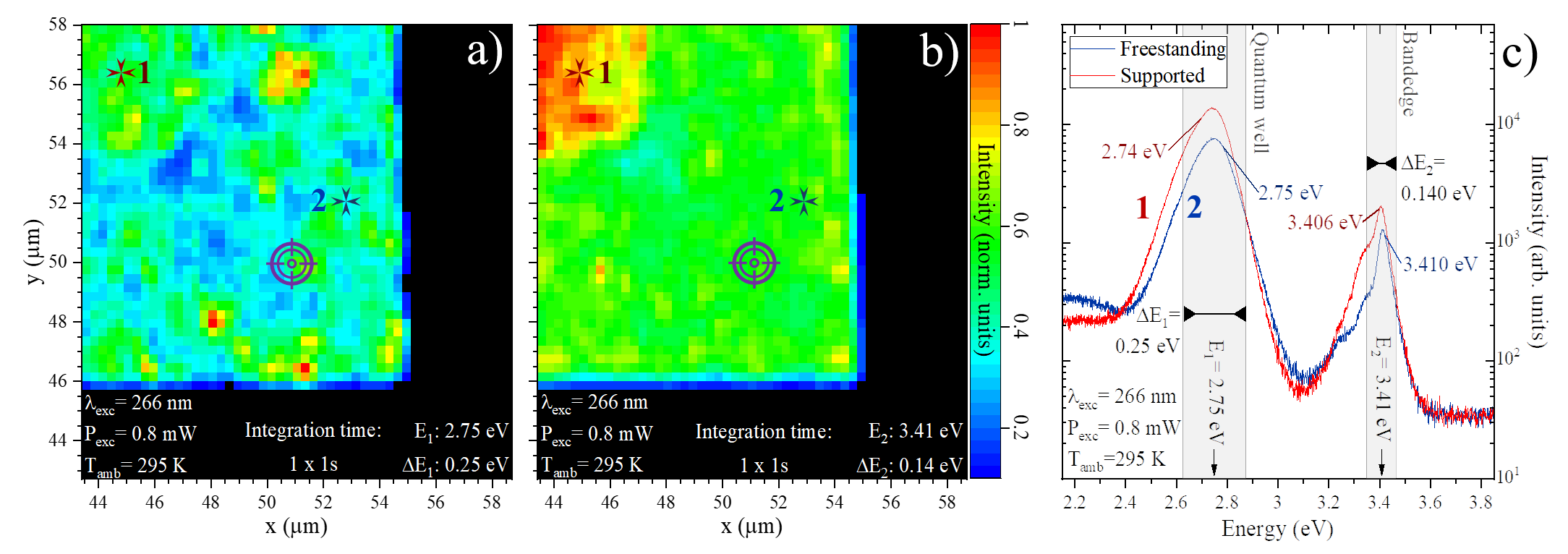}
    \caption{Characterization of the photonic membrane by $\mu\text{PL}$ spectroscopy at an ambient temperature $T_{amb}=\,295\,\text{K}$ under excitation with a cw UV laser with a wavelength $\lambda_{exc}=266\,\text{nm}$ at a power $P_{exc}\,=\,0.8\,\text{mW}$. (a) Intensity distribution of the QW emission centered around $E_{1}=2.75\,\text{eV}$ with a bandpass $\Delta E_{1}=0.25\,\text{eV}$, aiming to approximate the FWHM of the QW emission. Here, monolayer thickness and indium content fluctuations lead to intensity variations, which prominently appear as localization centers in the mapscan. (b) In contrast, the GaN bandedge emission at around $E_{2}=3.41\,\text{eV}$ shows a more homogeneous intensity distribution in the region of the freestanding GaN membrane with a bandpass matching its FWHM ($\Delta E_{2}=0.14\,\text{eV}$). Here, the out-coupled signal strength only significantly rises in the top left corner, where the underlying silicon support structure begins, cf. Fig.\,\ref{fig:exp_setup}(b). The target cross in (a) and (b) indicates the final desired position of the heating laser. (c) Comparison of two local $\mu\text{PL}$ spectra recorded at the color-matched indicators in (a) and (b).}
    \label{fig:PL}
\end{figure*}

Figure\,\ref{fig:PL} shows the $\mu$PL signal of the III-nitride epilayer introduced in Fig.\,\ref{fig:exp_setup}, which is required for step A. A UV laser ($\lambda_{exc}\,=\,266\,\text{nm}$), later described as the heating laser, is scanned over a corner region of the GaN pad and a $\mu$PL spectrum is recorded for each point, yielding polychromatic $\mu$PL mapscans as shown in Figs.\,\ref{fig:PL}(a) and \ref{fig:PL}(b) for the In$_{x}$Ga$_{1-x}$N ($x=0.15$) QW emission (bandpass centered at $E_{1}\,=\,2.75\,\text{eV}$, full width at half maximum (FWHM) $\Delta E_{1}\,=0.25\,\text{eV}$) and the GaN bandedge emission ($E_{2}\,=\,3.41\,\text{eV}$, $\Delta E_{1}\,=0.14\,\text{eV}$). Two exemplary local $\mu$PL spectra are shown in Fig.\,\ref{fig:PL}(c) for the two positions \textcolor{myRed}{\textbf{1}} (supported on silicon) and \textcolor{myBlue}{\textbf{2}} (freestanding) identified in Fig.\,\ref{fig:PL}(a) and (b). In addition to the two main emission features (QW and bandedge emission), the onset of GaN-related defect luminescence is visible below $\sim\,2.4\,\text{eV}$, which is of relevance for the 2LRT measurements introduced in Sec.\,\ref{sec:Heat}.

When comparing the intensity fluctuations from Fig.\,\ref{fig:PL}(a) and \ref{fig:PL}(b), it is apparent that the QW emission is less homogeneously distributed across the freestanding part of the GaN membrane compared to the GaN bandedge emission. For the QW intensity distribution, numerous, so-called, localization centers \cite{schomig_probing_2004} with higher emission intensity can be observed, which often can also be correlated to emission energy shifts (not shown). Here, local QW thickness fluctuations, structural defects (e.g., local assemblies of dislocations), and interrelated indium content fluctuations can lead to carrier localization \cite{schomig_probing_2004,hangleiter_suppression_2005,piccardo_localization_2017,callsen_probing_2019}. For instance, a QW thickness fluctuation of $\pm\,1$\,monolayer can already lead to emission energy shifts of $\approx\pm\,60\,-\,70\,\text{meV}$ for the given indium concentration \cite{bai_influence_2000} at room temperature and low carrier injection conditions that are met in Fig.\,\ref{fig:PL}. Clearly, such thickness, indium content, and defect concentration fluctuations naturally occur during growth of III-nitride epilayers on silicon (111), cf. Sec.\,\ref{sec:Sample}. Details regarding these carrier localization effects that are also linked to the high luminosity of this photonic membrane, despite the apparent structural defects, can, e.g., be found in Ref. \cite{Rousseau2018c}. 

The GaN bandedge emission from Fig.\,\ref{fig:PL}(b) shows a comparably homogeneous intensity distribution as the GaN thickness of $\approx\,200\,\text{nm}$ does not lead to any carrier confinement that would make thickness fluctuations a pronounced source of intensity variations. At the given ambient temperature $T_{amb}\,=\,295\,\text{K}$, both, excitons and band-to-band transitions already contribute to the bandedge emission of GaN as the thermal energy ($\approx\,26\,\text{meV}$) has already surpassed the binding energy of the free A-exciton and B-exciton of $\approx\,23\,\text{meV}$ \cite{callsen_excited_2018}. Here, the mapscans shown in Figs.\,\ref{fig:PL}(a) and \ref{fig:PL}(b) already enable a first estimation of an upper bound for the effective mean exciton and carrier diffusion length $l_{\text{diff}}\,\leq\,250\,\text{nm}$, as features with the size of individual pixels appear in the mapscan. Spatially resolved cathodoluminescence (CL) mapscans with higher spatial resolution recorded exactly in the corner region of the GaN membrane for low injection conditions are discussed in S-Sec.\,II. Here, an upper bound of $l_{\text{diff}}\lesssim\,115\,\text{nm}$ is derived, which will be of relevance for the interpretation of the thermal analyses discussed in Secs.\,\ref{sec:Comparison} and \ref{sec:DiscussionOutlook}.

The two local spectra \textcolor{myRed}{\textbf{1}} and \textcolor{myBlue}{\textbf{2}} shown in Fig.\,\ref{fig:PL}(c) illustrate the difference between the supported epilayer (1) and the freestanding  membrane (2).The emission intensity for these locations differs by approximately a factor of two. 
Since the underetching process (see Sec.\,\ref{sec:Sample}) is selective to the AlN layer, meaning that the InGaN/GaN layer stack remains unaffected, the reduction of intensity cannot be attributed to material damage.
The increase in intensity here is solely associated to the difference in refractive index contrast between AlN and silicon (111) or vacuum. In addition, this observation explains the motivation for a freestanding photonic membrane that requires most light to travel along in-plane directions for commonly applied one-\,\cite{trivino_continuous_2015} and two-dimensional \cite{vico_trivino_high_2012} cavity designs. Less apparently, but highlighted in Fig.\,\ref{fig:PL}(c), the QW emission and the bandedge emission exhibit an energy shift of $\approx\,10\,\pm\,3\, \text{meV}$ (QW, limited by the FWHM) and $4\,\pm\,1\text{meV}$ (bandedge), when comparing positions \textcolor{myRed}{\textbf{1}} and \textcolor{myBlue}{\textbf{2}}. This is a sign of tensile strain in the epilayer grown on silicon (111) due to lattice mismatch \cite{feltin_stress_2001}, which is released upon formation of the photonic membrane. The strain state of our III-nitride epilayer is further quantified based on $\mu$Raman spectroscopy in Sec.\,\ref{sec:Raman}.

After the $\mu$PL mapscan is recorded, the UV heating laser can be positioned at any desired position. We choose the position that is indicated by the purple target in Figs.\,\ref{fig:PL}(a) and \ref{fig:PL}(b), while the positioning precision is given by the stepping size (here 0.5\,$\mu$m). Once the first of the required two lasers for 2LRT$_0$ and 2LRT measurements has reached its position, the power can be varied to induce heating in the sample that will be probed by the second laser. Subsequent to this, the temperature probe laser needs to be positioned (step B) by recording an unheated, non-resonant Raman map as described in the following Sec.\,\ref{sec:Raman}.

\section{\label{sec:Raman}Recording of unheated, non-resonant Raman maps \& positioning of the probe laser (step B) \protect}

%
%
%
\begin{figure*}[]
    \includegraphics[width=\linewidth]{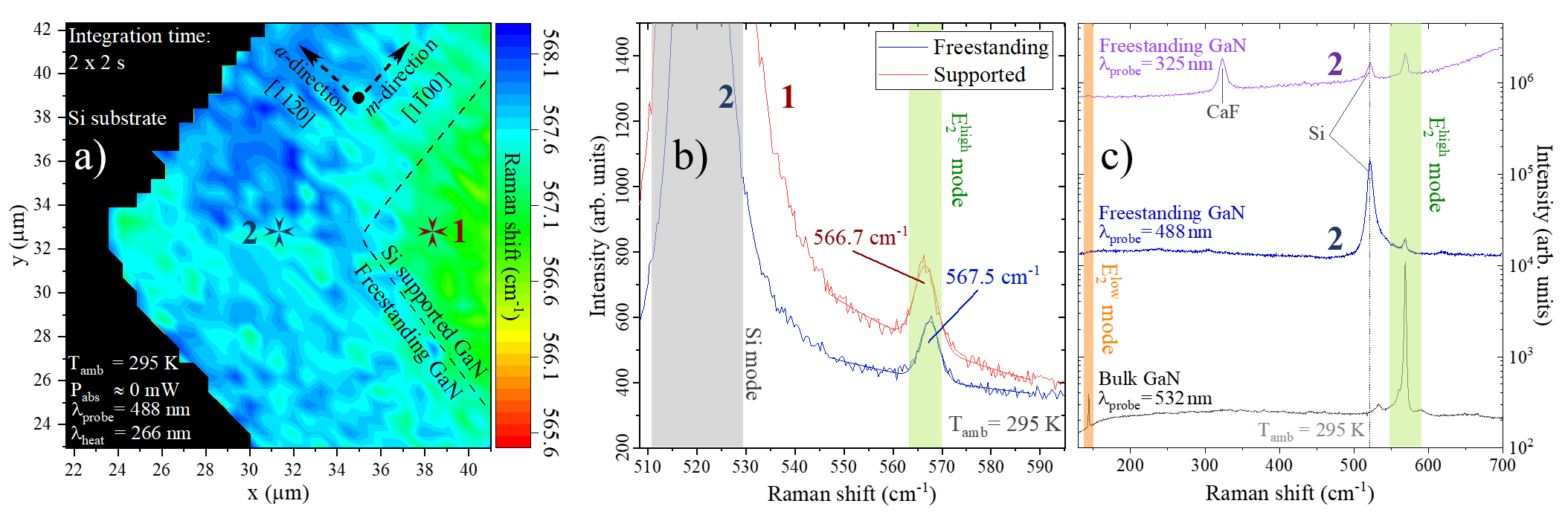}
    \caption{(a) Raman mapscan based on non-resonant Raman spectra, showing the energetic position of the $E_{2}^{\text{high}}$ Raman mode of wurtzite GaN for the corner region of the GaN pad. Two local non-resonant Raman spectra for the positions \textcolor{myRed}{\textbf{1}} and \textcolor{myBlue}{\textbf{2}} are shown in (b). Here, a small $E_{2}^{\text{high}}$ mode shift occurs due to the strain release in the freestanding part of the epilayer. Please note that for experimental reasons, this Raman mapscan is rotated by 135$^{\circ}$ in comparison to the $\mu\text{PL}$ mapscans from Figs.\,\ref{fig:PL}(a) and \ref{fig:PL}(b), showing exactly the same corner of the GaN pad. (c) Three Raman spectra of GaN, from bottom to top: non-resonant Raman spectrum of bulk GaN (black), equivalent non-resonant spectrum for the GaN pad (blue, position 2), and resonant Raman spectrum for the same GaN pad (purple, position 2). Any laser-induced heating was excluded for all Raman spectra by measuring at sufficiently low laser powers. The Raman mode of CaF in the resonant Raman spectrum (purple) originates from the microscope objective in use.}
    \label{fig:Raman}
\end{figure*}

A $\mu$Raman mapscan based on non-resonant Raman spectra ($\lambda_{\text{probe}}=488\,\text{nm}$) is shown in Fig.\,\ref{fig:Raman}(a). Here, exactly the same corner of the GaN membrane that was already probed by $\mu$PL is now analyzed by $\mu$Raman spectroscopy, cf. Fig.\,\ref{fig:PL}. As a result, control over the spatial displacement in between the laser spots of the heating and the probe laser with respect to the position of the structured GaN membrane (corner) is achieved (step B), representing a prerequisite for the Raman thermometry presented in Secs.\,\ref{sec:1LRTand2LRT} and \ref{sec:Heat} (steps C and D). To form the mapscan from Fig.\,\ref{fig:Raman}(a), first, non-resonant Raman spectra are recorded with a step resolution of 650$\,\pm\,50\,\text{nm}$ without any effect on the positioning of the disengaged heating laser that was achieved via the $\mu$PL mapscan from Fig.\,\ref{fig:PL}. Second, the position of the $E_{2}^{\text{high}}$ Raman mode of wurtzite GaN is extracted from the entire set of non-resonant Raman spectra. Plotting these Raman mode positions over the spatial coordinates lead to Fig.\,\ref{fig:Raman}(a). Here, again similar to Figs.\,\ref{fig:PL}(a) and \ref{fig:PL}(b), the freestanding and supported part of the GaN pad can clearly be distinguished as growth on silicon (111) leads to the formation of tensile strain in the entire epilayer, responsible for the additional Raman shift.

Two local spectra were recorded to illustrate the shift, at positions \textcolor{myRed}{\textbf{1}} (supported) and \textcolor{myBlue}{\textbf{2}} (freestanding) as indicated in Figs.\,\ref{fig:Raman}(a) and \ref{fig:Raman}(b). Here, in addition to a strong contribution from the silicon underneath the GaN pad, the $E_{2}^{\text{high}}$ Raman mode of GaN can be seen for the freestanding and supported part of the epilayer. Based on fitting such spectra with a suitable model, a small mode shift of $0.8\,\pm\,0.2\,\text{cm}^{-1}$ can be extracted, cf. Fig.\,\ref{fig:Raman}(b). For such fitting, we employed a Lorentzian peak with an exponential background. Such an exponential background arises from the tail of the silicon Raman mode visible in Fig.\,\ref{fig:Raman}(b).

In general, growth of GaN on silicon (111) leads to the build-up of tensile strain in the epilayer due to lattice mismatch, which can additionally be affected by the incorporation of AlN interlayers that also allow lowering the dislocation density \cite{feltin_stress_2001}. As such dislocations are a source of relaxation, the final tensile strain state of the epilayer is given by the balance between layer thickness and the dislocation density. Based on the bisotropic (i.e, biaxial and isotropic) phonon pressure coefficient of the $E_{2}^{\text{high}}$ Raman mode (2.86\,cm$^{-1}$/GPa) \cite{Callsen2011}, we find a minor stress release in our epilayer upon removal of the silicon (111) substrate of $\approx\,0.3\,\pm\,0.1\,\text{GPa}$. This minor stress release maintains the overall quality of the GaN membrane as, e.g., no cracks or additional structural defects are formed upon release. However, even though the assumption of a bisotropic strain situation is questioned by the particular geometry of the GaN pad corner, the value for the stress release still holds as an approximation. Only minor strain variations that are induced by the corner geometry of the freestanding GaN membrane can be observed in Fig.\,\ref{fig:Raman}(a). Please note that the color scale chosen in Fig.\,\ref{fig:Raman}(a) enables direct comparison with 2LRT mapscans introduced in Sec.\,\ref{sec:Heat}.

\subsection{\label{sec:RamanComparison}Non-resonant and resonant Raman spectra of GaN}

Three different non-resonant and resonant Raman spectra of bulk, \textit{c}-plane GaN and the \textit{c}-plane GaN membrane are shown in Fig.\,\ref{fig:Raman}(c) on a larger energy range that covers all first order Raman modes of GaN except for the longitudinal-optical (LO) modes. For a bulk piece of \textit{c}-plane GaN (bottom, black spectrum), the Raman selection rules \cite{Gil2014a} are obeyed as the $E_{2}^{\text{low}}$ and $E_{2}^{\text{high}}$ Raman modes dominate this spectrum (allowed modes) in the chosen non-polarized, backscattering geometry, cf. Fig.\,\ref{fig:exp_setup}(e). Thus, only minor traces of the transversal-optical (TO) Raman modes $A_{1}(TO)$ and $E_{1}(TO)$ can be observed on the lower wavenumber side of the $E_{2}^{\text{high}}$ Raman mode in addition to a broad background that arises from a defect-related $\mu$PL signal arising from sub-bandgap optical excitation. Analyzing the GaN membrane in position 2, i.e., in the freestanding part of the GaN epilayer, yields the blue and purple spectra shown in Fig.\,\ref{fig:Raman}(c). Despite of the large airgap ($\geq\,8\,\mu\text{m}$), for the middle spectrum, the Raman signal of silicon dominates the signal, while only the $E_{2}^{\text{high}}$ Raman mode of GaN remains visible as again the $\mu$PL background signal limits the dynamic range of the entire Raman spectrum, inhibiting the visibility of the $E_{2}^{\text{low}}$ Raman mode. Thus, only the $E_{2}^{\text{high}}$ Raman mode remains for Raman thermometry based on non-resonant spectra. In general, this situation can be improved by using a microscope objective with a numerical aperture (NA) in excess of 0.55. However, this choice is often not compatible with 2LRT thermometry employing UV heating lasers \cite{Rousseau2018c}. For 2LRT measurements, a dedicated UV-VIS microscope objective is required with a sufficiently large working distance for measurements in a vacuum chamber or cryostat to exclude thermal transport by convection. Furthermore, the focal planes for the two selected wavelengths need to overlap, while providing a sufficiently low curvature for the position mapping schemes for the probe laser spot as described in Sec.\,\ref{sec:Setup}. Consequently, all these experimental requirements hinder the utilization of microscope objectives with higher NA and consequently lower depth of field in our experimental setup.

A similar situation regarding the balance between Raman and $\mu$PL signals occurs for the resonant Raman spectrum shown in the top of Fig.\,\ref{fig:Raman}(c) in purple. As most of the probe light ($\lambda_{probe}\,=\,325\,\text{nm}$) is absorbed in the $\approx\,250\text{-nm-thick}$ membrane (light penetration depth $p^{325\,\text{nm}}_{\text{GaN}}\,=\,74\,\text{nm}$ \cite{kawashima_optical_1997}), the Raman signal of silicon is reduced when switching from the 488$\,$nm laser (non-resonant Raman) to the 325$\,$nm laser (resonant Raman). Please note that every value for $p_{\text{GaN}}$ given in this paper is based on spectroscopic ellipsometry. For this method, a suitable dispersion model is fitted to experimental data and we estimate the corresponding error of $p_{\text{GaN}}$ in the wavelength range of interest with $\pm\,5\%$. A strong luminescence background occurs, which is caused by band-to-band transitions in GaN and additional contributions belonging to the resonant Raman spectrum of GaN (e.g., LO Raman modes and multiple combinations of them). Therefore, resonant Raman spectroscopy on GaN represents a challenge, as often Raman modes are covered by inadvertent signal. In addition, the integration time of the resonant Raman spectrum (purple spectrum), surpasses the integration time of its non-resonant counterpart (red spectrum) in Fig.\,\ref{fig:Raman}(c) by a factor of $\approx\,100$. Clearly, this renders such non-resonant Raman spectra more promising for mapping applications, while also any accidental laser heating is less likely due to the high transparency of the entire GaN membrane at the laser probe wavelength used in this study, cf. Sec.\,\ref{sec:Sample}. 

\section{\label{sec:1LRTand2LRT}One- and two-laser Raman thermometry (step C) \protect}

After full positioning control is achieved for the probe and heat laser spots in regard to the corner of the GaN membrane based on steps A and B from Sec.\,\ref{sec:PL} and \ref{sec:Raman}, the 1LRT and 2LRT$_0$ measurements are performed at the same sample location. The position of the probe and heat laser spots coincides for 2LRT$_0$ measurements. In the spirit of a step-by-step approach, the 2LRT$_0$ measurements serve first as a direct comparison to the common 1LRT technique, before the introduction of spatial resolution through 2LRT measurements presented in Sec.\,\ref{sec:Heat}.

Generally, Raman spectra enable at least three methods to measure the temperatures that are required for Raman thermometry. One can either measure i) the Raman mode shift, ii) the Raman mode broadening, or iii) the intensity ratio of Stokes and Anti-Stokes Raman modes upon variations of the power of the heating laser. As approach iii) typically results in uncertainties of the measured temperature that are significantly larger compared to i) and ii) as described in Refs. \cite{beechem_micro-raman_2008,beechem_invited_2015}, we will discard an analysis of the Stokes/Anti-Stokes intensity ratios in this work. In general, an analysis of these Stokes/Anti-Stokes intensity ratios is not particularly promising to map temperatures, as we found that the required integration times exceed the integration times of the Raman spectra that exclusively focus on the Stokes side by a factor of $\sim\,100$ in our experimental setup around $T_{amb}$.

\subsection{\label{sec:1LRT}One-laser Raman thermometry on GaN}

%
%
%
\begin{figure}[]
    \includegraphics[width=\linewidth]{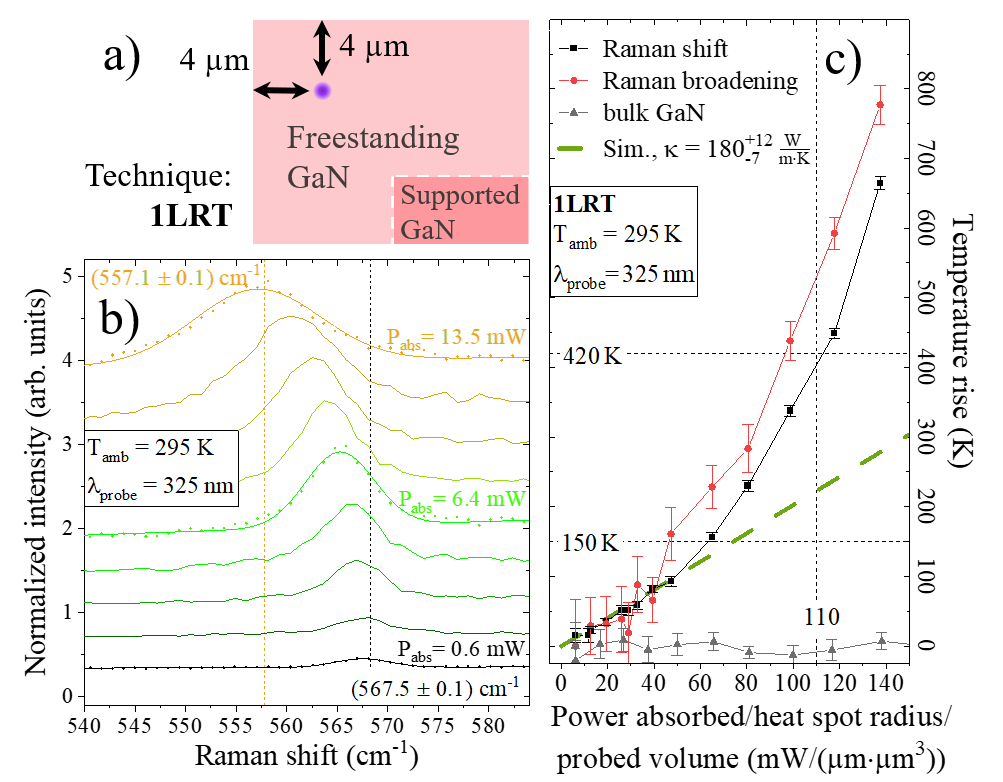}
    \caption{(a) Scheme of the 1LRT measurements showing the position of the heating laser spot (purple) on the corner of the GaN membrane. (b) Resonant Raman spectra showing the $E_{2}^{\text{high}}$ Raman mode of GaN with rising power of the heating laser ($\lambda_{heat}=\lambda_{probe}=325\,\text{nm}$) at an ambient temperature $T_{amb}\,=\,295\,\text{K}$. (c) Temperature rise induced by the heating laser plotted over its normalized absorbed power ($P_{abs}$). The reasoning behind this normalization of $P_{abs}$ is given in the main text. Here, the Raman mode shift and its broadening yield different temperature rises ($T_{rise}$) due to the build-up of thermally induced strain in the GaN membrane. For comparison, the corresponding $T_{rise}$ values for bulk GaN are also shown. The dashed green line illustrates our modeling for $T_{rise}\,\lesssim\,150\,\text{K}$ based on a  thermal conductivity $\kappa^{\,1LRT}_{\,\,memb}\,=\,180_{-7}^{+12}\,$\Wu. For more elevated powers of the heating laser, both temperature rises follow parabolas, as $\kappa$ continuously decreases. The dashed horizontal and vertical lines enable the comparison to the 2LRT$_0$ data, cf. Fig.\,\ref{fig:2LRT_wod}}
    \label{fig:1LRT}
\end{figure}

One-laser Raman thermometry is based on resonant Raman spectroscopy. The heating laser spot is positioned at the target position in the center of the freestanding GaN pad corner as sketched in Fig.\,\ref{fig:1LRT}(a) and the wavelength utilized for heating and temperature probing is the same ($\lambda_{heat}=\lambda_{probe}=325\,\text{nm}$). Resulting resonant Raman spectra are displayed in Fig.\,\ref{fig:1LRT}(b) for rising absorbed laser power $P_{\text{abs}}$. The $E_{2}^{\text{high}}$ Raman mode continuously shifts and broadens towards lower wavenumbers, indicating a local, laser-induced heating. The corresponding Raman mode positions and FWHM values are extracted by fitting the peaks with a Lorentzian peak over a linear background, as illustrated for three $P_{\text{abs}}$ values in Fig.\,\ref{fig:1LRT}(b) based on overlaying the experimental result (datapoints) with a solid line (fitting model). Position and FWHM values related to the $E_{2}^{\text{high}}$ Raman mode can then be translated into local temperatures based on an equally local temperature calibration. Details regarding this local temperature calibration are given in S-Sec.\,III. The resulting temperature rise ($T_{rise}$) values, either based on Raman mode position or FWHM values, are plotted in Fig.\,\ref{fig:1LRT}(c). Here, the horizontal axis represents $P_{abs}$ normalized by the product of the heat spot radius $r_{\text{abs}}$ and the temperature probe volume to enable direct comparison to the 2LRT$_0$ data presented in Fig.\,\ref{fig:2LRT_wod} from Sec.\,\ref{sec:2LRT}, where we employ a heating laser with $\lambda_{heat}\,=\,266\,\text{nm}$. The additional normalization by $r_{\text{abs}}$ originates from the approximation that $T_{rise} \sim \frac{P_{\text{abs}}}{r_{\text{abs}}}$ holds \cite{lax_temperature_1977}. The spot sizes of the probe and heat lasers are determined via the knife edge method (S-Sec.\,I\,A), and the probe volume is approximated by the product of the probe laser spot size and its penetration depth.

Up to $T_{rise}\,\approx\,150\,\text{K}$, the temperature rises originating from Raman mode positions and FWHM values overlap within the error bars. However, upon further increase in laser power, the FWHM of the $E_{2}^{\text{high}}$ Raman mode systematically indicates higher $T_{rise}$ values. As described by Beechem \textit{et\,al.} \cite{beechem_micro-raman_2008}, such discrepancy is caused by the built-up of thermally induced strain in the epilayer due to the local laser heating. 
As the heated volume tends to expand in the middle of the corner of the GaN membrane, the surrounding colder material hinders this expansion, mostly, for in-plane directions. The result is thermally induced compression in the plane of the GaN membrane around the heat spot, which is accompanied by a shift of the $E_{2}^{\text{high}}$ Raman mode to higher wavenumbers \cite{Callsen2011}, thus, counteracting its evolution towards lower wavenumbers due to laser-induced local heating. The temperature induced evolution of a Raman mode is not only caused by a volume effect (i.e. the thermal expansion), but also by phonon-phonon scattering events \cite{li_temperature_2000} that scale with the phonon population. Eventually, the Raman mode position tends to become a poor temperature sensor for $T_{rise}\,>\,150\,\text{K}$ in GaN at $T_{amb}\,=\,295\,\text{K}$, while the FWHM values are not affected by thermally-induced stress as a first order approximation \cite{beechem_micro-raman_2008,Callsen2011}. The quantification of the stress that is thermally induced by the local laser heating for $T_{rise}\,>\,150\,\text{K}$ of the GaN membrane remains a task for future work.

Consequently, $T_{rise}$ values extracted from the evolution of the FWHM values of the $E_{2}^{\text{high}}$ Raman mode are best suited for determining the thermal conductivity of the GaN membrane $\kappa^{\,1LRT}_{\,\,memb}$. For the given normalization of $P_{abs}$ in Fig.\,\ref{fig:1LRT}(c), the evolution of $T_{rise}$ should follow a linear function for a fixed $\kappa^{\,1LRT}_{\,\,memb}$ value. However, for $T_{rise}\,>\,150\,\text{K}$ the temperature trend follows a parabola, indicating a reduction of $\kappa^{\,1LRT}_{\,\,memb}$ with increasing $T_{rise}$ at the given $T_{amb}\,=\,295\,\text{K}$. Commonly, it is assumed that such an evolution of $\kappa^{\,1LRT}_{\,\,memb}$ can be described by a power law like $\kappa(T)=aT^{-b}$ \cite{graczykowski_thermal_2017} with the fitting parameters $a$ and $b$. This approach has already allowed the description of the decrease of the thermal conductivity $\kappa$ with increasing temperature in bulk silicon, bulk germanium \cite{glassbrenner_thermal_1964}, and silicon membranes \cite{graczykowski_thermal_2017}. 

However, in this work, we will not focus on this particular dependence of $\kappa(T)$. The reasoning for this is mostly related to the determination of the two fitting parameters $a$ and $b$. These need to be obtained from extended numerical simulations and fitting efforts beyond the scope of this work, as we do consider a more complex sample geometry, i.e., the corner of a GaN membrane and will constitute the basis of further investigations. In this work, we focus on the low $T_{rise}$ limit ($T_{rise}\,\lesssim\,150\,\text{K}$) to approximate a constant thermal conductivity $\kappa_{\,\,memb}$ for our GaN membrane, which we derive based on 1LRT and 2LRT$_{0}$, as well as 2LRT measurements. Already the comparison of these three values for $\kappa_{\,\,memb}$ bears interesting physical insight as described in Sec.\,\ref{sec:Comparison}.

\subsubsection{\label{subsubsec:ExtractKappa}Determination of the thermal conductivity}

As no straightforwardly accessible analytical solution exists for modeling the heating laser induced temperature distribution in the corner region of our GaN membrane \cite{carslaw_conduction_1959}, we directly turn towards finite element simulations based on the COMSOL Multiphysics\raisebox{0.2cm}{{\tiny\textcircled{R}}} software, which were already successfully applied to InP microcrystals with dimensions in the few- and sub-micrometer regime \cite{jaramillo-fernandez_raman_2018}. We operate COMSOL through scripts in MATLAB\raisebox{0.2cm}{{\tiny\textcircled{R}}} using LiveLink\raisebox{0.2cm}{{\tiny TM}}. In doing so, we enable the analysis of a large amount of simulations and are able to compute errors for all $\kappa$ values we derive. The procedure for extracting $\kappa^{\,1LRT}_{\text{\,\,memb}}$ based on the 1LRT data from Fig.\,\ref{fig:1LRT}(c) is described as follows. Details regarding the thermal model and the numerical simulations can be found in S-Sec.\,IV.

\hfill

\textbf{Step I - Experimental data analysis:}
We start by fitting a linear function to the datapoints from Fig.\,\ref{fig:1LRT}(c) for $T_{rise}\,\lesssim\,150\,\text{K}$. The least square regression is restricted to this range of $T_{rise}$ values for two reasons: First, here the mode shift and the mode broadening can be applied as valid equivalent thermometers. Second, limiting the fitting to lower values of $T_{rise}$ motivates the assumption of a constant $\kappa_{memb}$ value. As a result, we obtain the slope of T$_{rise}^{exp}$(P$_{abs})$ in addition to the symmetric slope error. 

\hfill

\textbf{Step II - Modeling the experiment:}
The numerical simulation aims to model the laser-induced heating, the thermal dissipation in the sample, and the temperature probing by a laser. To this end, the entire GaN pad along its silicon support is created inside of the simulation software with a size of 220\,$\mu$m\,$\times$\,220\,$\mu$m\,$\times$\,10.25\,$\mu$m. The heating laser is introduced as a surfacic Gaussian heating source with an integrated power density equal to P$_{abs}$, while the bottom of the silicon support acts as the heat sink remaining at ambient temperature (295\,K). When comparing a surfacic heating source with a volumetric heating source that considers the light penetration depth of the heating laser ($p_{abs}$), we found a maximal temperature deviation of less than 0.5\%. For the temperature probe volume, we define a cylindrical volume over which temperatures are averaged to $T_{rise}$, according to Eq.\,\ref{eq:avTemp}.
Here, the height of the cylinder in the cross-plane direction is given by the light penetration depth of the probe laser ($p_{abs}$). More details are provided in S-Sec.\,IV.
For the case of 1LRT measurements ($\lambda_{heat}\,=\,\lambda_{probe}\,=\,325\,\text{nm}$ and $p_{abs}\,=\,p_{probe}\,=\,74\,\text{nm}$) \cite{kawashima_optical_1997}, the heating and probe lasers are identical, hence, they are modeled with the same parameters. For the in-plane direction $T_{rise}$ is obtained from a Gaussian-weighted average of temperatures, considering the probe laser spot radius ($r_{probe}$) and its centering around the heat spot. Subsequently, the simulation is performed for all experimental $P_{abs}$ values, yielding a linear trends T$_{rise}^{sim}$(P$_{abs})$. In addition, these simulations are repeated for a suitable range of $\kappa^{\,1LRT}_{\,\,memb}$. The spacing between two values of $\kappa_{memb}^{1LRT}$ is mainly determined by the slope error of the linear fit function obtained in step I.

\hfill

\textbf{Step III - Deducing the thermal conductivity:} 
We deduce $\kappa_{memb}^{1LRT}$ by comparing the slopes of the simulated and measured trends for T$_{rise}$(P$_{abs})$. 
Once the best matching simulation is found, the error of $\kappa_{memb}^{1LRT}$ is determined via the experimental slope error obtained under step I.
For a given slope and slope error of T$_{rise}^{exp}$(P$_{abs})$, one obtains the lower bound ($\kappa_{memb}^{1LRT}\,-\, \Delta \kappa_1$) and the upper bound ($\kappa_{memb}^{1LRT} \,+\, \Delta \kappa_2$) of the thermal conductivity with $\Delta \kappa_1 \neq \Delta \kappa_2$.
Hence, we adopt the following notation for asymmetric errors: $\kappa_{-\Delta\kappa_1}^{+\Delta\kappa_2}$ \cite{barlow_asymmetric_2006}.
These asymmetries on the error intervals can be understood based on the standard analytical relation for bulk material that relates $T_{rise}$ to the inverse of $\kappa^{\,1LRT}_{\,\,bulk}$ as follows: $\frac{T_{rise}}{P_{abs}}\,\propto\,1/(r\,\cdot\,\kappa^{\,1LRT}_{\,\,bulk})$. Here, the precise prefactor of this proportionality depends on the experimental conditions as described by Lax \cite{lax_temperature_1977}.

\hfill

As a result of our modeling procedure  based on step I\,-\,III, we derive $\kappa^{\,1LRT}_{\,\,memb}\,=\,180_{-7}^{+12}\,\text{W/mK}$, illustrated by the linear function (dashed, green) shown in Fig.\,\ref{fig:1LRT}(c). This value of $\kappa^{\,1LRT}_{\,\,memb}$ that we extracted here for our thin photonic membrane, comprising around 200\, nm of GaN and a 50-nm-thick AlN interlayer, is highly doubtful. Only GaN samples of very high crystalline quality (low structural and point defect concentrations) with a thickness of several micrometers grown on bulk GaN substrates reach such high $\kappa$ values \cite{beechem_size_2016}. A careful analysis regarding the reliability of $\kappa^{\,1LRT}_{\,\,memb}$ follows in Sec.\,\ref{sec:Comparison} along with a comparison to the results of 2LRT$_0$ and 2LRT measurements. Here, we will reveal a systematic problem of 1LRT measurements for the given heating and probe laser spot sizes and laser penetration depths. As a precise knowledge of $P_{abs}$ is of high importance for all 1LRT and 2LRT measurements, the following Sec.\,\ref{subsubsec:Pabs} is dedicated to its determination. Furthermore, also the case of 1LRT measurements on bulk GaN is briefly shown in Sec.\,\ref{subsubsec:1LRTonBULK}, before 2LRT measurements are introduced in Sec.\, \ref{sec:2LRT}, following a rigorous step-by-step approach.

\subsubsection{\label{subsubsec:Pabs}Determination of the absorbed and the heating power}

Determining a robust value for the power absorbed by the layer $P_{abs}$ and its fraction that goes into heating $P_{heat}$ is of high relevance for the Raman thermometry presented in this work. Generally, for the acquisition of Raman spectra, we measure the power of the heating laser between the two scanning lenses illustrated in Fig.\,\ref{fig:PL}(e). This position for the power measurements proved most convenient during the operation of the experimental setup. However, the power of the heating laser that reaches the surfaces of our GaN pad can only be derived based on a precise knowledge of all losses introduced by the optical elements in the beampath (lenses, mirrors, microscope objective, cryostat window). Additionally, the power of the heating laser that is reflected from the surface of the photonic membrane is determined to provide the power of the heating laser entering into the sample. In our experimental setup, the power of the reflected heating laser $P_{refl}$ can directly be measured for every step in power of the heating laser. We then cross-checked this method by both direct, angular-dependent reflection measurements at $\lambda_{heat}\,=\,325\,\text{nm}$ and $266\,\text{nm}$ in a dedicated and calibrated UV-VIS spectrophotometer and by straightforward calculations based on Fresnel equations. Even for elevated excitation powers of the heating laser, we never observed any significant thermally induced change in reflectivity. Any reflectivity changes we measured were always directly linked to surface damage of the GaN material that only occurred at excitation powers beyond the values presented in this work. Any power that is attributed to elastic (Rayleigh scattering) and inelastic (Brillouin and Raman scattering) light scattering $P_{scat}$ is neglected in this work as in previous works on 2LRT \cite{reparaz_novel_2014,graczykowski_thermal_2017}. Commonly, inelastic scattering is orders of magnitude weaker than elastic scattering, while the latter is strongly reduced in our sample due to its monocrystallinity.

After the determination of the power entering the GaN pad, its transmitted fraction must be estimated to find the absorbed laser power $P_{abs}$. For the sake of simplicity, we treat the GaN/InGaN/GaN membrane as one effective GaN membrane with a thickness of 203\,nm (3-nm-thick InGaN QW). At such GaN layer thicknesses, only 0.5\% (325\,nm) or 0.3\% (266\,nm) of the light of the heating laser are transmitted through the membrane. Thus, even though the backside of our sample is not experimentally accessible as in measurements on silicon membranes \cite{reparaz_novel_2014}, we can safely assume that the power of the heating laser that enters the sample is also fully absorbed, enabling the direct determination of $P_{abs}$.

Furthermore, it shall be noted that the absorption of GaN at 325\,nm or 266\,nm is so high \cite{Muth1997a} that Fabry-Perot interference effects can be neglected for our GaN membrane despite its sub-wavelength thickness. In contrast, for 1LRT measurements on silicon membranes such interference effects can complicate the calculation of the absorbed power, rendering direct transmission measurements at the wavelength of the heating laser of high importance \cite{chavez-angel_reduction_2014}. This is also especially relevant for 2LRT measurements on silicon membranes that use visible lasers for the heating (405\,nm) and the temperature probing (488\,nm) as demonstrated in Ref.\,\cite{reparaz_novel_2014}. At such wavelength in the visible spectral range, the absorption of silicon does not suffice to suppress light propagation for in-plane directions in silicon membranes. Depending on the thickness of such silicon membranes, Fabry-Perot interference effects can occur that strongly alter the overall membrane transmission. Such silicon membranes therefore require precise transmission measurements when heated with visible lasers, which in turn requires full optical access from the front- and backside of the membrane. Alternatively, a suitable modeling can be applied \cite{chavez-angel_reduction_2014}, however, this approach requires precise knowledge about the top and bottom surface roughnesses.
\begin{figure}[]
    \includegraphics[width=\linewidth]{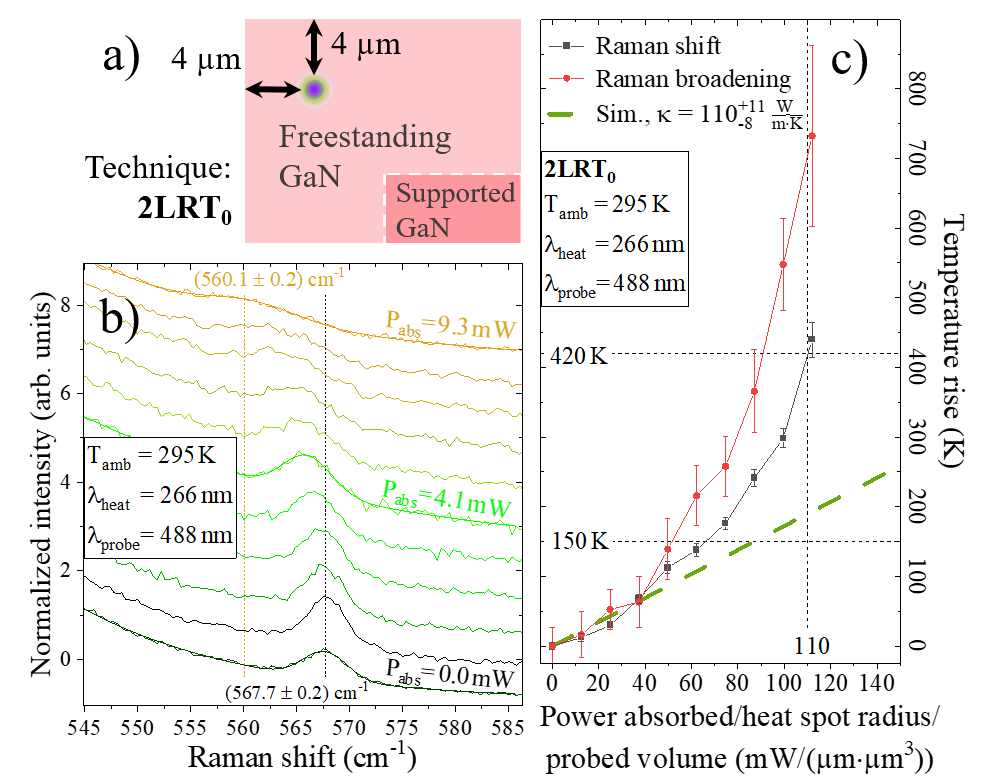}
    \caption{(a) Scheme of the 2LRT$_0$ measurements without a spatial displacement in between the heating laser spot (purple) and the probe laser spot (green), both positioned in the corner region of the GaN pad. (b) Non-resonant Raman spectra showing the $E_{2}^{\text{high}}$ Raman mode of GaN with rising power of the heating laser ($\lambda_{heat}\,=\,266\,\text{nm}$) at an ambient temperature $T_{amb}\,=\,295\,\text{K}$. Here a probe laser wavelength $\lambda_{probe}\,=\,488\,\text{nm}$ was used. (c) Temperature rise induced by the heating laser plotted over normalized absorbed laser power ($P_{abs}$). Here, the given normalization of $P_{abs}$ allows direct comparison of this plot with Fig.\,\ref{fig:2LRT_wod}(c). The Raman mode shift and its broadening yield different temperature rises due to the build-up of thermally induced strain in the GaN membrane. The dashed green line illustrates our modeling for $T_{rise}\,\lesssim\,150\,\text{K}$ and a fixed thermal conductivity yielding $\kappa^{\,2LRT_{0}}_{\,\,memb}\,=\,110_{-8}^{+11}\,\text{W/mK}$. For more elevated powers of the heating laser, both temperature rises follow parabolas, as $\kappa$ continuously decreases. The intersection of the dashed horizontal and vertical lines enable the comparison to the 1LRT data, cf. Fig.\,\ref{fig:1LRT}.}
    \label{fig:2LRT_wod}
\end{figure}
The last step is to determine which fraction of $P_{abs}$ leads to a local heating of the GaN membrane. In principle, one can assume that a certain fraction of $P_{abs}$ leads to the formation of, e.g., excitons and/or electron-hole pairs, which in turn can emit light with a radiated power $P_{rad}$. Clearly, such light emission is one of the main aims of a photonic membrane comprising a QW as built-in light source. In the best case scenario, the external quantum efficiency $\eta_{EQE}$ of the GaN membrane would be available to evaluate the following equality: $P_{abs}\,=\,P_{heat}\,+\,P_{rad}$, with $P_{heat}$ being the fraction of the absorbed power that went into heating. 

In a light emitting diode $\eta_{EQE}$ is defined by the number of photons emitted to free space per second divided by the number of injected electrons per second. However, despite the numerous $\mu$PL results shown in Fig.\,\ref{fig:PL}, $\eta_{EQE}$ of the present sample is low at $T_{amb}\,=\,295\,K$. In Ref.\,\cite{Rousseau2018c} time-resolved PL measurements were performed between 5\,K and 295\,K to measure the internal quantum efficiency $\eta_{IQE}$ of exactly the GaN membrane structure used for the present work. Here $\eta_{IQE}$ is again defined in a light emitting diode by the number of photons emitted from the active region per second divided by the number of injected electrons per second. Consequently, the time-resolved PL analysis yielded $\eta_{EQE}$ < $\eta_{IQE}\,\approx\,0.1\%$ for $T_{amb}\,=\,295\,\text{K}$. Thus, we can safely assume $P_{abs}\,\approx\,P_{heat}$ for the present sample at room temperature and beyond. However, this situation would change, if the sample was cooled towards, e.g., 5\,K or if in general a sample with less non-radiative defects would be analyzed, which remains a task for future work. Adding an InGaN underlayer to our structure grown on silicon as described by Haller \textit{et\,al.} would already drastically boost $\eta_{IQE}$ at room temperature \cite{haller_burying_2017}. Nevertheless, already the present sample's PL signal rises by up to four orders of magnitude when transitioning from 295\,K to 5\,K, which is accompanied by $\eta_{IQE}$ reaching several tens of percent at 5\,K.

\subsubsection{\label{sec:Tmeas}Experimental determination of the temperature}
The temperature derived from our experiments are either based on the Raman mode shift or broadening, which are in turn acquired from every point under the laser spot. Strictly speaking, the technique does not probe the local temperature but its average under the laser spot, weighted by the beam profile intensity and its penetration depth \cite{beechem_invited_2015,ocenasek_raman_2015}:

\begin{equation} \label{eq:avTemp}
    T_{meas} = \int_V\,dr\,d\theta\,dz \,\,r\cdot T(r,\,\theta,\,z) \cdot f_{probe}(r,\,\theta,\,z)
\end{equation}

with $f_{probe}(r,\,\theta,\,z)$ being the weighting function of the probe laser and $T(r,\,\theta,\,z)$ the local temperature of the sample. The weight function $f_{probe}(r,\,\theta,\,z)$ can be written as product of the probe laser beam profile (assumed as a Gaussian) and the attenuation of the back-scattered probe laser beam:
\begin{equation}
    f_{probe}(r,\,\theta,\,z) = \frac{1}{w_e\sqrt{\pi}}\cdot e^{-\frac{(r-r_0)^2}{w_e^2}} \cdot e^{-\frac{2z}{p_{probe}}}
\end{equation}
with $w_e$ the beam waist as the beam intensity is reduced to $\frac{1}{e}$ of its initial value, $r_0$ the location of the probe beam, $p_{probe}$ the penetration depth of the laser, and $r, \,\theta$, and $z$ a cylindrical coordinate system centered at the probe beam spot location. The factor of $2$ in the attenuation term accounts for the probe beam travel to depth $z$ into the material and its return to the material surface. 
To extract the coefficient of thermal conductivity $\kappa$, we compare $T_{meas}$ to numerical simulations. As the outcome of such simulations is usually the simulated local temperature $T_{sim}(r,\,\theta,\,z)$, it needs to be averaged by a function that models the probe laser:  
\begin{equation}
    f_{probe}^{sim}(r_{sim},\,\theta_{sim},\,z_{sim}) = \frac{1}{w_e\sqrt{\pi}}\cdot e^{-\frac{(r_{sim}-r_0)^2}{w_e^2}} \cdot e^{-\frac{2z_{sim}}{p_{probe}}}
\end{equation}
with $r_{sim},\,\theta_{sim}$ and $z_{sim}$ being the cylindrical \textit{simulated} coordinates, to distinguish from the physical ones. $w_e$ and $r_0$ are measured and inserted in the model and $p_{probe}$ comes from Ref.\,\cite{kawashima_optical_1997}. 

\subsubsection{\label{subsubsec:1LRTonBULK}One-laser Raman thermometry on bulk GaN}

1LRT measurements are frequently reported in literature and were, e.g., applied to bulk germanium \cite{jaramillo-fernandez_raman_2018}, bulk silicon \cite{stoib_spatially_2014}, and related membranes \cite{chavez-angel_reduction_2014}. However, performing such measurements on bulk GaN already represents a challenge for the reasons given in Sec.\,\ref{sec:RamanComparison}. Nevertheless, we recorded the required resonant Raman spectra of bulk GaN with $\lambda_{heat}=\lambda_{probe}=325\,\text{nm}$, showing the $E_{2}^{\text{high}}$ Raman mode of GaN, cf. S-Sec.\,IV. Even when raising the laser power to the maximum laser power available, no Raman mode shift or Raman mode broadening as a sign of local heating occurred. The corresponding temperatures are shown for bulk GaN in Fig.\,\ref{fig:1LRT}(c). Thus, the thermal conductivity of high quality bulk GaN $\kappa_{bulk}$ cannot be measured by 1LRT in our setup as described in S-Sec.\,IV. Even a further increase of the power of the heating laser does not represent a promising pathway for 1LRT measurements on bulk GaN with very high thermal conductivities, as also any heating laser-induced material damage must always be avoided.

\subsection{\label{sec:2LRT}Two-laser Raman thermometry on GaN}
%
%
%
\begin{figure*}[]
    \includegraphics[width=\linewidth]{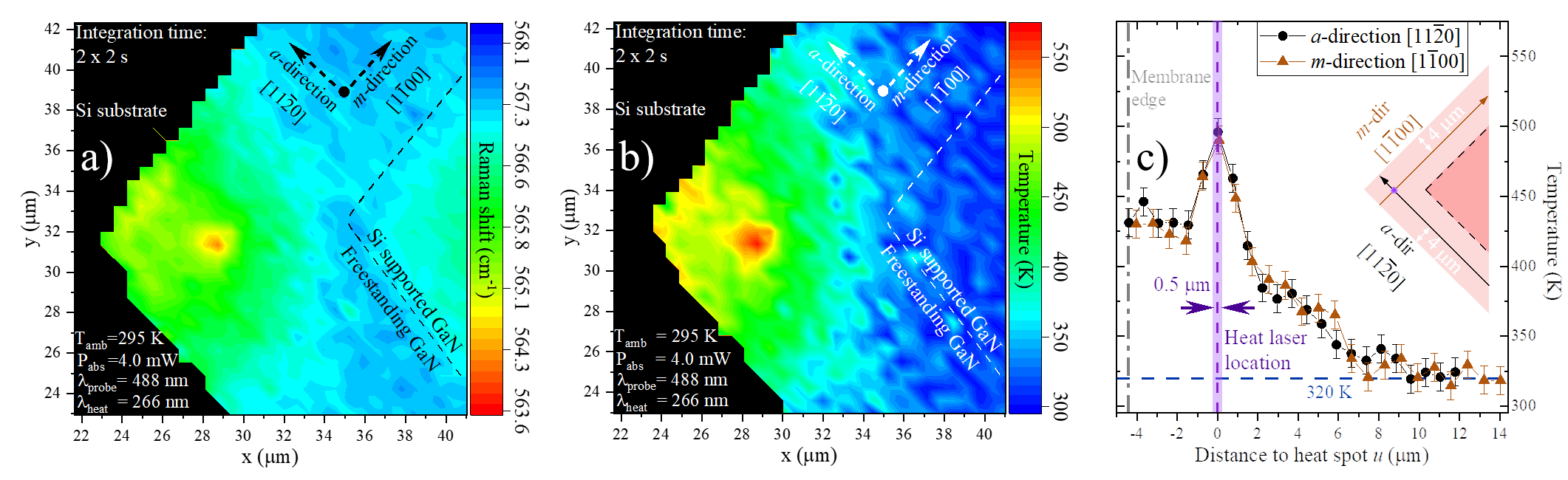}
    \caption{(a) Raman map illustrating the position of the $E_{2}^{high}$ Raman mode extracted from non-resonant Raman spectra ($\lambda_{probe}\,=488\,\text{nm}$) at an ambient temperature of 295\,K. The effect of the simultaneously engaged heating laser ($\lambda_{probe}\,=266\,\text{nm}$) can directly be observed close to the corner of the GaN membrane. There, an additional Raman mode shift can be seen, which is induced by the heating laser. See Fig.\,\ref{fig:Raman}(a) for the same, but non-locally heated Raman map. (b) Based on directly the unheated Raman map from the locally heated map, one can derive a temperature map based on a suitable calibration (see main text). (c) Cut lines showing the evolution of the temperature rise along the \textit{a}- (circles) and \textit{m}-direction (triangles) in the \textit{c}-plane of the GaN membrane. Here, the sample edge as well as the location and spot size of the heating laser are illustrated.}
    \label{fig:2LRT}
\end{figure*}

The results of 2LRT$_0$ measurements without a spatial displacement are shown in Fig.\,\ref{fig:2LRT_wod}. Again the heating laser (purple, $\lambda_{heat}\,=\,266\,\text{nm}$) is positioned at the same position as for the 1LRT measurements, cf. Fig.\,\ref{fig:1LRT}(a). However, now this heating laser is overlapped with the probe laser (green, $\lambda_{probe}\,=\,488\,\text{nm}$) as sketched in Fig.\,\ref{fig:2LRT_wod}(a). Upon varying the power of the heating laser, non-resonant Raman spectra are recorded with the probe laser and are shown for the wavenumber range around the $E_{2}^{high}$ Raman mode in Fig.\,\ref{fig:2LRT_wod}(b). Again a suitable fitting model is applied to the Raman spectra to extract the position and FWHM of the $E_{2}^{\text{high}}$ Raman mode of GaN with rising power of the heating laser. A Lorentz peak with an exponential background is best suited for fitting the $E_{2}^{high}$ Raman mode in close energetic vicinity of the omnipresent Raman mode of silicon. Resulting fitting functions are illustrated in Fig.\,\ref{fig:2LRT_wod}(b) for selected powers of the heating laser.

From the Raman mode positions and FWHM values, two local temperatures can be extracted based on our local temperature calibration, cf. S-Sec.\,III. Similar to the results of the 1LRT measurements, the two trends for the local temperatures shown in Fig.\,\ref{fig:2LRT_wod}(c) start to deviate from each other at a $T_{rise}$ value of about 150\,$\text{K}$. Our subsequent method to extract the thermal conductivity from our 2LRT$_0$ measurements $\kappa^{\,2LRT_0}_{\,\,memb}$ is similar to our evaluation of the 1LRT measurements described in Sec.\,\ref{sec:1LRT} based on steps I\,-\,III. The only major difference is now related to the heat and probe volume due to the different wavelengths of the heating and probe lasers. As the GaN membrane is transparent to the probe laser, the corresponding temperature probe volume for our simulations now corresponds to a cylinder with a height of 203\,nm and a radius of 600\,nm, which includes the QW for sake of simplicity. See S-Sec.\,IV for details regarding the modeling.

For $T_{rise}\,\lesssim\,150\,\text{K}$ we derive $\kappa^{\,2LRT_0}_{\,\,memb}\,=\,110_{-8}^{+11}\,\text{W/mK}$ as illustrated by the linear function (dashed, green) illustrated in Fig.\,\ref{fig:2LRT_wod}(c). Again, here the sub- and superscript of this value denote the lower and upper bound of the asymmetric error interval. The reliability of this value for $\kappa^{\,2LRT_0}_{\,\,memb}$ and its obvious discrepancy to the higher value of $\kappa^{\,1LRT}_{\,\,memb}\,=\,180_{-7}^{+12}\,\text{W/mK}$ will be discussed in Sec.\,\ref{sec:Comparison}. As the increase in the temperature probe volume will be key to an understanding of $\kappa^{\,2LRT_0}_{\,\,memb} < \kappa^{\,1LRT}_{\,\,memb}$, it is promising to further increase the temperature probe volume by scanning the probe laser across the focus of the heating laser during 2LRT measurements as described in the following Sec.\,\ref{sec:Heat}. 

Please note that the slope of the green dashed line (simulated $T_{rise}$ values for a fixed $\kappa$) in Fig.\,\ref{fig:1LRT}(c) is steeper than its equivalent in Fig.\,\ref{fig:2LRT_wod}(c). This could appear counter-intuitive as in general one expects higher values of the thermal conductivity $\kappa$ to result in smaller temperature rises. However, the slope additionally depends on the heat laser spot radius and the temperature probe volume. The heat laser spot radius is of 600\,nm for the 1LRT measurements from Fig.\,\ref{fig:1LRT}(c), which surpasses the corresponding value for the 2LRT$_0$ measurements (250\,nm), which explains the aforementioned slope discrepancy. See Sec.\,\ref{sec:1LRT} for details regarding the normalization of the \textit{x}-axis in Figs.\,\ref{fig:1LRT}(c) and \ref{fig:2LRT_wod}(c).

\section{\label{sec:Heat}Spatially resolved two-laser Raman thermometry (step D) \protect}

To derive spatially resolved temperature mapscans by 2LRT measurements, it is necessary to displace the probe laser ($\lambda_{probe}\,=488\,\text{nm}$) from the heating laser ($\lambda_{heat}\,=266\,\text{nm}$). One then obtains Raman mapscans showing, e.g., the position of the $E_{2}^{high}$ Raman mode extracted from non-resonant Raman spectra, depending on the spatial coordinates as shown in Fig.\,\ref{fig:2LRT}(a). Here, the position of the focus spot of the heating laser is clearly visible close to the corner of the membrane, as locally an additional Raman mode shift towards lower wavenumbers is induced. This additional Raman mode shift was missing in the unheated Raman mapscan from Fig.\,\ref{fig:Raman}(a). By subtracting the locally heated Raman mapscan in Fig.\,\ref{fig:2LRT}(a) from this unheated Raman mapscan in Fig.\,\ref{fig:Raman}(a), one can now determine the spatial distribution of the Raman mode shift that is exclusively induced by the local heating. By applying a suitable local temperature calibration (see S-Sec.\,III), we can translate these Raman mode shifts to temperatures as shown in Fig.\,\ref{fig:2LRT}(b). Strain variations that induced Raman mode shifts in the not locally heated Raman mapscan are eliminated by this approach. This is observable in Fig.\,\ref{fig:2LRT}(a), when looking at the area enclosed by a black dashed line. Here, the area of fully supported GaN is visible due to the occurrence of tensile strain. Exactly this area is no longer visible in the temperature mapscan of Fig.\,\ref{fig:2LRT}(b) due to the subtraction of mapscans. Please note that the zig-zag shape of the upper edge of the corner shown in this temperature map is exclusively caused by the step resolution in our Raman mapscans of $\approx\,650\,\text{nm}$. From a temperature mapscan as shown in Fig.\,\ref{fig:2LRT}(b), one can now extract temperature profiles for any in-plane direction. Corresponding temperature profiles are shown in Fig.\,\ref{fig:2LRT}(c) for the \textit{a}- and \textit{m}-directions in the \textit{c}-plane of the GaN membrane. Exactly such cut lines through a temperature mapscan resulting from 2LRT measurements can form the basis for extracting the thermal conductivity for all in-plane directions.

\begin{figure}[]
    \includegraphics[width=8.0cm]{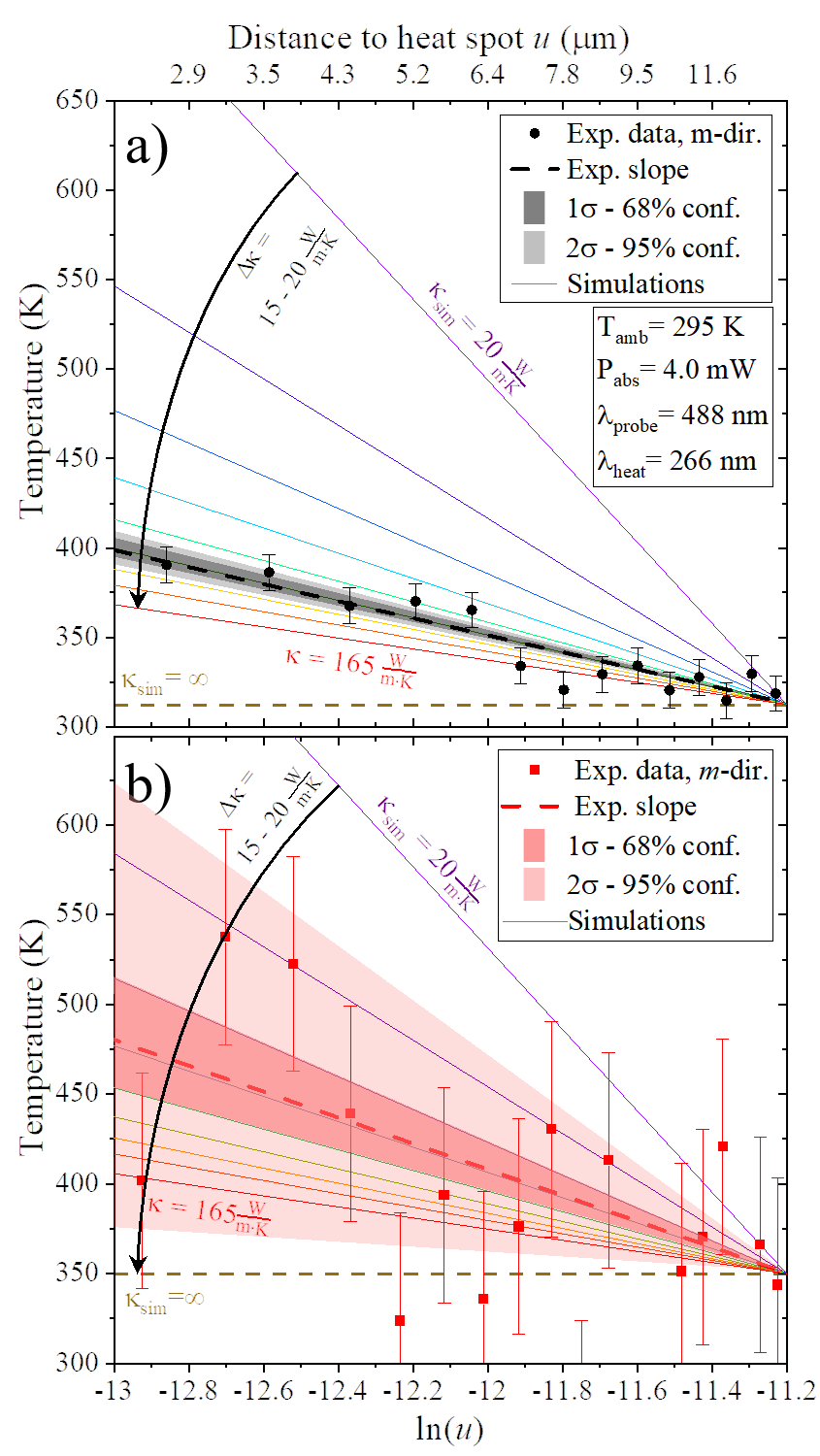}
    \caption{ (a) Temperature profile plotted over the logarithmized distance to the center of the heat spot $u$ (black rectangles). This experimental data matches the temperature data relying on the Raman mode shifts introduced in Fig.\,\ref{fig:2LRT}(c) for the \textit{m}-direction. A  linear least square fit to the data is shown by the dashed, black line, while also the corresponding 1$\sigma$- and 2$\sigma$ confidence intervals are shown. The solid, color-coded lines show a subset of the the simulated temperature trends for different $\kappa^{\,2LRT}_{\,\,memb}$ values ranging from $20\,-\,165\,\text{W/mK}$ in steps of $15\,-\,20$\,\Wu. We do not show the full set of simulated curves, in steps of 5\,\Wu, to lighten the image. Here, the horizontal dashed line illustrates an infinite thermal conductivity. (b) Similar data for the temperature profile extracted from the Raman mode shift broadening. Here, a much larger fluctuation is visible for the experimental data (red rectangles), which challenges any meaningful determination of $\kappa^{\,2LRT}_{\,\,memb}$ at the given value of $P_{abs}\,=\,4.0\,\text{mW}$.}
    \label{fig:2LRTfit}
\end{figure}

In Fig.\,\ref{fig:2LRTfit} we exemplify the determination of the thermal conductivity $\kappa^{\,2LRT}_{\,\,memb}$ based on 2LRT measurements. Therefore, Fig.\,\ref{fig:2LRTfit}(a) shows the spatial evolution of temperatures extracted from the shift of the $E_{2}^{high}$ Raman mode for $P_{abs}\,=\,4.0\,\text{mW}$. The data shown matches the experimental one depicted in Fig.\,\ref{fig:2LRT}(c) for the \textit{m}-direction in our \textit{c}-plane GaN membrane. First, we obtain a linear least square fit (dashed black line) to the experimental temperature values (black rectangles), which are plotted over logarithmized distances $u$ to the heat spot in Fig.\,\ref{fig:2LRTfit}(a). Here a linear fit to the experimental data is adequate due to the analytical solution of the Fourier equation for an infinite two-dimensional membrane \cite{reparaz_novel_2014}. In addition, we also obtained corresponding 1$\sigma$- and 2$\sigma$ confidence intervals for the slopes from the linear least square fit, which are illustrated as gray-shaded areas in Fig.\,\ref{fig:2LRTfit}(a). Note that the intercept in the fit is not considered for the analysis.

Subsequently, we fit our already spatially resolved thermal model described in Sec.\,\ref{subsubsec:ExtractKappa} and S-Sec.\,IV to the experimental temperatures by scanning through a range of $\kappa^{\,2LRT}_{\,\,memb}$ values in steps of 5\,\Wu. This stepping matches half of the error of the final value of $\kappa^{\,2LRT}_{\,\,memb}$, aiming to avoid the introduction of uncertainties from the data analysis. In Fig.\,\ref{fig:2LRTfit}(a) we illustrate this procedure by showing the simulated temperature trends for different $\kappa^{\,2LRT}_{\,\,memb}$ values ranging from $20\,-\,165$\,\Wu in steps of $15\,-\,20$\,\Wu by the solid, color-coded lines. 

To determine the error of the $\kappa^{\,2LRT}_{\,\,memb}$ value that reaches the best agreement with the experimental data, we match the slopes of the numerical temperature trend to the boundaries of the 2$\sigma$ confidence interval shown in Fig.\,\ref{fig:2LRTfit}(a). As a result, we obtain asymmetric error bars for $\kappa^{\,2LRT}_{\,\,memb}$ by our numerical fitting model similar to the error bars belonging to $\kappa^{\,2LRT_0}_{\,\,memb}$ and $\kappa^{\,1LRT}_{\,\,memb}$, cf. Sec. \ref{sec:1LRTand2LRT}. This approach resembles the fitting procedure described in Sec.\,\ref{sec:1LRTand2LRT} under step III. As a result, for the experimental temperature values that were extracted from Raman mode shifts, we obtain $\kappa^{\,2LRT}_{\,\,memb}\,=\,95_{-7}^{+11}$\,\Wu based on our numerical fitting approach. Please note that we restrict our numerical fitting approach to $T_{rise}\,\lesssim\,150\,\text{K}$ to avoid a thermally-induced lowering of $\kappa^{\,2LRT}_{\,\,memb}$, similar to the data analysis of the 1LRT and 2LRT$_0$ measurements presented in Sec.\,\ref{sec:1LRTand2LRT}.

\begin{figure}[]
    \includegraphics[width=\linewidth]{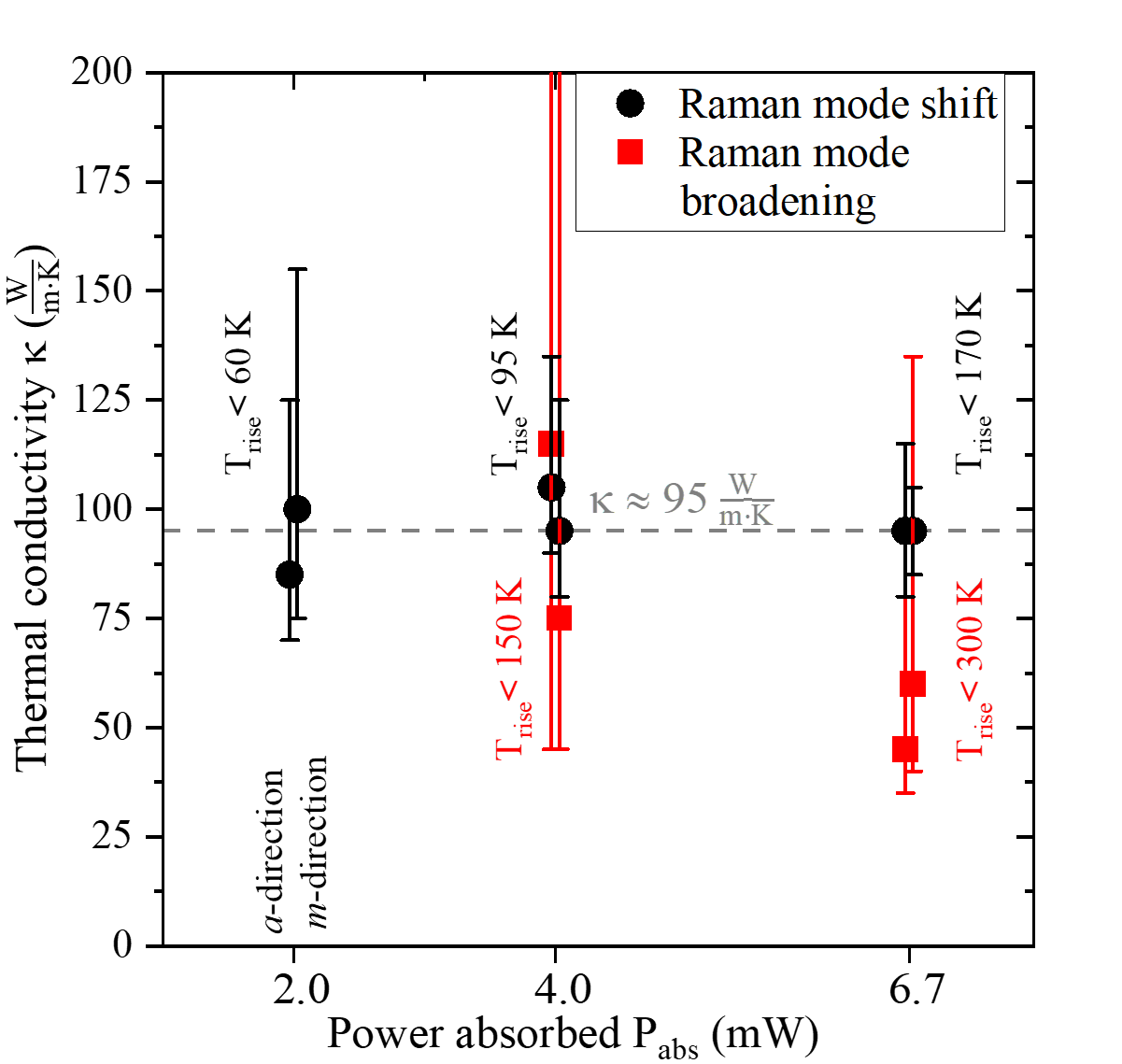}
    \caption{Summary of all $\kappa^{\,2LRT}_{\,\,memb}$ values found for different values of $P_{abs}$ based on 2LRT measurements either relying on Raman mode shifts (black circles) or Raman mode broadenings (red squares). Here, the \textit{a}- and \textit{m}-direction in the \textit{c}-plane of our photonic membrane are considered, as illustrated in the inset of Fig.\,\ref{fig:2LRT}(c). In addition, the upper bound for the $T_{rise}$ interval is given, which formed the basis for the extraction of $\kappa^{\,2LRT}_{\,\,memb}$. Comparably large error bars for the $\kappa^{\,2LRT}_{\,\,memb}$ values based on Raman mode broadening (red symbols) originate from the large scatter among the spatially resolved temperatures as shown in Fig.\,\ref{fig:2LRTfit}(b). This scatter is so large that it made it impossible to extract a meaningful thermal conductivity when $P_{abs}\,=\,2.0$mW, as the maximal $T_{rise}$ observed was almost completely within the noise level. Finally, we determine $\kappa^{\,2LRT}_{\,\,memb}\,=95^{+11}_{-7}$\,\Wu from our experimental temperature trends solely based on the Raman mode shift.}
    \label{fig:2LRT_kappa_graph}
\end{figure}

Similar experimental data along with the corresponding modeling is shown in Fig.\,\ref{fig:2LRTfit}(b) for temperature values extracted from the broadening of the $E_{2}^{high}$ Raman mode with temperature. However, for the given $P_{abs}\,=\,4.0\,\text{mW}$ value, we observe a large scatter of the temperature values in Fig.\,\ref{fig:2LRTfit}(b) and consequently in the thermal conductivities presented in Fig.\,\ref{fig:2LRT_kappa_graph}. That is because the determination of the mode width by fitting relies on finding the maximum of the peak and its baseline. Each of these steps introduces errors that rely on the signal-to-noise ratio of the Raman signal, while temperatures obtained from the Raman mode shift only rely on finding the maximum of the peak, cf. Fig.\,\ref{fig:2LRTfit}(a). Thus, the numerical fitting of cut lines extracted from the mode broadening (Fig \ref{fig:2LRTfit}(b)), leads to an uncertainty on the thermal conductivity so large that within the standard 2-$\sigma$ confidence interval, the $T$\,vs.\,$\text{log}(u)$ slope can be equal to zero. This results in an upper bound for the thermal conductivity of $+\infty$, cf. Fig.\,\ref{fig:2LRT_kappa_graph}.

In Fig.\,\ref{fig:2LRT_kappa_graph} we summarize all our $\kappa^{\,2LRT}_{\,\,memb}$ values for three different values of $P_{abs}$ and the two selected crystal directions in the \textit{c}-plane of our GaN membrane. Here, we also state the maximal value of $T_{rise}$ among the range of the experimental temperatures that were fitted by our numerical approach. As Fig.\,\ref{fig:2LRT_kappa_graph} shows, values for $\kappa^{\,2LRT}_{\,\,memb}$ that are based on the Raman mode broadening (red symbols) show significantly larger error bars compared to their counterparts arising from the Raman mode shift (black symbols). For this reason for $P_{abs}\,=\,2$\,mW we can only obtain meaningful values for $\kappa^{\,2LRT}_{\,\,memb}$ based on the mode shift. Interestingly, all $\kappa^{\,2LRT}_{\,\,memb}$ values that originate form Raman mode shifts in Fig.\,\ref{fig:2LRT_kappa_graph} indicate thermal conductivities around $95^{+11}_{-7}$\,\Wu. In contrast, while for $P_{abs}\,=\,4.0\,\text{mW}$ the $\kappa^{\,2LRT}_{\,\,memb}$ values  originating from Raman mode broadenings still overlap with their counterparts originating from Raman mode shifts, an increasing deviation can tentatively be observed towards $P_{abs}\,=\,6.7\,\text{mW}$. This is connected to the corresponding $T_{rise}$ values at our highest heating power, which yield $T_{rise}\,\leq\,170\,\text{K}$ for the Raman mode shift (black symbol) and $T_{rise}\,\leq\,300\,\text{K}$ for the Raman mode broadening. For the latter case, such drastic increase in temperature yields a decline of the measured $\kappa^{\,2LRT}_{\,\,memb}$ value, which matches our observation from Figs.\,\ref{fig:1LRT} and \ref{fig:2LRT_wod}, where we observed a parabolic evolution of $T_{rise}$ with increasing power of the heating laser for the same reason. Furthermore, within the given error bars, no directional anisotropy can be determined for $\kappa^{\,2LRT}_{\,\,memb}$ in the \textit{c}-plane of our GaN membrane.

%
%
%

\section{\label{sec:Modeling}Ab initio computation of phonon transport}

To theoretically compute the thermal conductivity $\kappa$, we use an \textit{ab initio} solution of the linearized phonon Boltzmann transport equation (BTE), which furnishes the mode (branch $s$, wave vector $\mathbf{q}$) resolved steady state distribution function 
\begin{equation} \label{eq:phdistrib}
    n_{s\mathbf{q}} = n^{0}_{s\mathbf{q}} \left[ 1 - (1 + n^{0}_{s\mathbf{q}}) \dfrac{\mathbf{\nabla}T\cdot\mathbf{F}_{s\mathbf{q}}}{k_{\text{B}}T} \right],
\end{equation}
where $n^{0}$ is the equilibrium Bose-Einstein distribution, $T$ is the crystal temperature, $\mathbf{\nabla}T$ is the applied temperature gradient, $k_{\text{B}}$ is the Boltzmann constant, and $k_{\text{B}}^{-1}\mathbf{F}_{s\mathbf{q}}$ is a vectorial ``mean free displacement'' \cite{li_shengbte_2014} carrying the unit of length. This quantity is essentially a measure of how far each phonon mode is moved out of equilibrium due to the presence of the temperature gradient field.

The phonon BTE can be conveniently cast in the following form: 
\begin{equation} \label{eq:phBTE}
    \mathbf{F}_{s\mathbf{q}} = \mathbf{F}^{0}_{s\mathbf{q}} + \Delta \mathbf{F}_{s\mathbf{q}}^{\text{S}}[\mathbf{F}],
\end{equation}
where the first term on the right hand side is the mode-resolved relaxation time approximation (RTA) expression and the second, non-local term gives the in-scattering corrections. The expressions for these terms have been given before in, e.g., Ref. \cite{protik_elphbolt_2022}. A full self-consistent solution of this equation takes us beyond the RTA.

From the solution of the BTE, a scalar, phonon mean free path (MFP) is constructed in the following manner:
\begin{equation}
    l^{\text{MFP}}_{s\mathbf{q}} = \dfrac{\mathbf{F}_{s\mathbf{q}} \cdot \mathbf{v}_{s\mathbf{q}}}{k_{\text{B}}v_{s\mathbf{q}}},
\end{equation}
where $\mathbf{v}$ is the phonon group velocity.

The phonon thermal conductivity tensor is given by
\begin{equation}
    \kappa = \dfrac{1}{Vk_{\text{B}}T}\sum_{s\mathbf{q}}\hhbar\omega_{s\mathbf{q}}n^{0}_{s\mathbf{q}}(1 + n^{0}_{s\mathbf{q}})\mathbf{v}_{s\mathbf{q}}\otimes\mathbf{F}_{s\mathbf{q}},
\end{equation}
where $\hhbar$ is the reduced Planck constant, $\omega$ is the phonon angular frequency, and $V$ is the crystal volume.

In wurtzite crystals, the thermal conductivity has the following structure: $\kappa_{xx} = \kappa_{yy} \neq \kappa_{zz}$, with the off-diagonal terms identically zero. In the harmonic approximation, phonons are infinitely long-lived collective excitations of the crystal. Anharmonicity is captured via the quantum mechanical interactions between the phonons (ph). In this work, we included 3ph and 4ph interactions. We have also included the ph-(iso)tope scattering within the Tamura model \cite{tamura_isotope_1983}. 
Phonon-boundary scattering, hereafter referred to as \emph{ph-thin-film}, is included empirically via~\cite{li_gan_2020}
\begin{equation}\label{scattering}
\tau^{-1}_{s\mathbf{q},\text{ph-thin-film}} = \dfrac{2v^{\perp}_{s\mathbf{q}}}{h},
\end{equation}
where $v^{\perp}$ is the phonon group velocity component perpendicular to the plane of the film and $h$ is the film thickness. Equation ~\ref{scattering} includes diffuse scattering with the top and bottom facets of the GaN membrane (see Fig.\,\ref{fig:exp_setup}). We note that it is possible to derive more refined expressions for boundary scattering, including in-plane~\cite{carrete_almabte_2017} (commonly known as the Fuchs-Sondheimer model) and cross-plane directions \cite{vermeersch_cross-plane_2016}. The development of these models is beyond the scope of this work, and will be considered in future works. 

While the real-life sample contains dislocation defects and charge carriers, we have not included the effects of the presence of these in our current calculations. Each of these excluded phonon scattering channels degrades the thermal current further and, as such, the theoretical prediction made here should be interpreted as an upper limit of the actual $\kappa$. Nevertheless, our computation of $\kappa$ and the $l^{MFP}$ values will prove valuable for the qualitative comparison to the experimental results that we obtain from our different thermometric techniques, cf. Sec.\,\ref{sec:Comparison}.

\subsection{\label{subsec:CompDet}Computational details}

We use the \texttt{Quantum Espresso} suite \cite{giannozzi_quantum_2009,giannozzi_advanced_2017,giannozzi_quantum_2020} for our density functional theory (DFT) and density functional perturbation theory (DFPT) calculation. We employ the Optimized Norm-Conserving Vanderbilt (ONCV) \cite{hamann_optimized_2013} pseudopotential with the local density approximation for the exchange correlation function. The relaxed lattice constants are $a = 3.15$ \AA{} and $c = 5.14$ \AA.

The 2nd order force constants (2FCs) required for the harmonic properties are computed using DFPT with a $6\times 6\times 6$ wave vector mesh. The 3rd order force constants (3FCs) needed for the 3-phonon scattering are calculated using the displaced supercell method. For this, the \texttt{thirdorder} code \cite{li_shengbte_2014} is used to generate an irreducible set of $4\times 4\times 4$ ($256$ atoms) supercells with two displaced atoms. Following the self-consistent-field (SCF) DFT calculations on these supercells using a $\Gamma$-point sampling of the electronic reciprocal space, the 3FCs are read-out using again the \texttt{thirdorder} code. In this work, we include 4-phonon scattering since, in general, it is not possible to know \textit{a priori} whether this will significantly limit the phonon transport or not. To calculate the 4-phonon scattering rates, we need the 4th order force constants (4FCs). For this, we use the \texttt{fourthorder} \cite{han_fourphonon_2022} code to generate $256$-atom supercells with $3$ atoms displaced in each. Following the SCF calculations, the same code is used to read-out the 4FCs. Furthermore, the \texttt{FourPhonon} code \cite{han_fourphonon_2022} is used to generate the 4-phonon scattering rates on a coarse $12\times 12\times 12$ phonon wave vector ($\mathbf{q}$) mesh, $\tau^{-1}_{\text{4ph, coarse}}$.

The 2FCs, 3FCs, and $\tau^{-1}_{\text{4ph, coarse}}$ are fed into \texttt{elphbolt} \cite{protik_elphbolt_2022} which calculates the phonon transport properties by solving the BTE on a converged $36\times 36\times 36$ $\mathbf{q}$-mesh. The code calculates harmonic properties and 3-phonon scattering rates using the 2FCs and 3FCs, respectively. The energy conserving delta functions are calculated using the analytic tetrahedron method \cite{lambin_computation_1984}. The $\tau^{-1}_{\text{4ph, coarse}}$ are calculated on the fine transport $\mathbf{q}$-mesh using an (in general tri-) linear interpolation method. We use this indirect method for computing the 4ph scattering rates on the transport $\mathbf{q}$-mesh since the direct computation of this quantity on such a fine mesh is prohibitively expensive at the moment. The RTA term, $\mathbf{F}^{0}$, is constructed using Mathiessen's rule where the total scattering rate is given by $\tau^{-1}_{\text{total}} = \sum_{x}\tau^{-1}_{x}$, where $x = \{\text{ph-iso, ph-thin-film, 3ph, 4ph}\}$. In the full solution of the BTE, the in-scattering correction due to the 3ph interactions is added back in iteratively. The thin-film is modeled as having a $250$ nm height along the crystal $c$-axis and an infinite extent in the $c$-plane.

\subsection{\label{subsec:CompDet}Presentation of the theoretical results}

\begin{figure}[]
    \includegraphics[width=8.4cm]{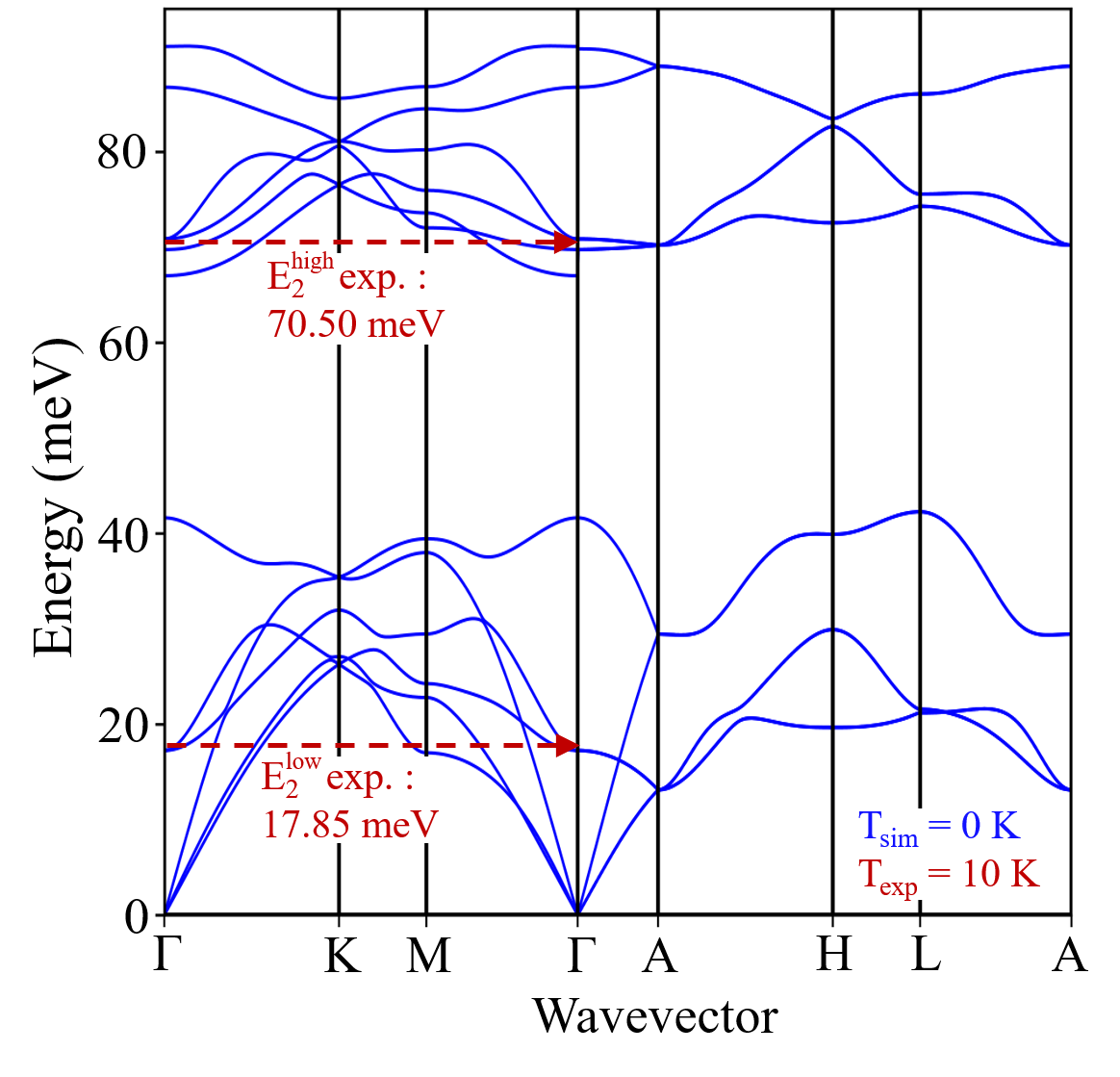}
    \caption{Calculated phonon dispersions along high-symmetry paths in the Brillouin zone for GaN at 0\,K. We observe a total of 12 branches: 3 acoustic + 3 low-lying optical phonon modes separated by an energy gap from the 6 high-lying optical phonon modes. The experimental energies of the $E_2^{high}$ and $E_2^{low}$ optical modes measured at 10\,K are indicated by the red dashed arrows close to the $\Gamma$ point.}
    \label{fig:phdisp}
\end{figure}



We present in Fig.\,\ref{fig:phdisp} the calculated phonon dispersions at 0\,K. The 3 acoustic + 3 low-lying optical phonon branches are separated from the 6 high-lying optical phonon branches by a large energy gap which, as we shall see shortly, has important consequences for the thermal transport properties of this material. The two red arrows indicate the experimental energies of the $E_2^{high}$ and $E_2^{low}$ optical modes, measured at 10\,K, in comparison to the calculated ones.

\begin{figure}[]
    \includegraphics[width=\linewidth]{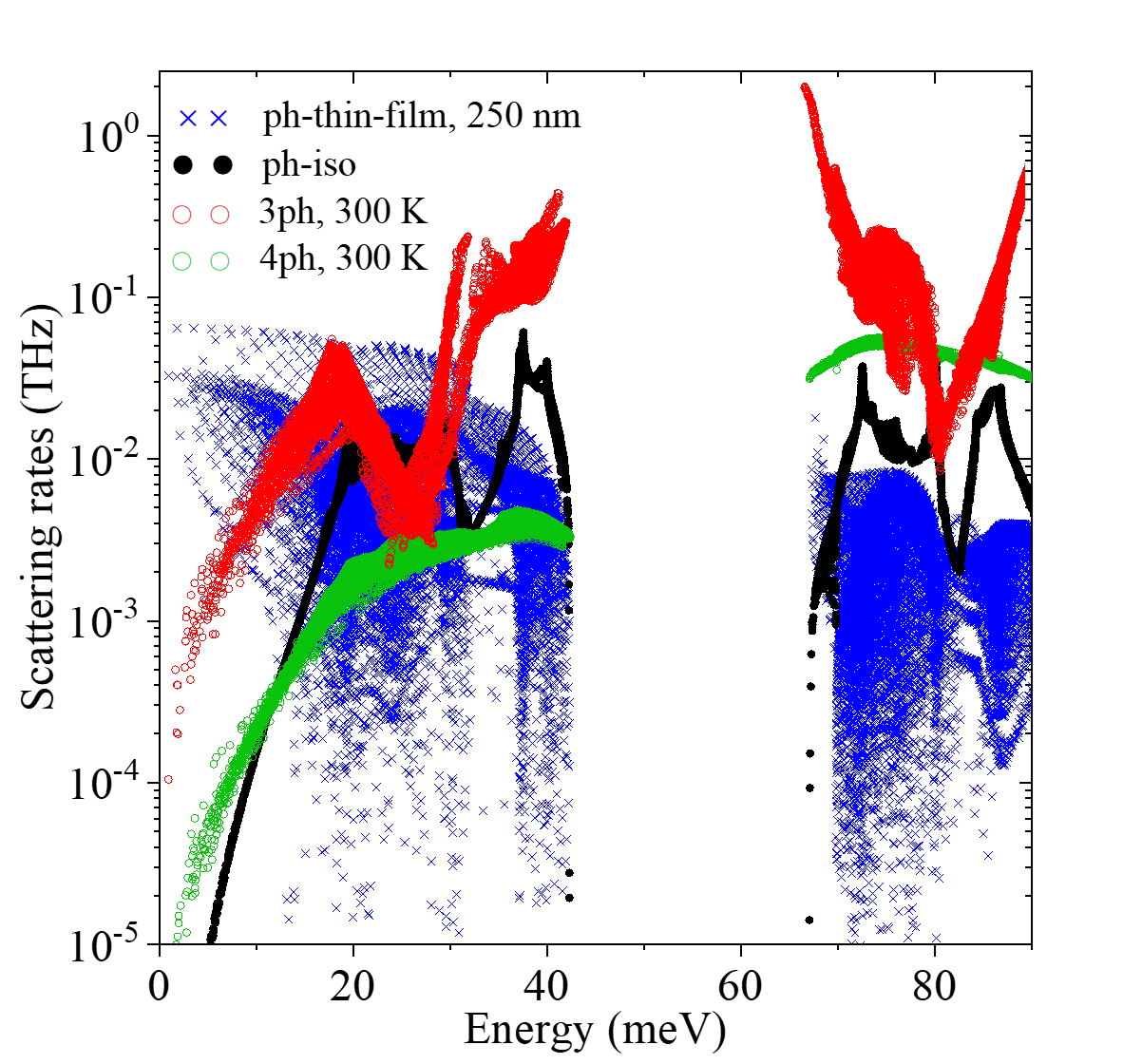}
    \caption{Room temperature phonon scatterings rates in the RTA for the various scattering channels included in the simulation. Because of the energy gap around 50\,meV, two phonons with energies around 25\,meV cannot coalesce together, creating a dip around 25\,meV in the 3ph scattering channel. This dip is completely absent in the 4ph scattering process since no such restriction exists.}
    \label{fig:scattering}
\end{figure}

Figure\,\ref{fig:scattering} shows the 300\,K, RTA scattering rates of phonons due to their interactions in the various scattering channels included in the simulation. We discuss here some salient features of the acoustic scattering rates that highlight the roles of the different competing interactions. First, the 3ph scattering rates (empty red circles) show a pronounced dip in the 20\,-\,30\,meV energy range. As such, the phonon-isotope scattering (black circles) becomes dominant in this energy window. The large dip can be connected to the phonon dispersions of the material. Specifically, the presence of the large gap between the acoustic+low-lying optic and the high-lying optic sectors results in a reduction in the 3ph scattering phase space, explaining the corresponding dip in the scattering rates. One can understand this by noticing in Fig.\,\ref{fig:phdisp} that two $\sim$\,25\,meV phonons cannot coalesce since there is no 50\,meV state available in the spectrum. In contrast, three $\sim$\,25\,meV phonons can indeed coalesce, explaining that fact that no such dip in the 4ph scattering rates (empty green circles) is present. As such, the 4ph scattering rates become important, even at 300\,K. This suggests that at higher temperatures, the 4ph interaction will play a progressively stronger role and will cause a stronger than $T^{-1}$ scaling of $\kappa$. Observation of such a deviation from the $T^{-1}$ scaling of $\kappa$ has been reported in Ref. \cite{zheng_thermal_2019}. Next, we focus on the high energy optic phonons. As for the low energy optic phonons, because of scattering phase-space restrictions of the 3ph scattering rates, the 4ph scattering rates become important in the 70\,-\,85\,meV energy range. However, these phonons do not directly transport heat since they have high scattering rates and/or have low group velocities. Lastly, The ph-thin-film scattering (blue crosses) rates are weakly dependent on energy and are high for the heat carrying phonons. This, as we shall see below, makes them a strong limiter of the $\kappa$ even at 300\,K.

\begin{figure}[]
    \includegraphics[width=\linewidth]{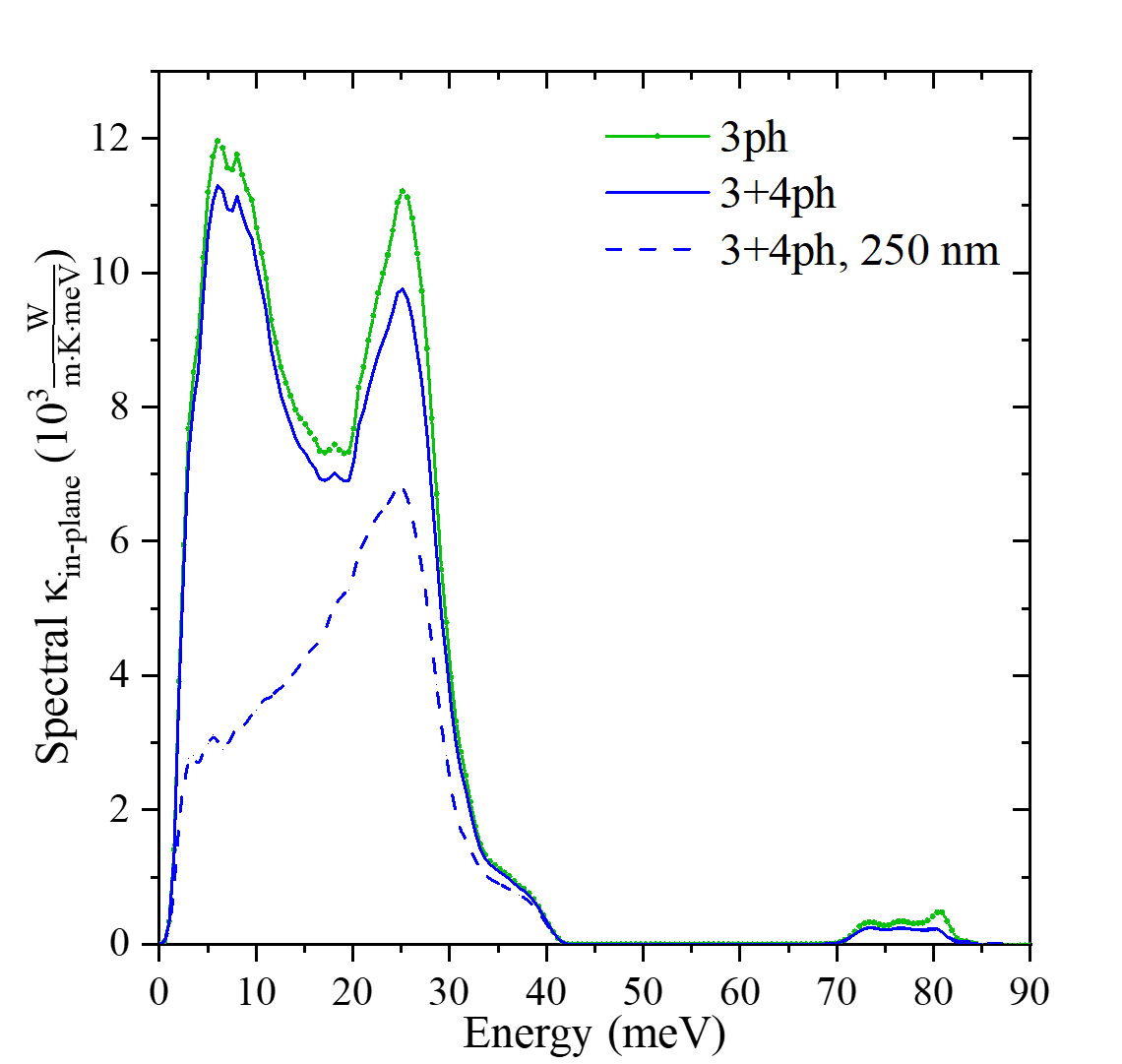}
    \caption{Room temperature spectrum of $\kappa_{\text{in-plane}}$ for an infinite GaN crystal and a 250-nm-thick GaN membrane. Phonon-isotope scattering is included in all three cases. Of the two peaks visible for the infinite crystal (solid green and blue line), the one corresponding to the acoustic modes is the most affected by the introduction of the top and bottom facets boundary scattering caused by our membrane (dashed blue line).}
    \label{fig:kxx_spec}
\end{figure}

Next, we plot the energy spectrum of the $\kappa_{\text{in-plane}}$ in Fig. \ref{fig:kxx_spec}. The ph-iso scattering is included in all the cases presented. For the 3ph limited case (green dot-line), we see two dominant peaks in the 3\,-\,12 and 20\,-\,30 meV range. The lower energy peak is due to high velocity acoustic phonons while the second peak is owing to the dispersive low-lying optic phonons. The low energy peak follows from the classical Slack criteria for high $\kappa$ materials \cite{lindsay_first-principles_2013}. The second peak, on the other hand, is enabled by the fact that the 3ph scattering rates show an anomalous dip, as explained in the previous paragraph, reminiscent of the material BAs \cite{lindsay_first-principles_2013}. The optic phonons across the energy gap contribute negligibly to the thermal conductivity. Next, with the 4ph scattering turned on, we see a general reduction in the $\kappa$ contribution (blue curve), especially in the 20-30 meV energy range. Lastly, the thin-film scattering causes a significant reduction in the spectral $\kappa$ (dashed blue line), especially for the low energy region where this is the dominant scattering channel. The second peak is also significantly reduced, albeit to a lesser extent than the first, evidently owing to the anomalous weakening of the 3ph scattering rates. The strong suppression of the low energy peak due to the thin-film scattering suggests that the low energy acoustic phonons are carrying heat ballistically. In contrast, the moderate suppression of the second peak suggests diffusive transport of the phonons in that energy range since the isotope, 3ph, and 4ph, all play an important role in degrading the heat current.

\begin{figure}[]
    \includegraphics[width=\linewidth]{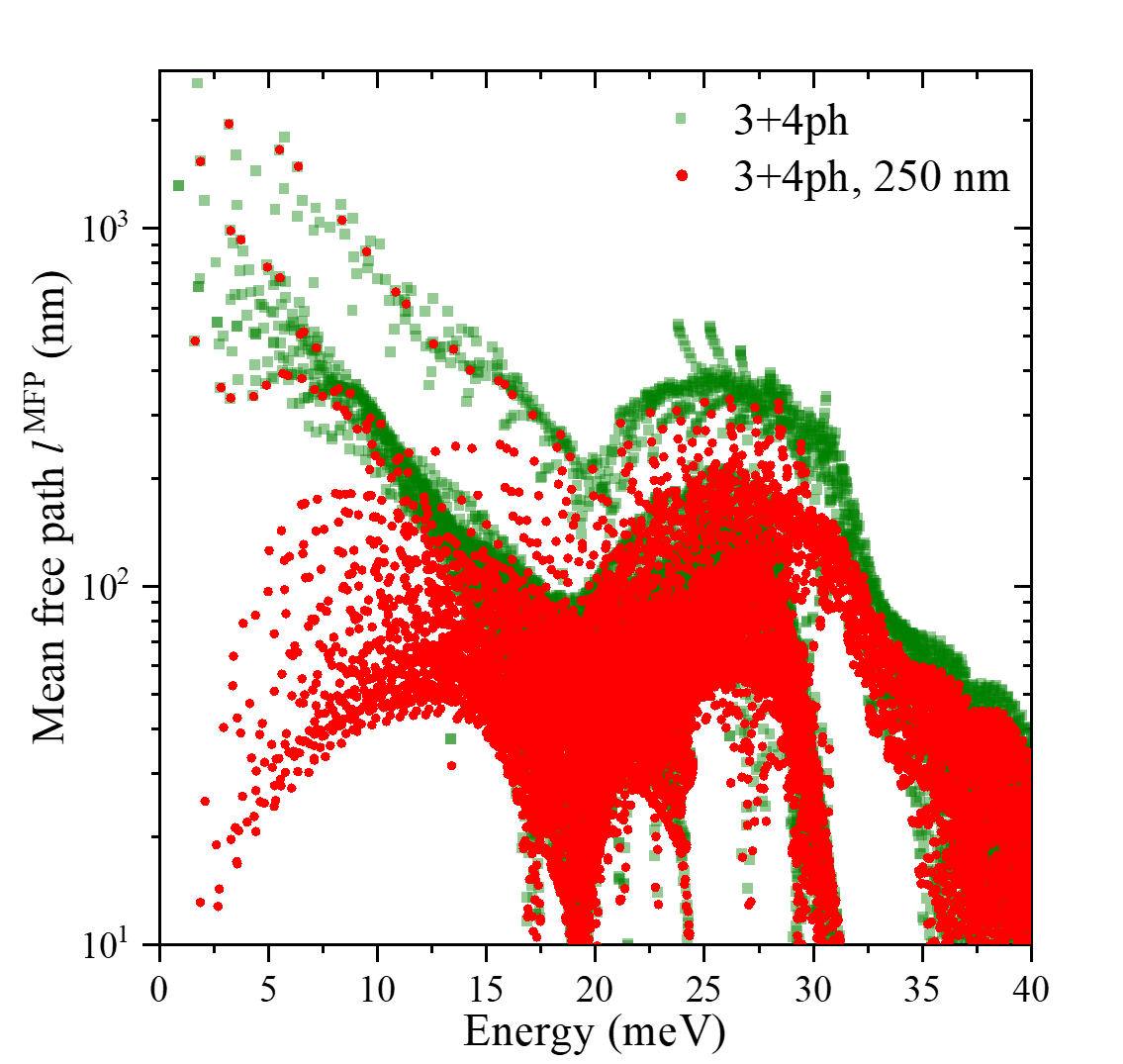}
    \caption{Room temperature phonon mean free path $l^{MFP}$ for the acoustic\,+\,low-lying optical phonons, with and without the introduction of the top and bottom facets boundary scattering. Phonon-isotope scattering is included in both cases. We observe a strong reduction of the acoustic modes mean free paths, due the large cross-plane component of their mean free paths. Some phonons are however barely affected by the introduction of boundary scattering; these are phonons with large cross-plane component of their velocities.}
    \label{fig:mfp}
\end{figure}

The effect of the thin-film on the phonon transport can be seen further in Fig.\,\ref{fig:mfp} where we plot the phonon MFP vs. phonon energy for the transport active acoustic+low-lying optic states. When the 250 nm thin-film scattering is turned on, a significant reduction in the MFP occurs, especially for the low energy acoustic phonons. This is the reason for the corresponding reduction in the spectral $\kappa$ in that energy range. We notice that for some states, the thin-film scattering does not affect the MFP. These are the states with a vanishing velocity component along the $c$-axis. Consequently, these phonons continue to carry heat along the $c$-plane despite the imposition of the thin-film boundary.

\begin{figure}[]
    \includegraphics[width=\linewidth]{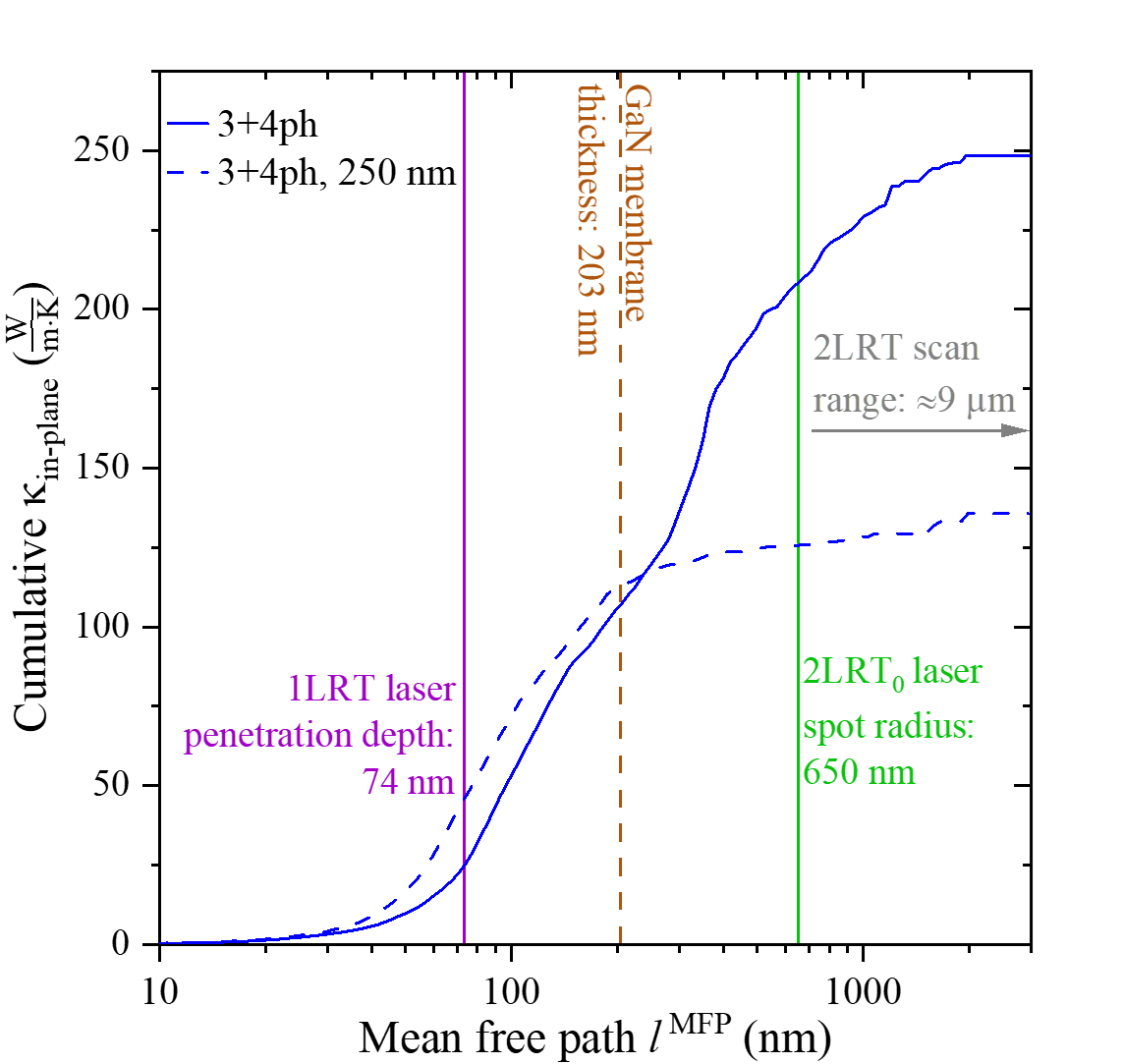}
    \caption{$\kappa_{\text{in-plane}}$ accumulation with respect to phonon mean free path $l^{MFP}$ for an infinite crystal and a 250-nm-thick membrane. Phonon-isotope scattering is included in both cases (3+4ph and 3+4ph including boundary scattering). Vertical lines mark the relevant length scales $r_{lim}$ of the 1LRT (purple) and 2LRT$_{0}$ (green) measurement techniques, while the gray arrow indicates the corresponding $r_{lim}$ of 2LRT measurements. As $r_{lim}$
    of the 2LRT measurement is $9$\,$\mu$m, all relevant phonon mean free paths are encompassed by the temperature probe volume in the GaN membrane. Additionally, a dashed vertical brown line indicates the GaN (without the AlN) membrane thickness.}
    \label{fig:kxx_accum_mfp}
\end{figure}

Lastly, in Fig. \ref{fig:kxx_accum_mfp} we show the accumulation of the \textit{c}-plane $\kappa$ as a function of MFP. The ph-iso scattering is included in all these calculations. The inclusion of the 4ph scattering on top of the 3ph scattering leads to a reduction of the total $\kappa_\text{in-plane}$ from $270$ to $248$\,\Wu, the last value being in good agreement with earlier experimental measurements on high purity, bulk samples with a natural isotopic mix \cite{jezowski_thermal_2003}. For reference, we also give the corresponding $\kappa_\text{cross-plane}$ values for the 3ph and 3+4ph cases: $275$ and $254$\,\Wu, respectively. The anisotropy between the bulk thermal conductivities is low, with the in-plane value smaller than the cross-plane value by 1.8\% and 2.4\% for the 3ph and 3+4ph calculations, respectively. The 3+4ph calculation for the $250$-nm-thin film case (dashed blue curve) is more relevant to our experimental observations. The calculated total $\kappa_\text{in-plane}$ in this case is $136$\,\Wu. Again, for reference, the calculated total $\kappa_\text{cross-plane}$ is $89$\,\Wu in this case. The significantly stronger reduction of the $\kappa_\text{cross-plane}$ is due to the fact that the thin-film preferentially scatters phonons with large $c$-axis velocity component.

In Fig.\,\ref{fig:kxx_accum_mfp} we mark with vertical lines the various limiting length scales $r_{lim}$ of our experimental temperature probe volumes. This either corresponds to the penetration depth of the laser ($p_{abs}$) for the 1LRT measurements, the probe spot radius ($r_{probe}$) for the 2LRT$_0$ measurements, and the scanning range ($r_{range}$) of the 2LRT measurements beyond the limits of Fig.\,\ref{fig:kxx_accum_mfp}. Physically, in our experimental methods, these various lengths are phonon mean free path visibility cut-offs. That is, the heat dissipation effects of any phonon with a mean free path larger than $r_{lim}$ cannot be detected. Thus, any measurement with $r_{lim}$ shorter than the maximum phonon mean free path will measure too large values of $\kappa$. Our theoretical calculations predict that in order to capture the heat dissipation effects of all the phonons in this material, $r_{lim}$ must be at least $3$\,$\mu$m, which can only be met by 2LRT measurements as summarized in Sec.\,\ref{sec:Comparison} via Fig.\,\ref{fig:1LRT2LRTcomparison}.

\section{\label{sec:Comparison}Comparison of 1-laser and 2-laser Raman thermometry\protect}

Interestingly, based on 1LRT, 2LRT$_0$, and 2LRT thermometry we determined a continuous decline of the measured thermal conductivities from $\kappa^{\,1LRT}_{\,\,memb}\,=\,180^{+12}_{-7}\,$\Wu, over $\kappa^{\,2LRT_{0}}_{\,\,memb}\,=\,110^{+11}_{-8}\,$\Wu, to $\kappa^{\,2LRT}_{\,\,memb}\,=\,95^{+11}_{-7}\,$\Wu as summarized in Fig.\,\ref{fig:1LRT2LRTcomparison}. In the following, we will first discuss the relevance of the absolute values that we determined for $\kappa_{memb}$ in Sec.\,\ref{subsec:absolutekappa}, before the evolution of $\kappa_{memb}$ that depends on the Raman thermometry technique in use is discussed in Sec.\,\ref{subsec:scalingkappa}.

\subsection{\label{subsec:absolutekappa}Absolute value of the thermal conductivity\protect}

Based on our theoretical results presented in Sec.\,\ref{sec:Modeling} and literature values, only $\kappa^{\,2LRT}_{\,\,memb}$ represents a realistic value for our 253-nm-thick photonic membrane due to the dominance of phonon boundary scattering \cite{beechem_size_2016} that inhibits thermal transport. For GaN layers with a similar thickness, Beechem \textit{et\,al.} \cite{beechem_size_2016} predicted a thermal conductivity of $\approx\,70\,$\Wu if phonon boundary scattering represents the main limiting factor for $\kappa$. However, even drastically lower $\kappa$ values were predicted for the dislocation density in our photonic membrane of $3\,-\,4\,\times\,10^{10}\,\text{cm}^{-2}$. In our membrane, anharmonic scattering, phonon-boundary-scattering, i.e., size effects, isotope-scattering, and phonon-dislocation scattering are potentially relevant limitations for thermal transport. Phonon-impurity-scattering can only be of minor relevance due to the low impurity density ($<\,10^{17}\,\text{cm}^{-3}$) in our sample, cf. Sec.\,\ref{sec:Sample}. Regarding the phonon-dislocation-scattering, Li \textit{et\,al.} \cite{li_gan_2020} have experimentally and theoretically demonstrated that the role of phonon-dislocation-scattering is often overestimated in literature for GaN layers, as even for a dislocation density of $1.80\,\times\,10^{10}\,\text{cm}^{-2}$ they measured $\kappa\,\approx\,175\,\text{W/mK}$ for a 3.19-$\mu$m-thick GaN sample grown on AlN and sapphire. Nevertheless, as our photonic membrane also includes 50\,nm of AlN, we can expect that this interlayer can further boost thermal transport in our entire photonic membrane. The thermal conductivity of bulk AlN is commonly around 50\% higher compared to bulk GaN at room temperature, when considering the natural abundance of isotopes \cite{dagli_thermal_nodate}, while the AlN in our epilayer contributes $\approx\,20\%$ to the overall volume.

However, as the AlN buffer layer is grown on silicon (111), it is particularly rich in structural (e.g., dislocations) and point defects (e.g., silicon originating from diffusion from the substrate), which will further limit its thermal conductivity advantage compared to the GaN layer grown on top. In addition, with 50\,nm the AlN layer is comparably thin, meaning that phonon-boundary scattering at the AlN/vacuum interface and phonon-interface scattering at the AlN/GaN interface, will significantly lower its thermal conductivity. Here, only future work can provide insight into the particular contribution of this AlN interlayer to the overall $\kappa^{\,2LRT}_{\,\,memb}$ value of $95^{+11}_{-7}\,\text{W/mK}$, which we determined for our photonic membrane. Either depth-resolved thermometry based on frequency-domain thermal reflectance can be performed, or the AlN bottom layer must be removed by chemically selective wet etching. However, Rousseau \cite{Rousseau2018c} has shown that the removal of this tensilely strained AlN layer leads to the bowing of the rest of the photonic membrane, which renders it unusable for any photonic applications. Thus, such modification of the photonic membrane would counteract the motivation of this work to study structures with a high relevance for photonic applications from an optical and thermal viewpoint. In addition, the particular smooth top and bottom facets of our photonic membrane described in Sec.\ref{sec:Sample}, potentially lower the impact of phonon-boundary scattering on $\kappa^{\,2LRT}_{\,\,memb}$.

Regarding the absolute comparison between our experimental value  $\kappa^{\,2LRT}_{\,\,memb}\,=\,95^{+11}_{-7}\,$\Wu and its theoretical counterpart $\kappa_\text{in-plane}\,=\,136$\,\Wu from Sec.\,\ref{sec:Modeling}, it is important to mention that our theory only predicts an upper bound for the thermal conductivity. In this work, we refrain from implementing the impact of dislocations for the following reasons:
a) The impact of dislocations on the thermal conductivity of GaN was lately strongly debated in literature \cite{beechem_size_2016,li_gan_2020} and theoretical work by Wang \textit{et\,al.} \cite{wang_phonon_2019} has predicted a minor impact of dislocations in GaN layers up to $\approx\,10^{10}\,\text{cm}^{-2}$. Thus, in a photonic membrane with a thickness of $\approx$\,250\,nm one can expect that phonon-boundary scattering will dominate scattering in samples with a dislocation density of $3\,-\,4\,\times\,10^{10}\,\text{cm}^{-2}$, cf. Sec\,\ref{sec:Experiment}.
b) Additionally the preferential alignment of dislocations in our \textit{c}-plane GaN membrane would need to be considered in a more advanced theory, which represents an interesting task for a future sample series with varying dislocation densities. Furthermore, we can expect that not only the consideration of phonon-dislocation scattering but also the theoretical treatment of thermal transport across the interfaces of our pad structure (GaN/InGaN, GaN/AlN, AlN/Si interfaces) would further close the gap between our theoretical value $\kappa_\text{in-plane}$ and its experimental counterpart $\kappa^{\,2LRT}_{\,\,memb}$. Here a model based on Green's function \cite{dai_rigorous_2020} could be employed to model the interfacial thermal transport. Future work may be devoted to understand and quantify the relative contributions of these effects.

\subsection{\label{subsec:scalingkappa}Scaling of the thermal conductivity\protect}

\begin{figure}[]
    \includegraphics[width=\linewidth]{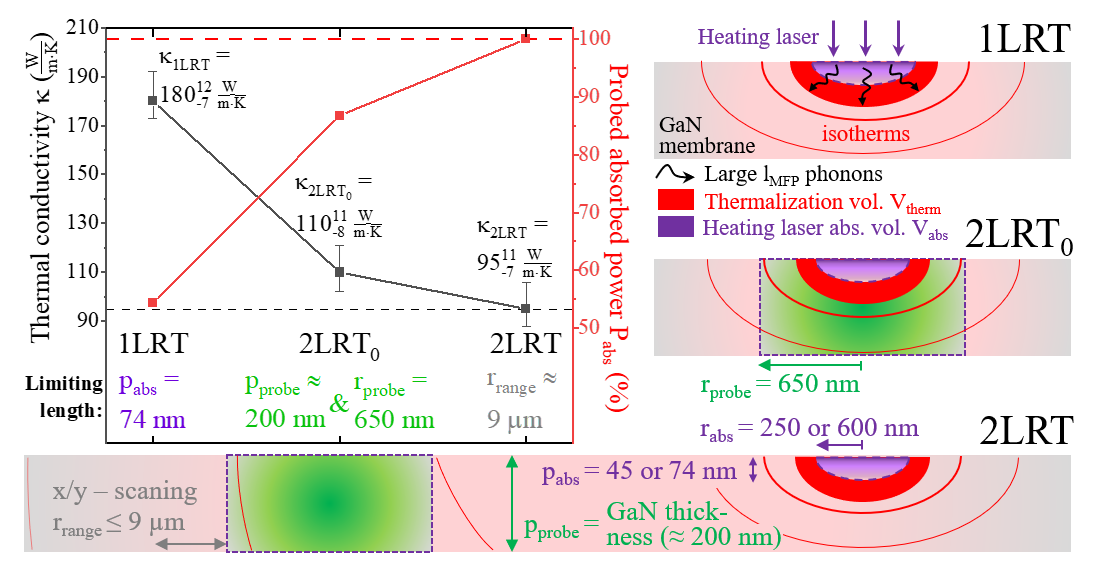}
    \caption{Comparison of all thermal conductivities $\kappa$ that were determined in this work. When increasing the probe volume, the measured $\kappa$ values continuously decrease, starting with the highest value for 1LRT and ending with the lowest value for 2LRT measurements. The corresponding experimental situation for all three experimental techniques, namely 1LRT, 2LRT$_0$, and 2LRT, is sketched along with the related laser spot radii for the heating ($r_{abs}$) and the probe ($r_{probe}$) lasers. As the balance between the $l^{MFP}$ values and the limiting length scales ($p_{abs}$, $r_{probe}$, and $r_{range}$) of the associated temperature probe volumes varies, the measured $\kappa$ values are altered. In the end, only 2LRT measurements allow enclosing most $l^{MFP}$ values in the thermalization volume, which leads to a realistic measure of $\kappa^{\,2LRT}_{\,\,memb}$. Due to the limited temperature probe volume, 1LRT and 2LRT$_0$ only probe effective and artificially enhanced $\kappa$ values as only a fraction of $P_{abs}$ is probed.}
    \label{fig:1LRT2LRTcomparison}
\end{figure}

The particular scaling for the derived $\kappa$ values as one of the main results of this work can be explained by considering the temperature probe volume. For this matter, Fig.\,\ref{fig:1LRT2LRTcomparison} summarizes the experimental situation for 1LRT, 2LRT$_0$, and 2LRT measurements in three sketches that highlight the main length scales that limit the temperature probe volume. 

\subsubsection*{\label{subsubsec:1LRT_vol}1-laser Raman thermometry\protect}

For 1LRT measurements the volume in which the power of the heating laser is absorbed coincides with the temperature probe volume as a first approximation. This volume is the origin of the resonant Raman spectra that are used to determine the local temperature. Here, the heating laser spot radius ($r_{abs}$ measures $600\,\pm\,50\,\text{nm}$, while the light penetration depths $p_{abs}$ yields $74\,\text{nm}$ \cite{kawashima_optical_1997}. Thus, the main size limit for the temperature probe volume for 1LRT measurements is oriented in the cross-plane direction, as a significant fraction of heat carrying phonons with $l^{MFP}\,>\,p_{abs}$ exists that still contributes to the overall $\kappa$ of the sample, cf. Fig.\,\ref{fig:kxx_accum_mfp}. A certain fraction of all the generated heat carrying phonons travels quasi ballistically and the first scattering event that starts their thermalization occurs outside the 1LRT temperature probe volume $V_{probe}^{exp}$, creating a thermalization volume $V_{therm}^{exp}$ that is larger than the light absorption volume $V_{abs}^{exp}$, with $V_{abs}^{exp}\,=\,V_{probe}^{exp}$ for 1LRT. 

The problem occurs when the thermal conductivity $\kappa^{1LRT}_{memb}$ is extracted based on the comparison to our Fourier-based model described in Sec.\,\ref{sec:1LRT}. Even though we precisely consider all experimental parameters (e.g., $p_{abs}$, $r_{probe}$, $P_{abs}$) for such a model, the calculated $T_{rise}$ will always exceed the experimental $T_{rise}$ values due to the reduced thermalization volume $V_{therm}^{sim}$ in the simulation of 1LRT measurements. Since no model exists that accounts for the extended $V_{therm}^{exp}$ (even based on the results presented in Sec.\,\ref{sec:Modeling}), our analysis of the 1LRT data is always limited to:
\begin{equation*}
1LRT:\,\,\,\,\,V_{abs}^{sim,exp}\,=\,V_{probe}^{sim,exp}\,=\,V_{therm}^{sim}\,\ll\,V_{therm}^{exp}.
\end{equation*}
Consequently, based on our model for 1LRT measurements that aims to match the calculated $T_{rise}$ values to the experimental ones, we always determine artificially enhanced $\kappa^{\,1LRT}_{\,\,memb}$ values, cf. Fig.\,\ref{fig:1LRT2LRTcomparison}.

\subsubsection*{\label{subsubsec:2LRT_0_vol}2-laser Raman thermometry\protect}

The situation is improved as soon as we transition from 1LRT to 2LRT$_0$ measurements that probe the entire GaN membrane thickness through non-resonant Raman scattering. Here the temperature probe volume in the cross-plane direction is limited by the GaN fraction of the membrane and for all in-plane directions by the radius $r_{probe}\,=\,600\,\pm\,50\,\text{nm}$ of the temperature probe laser spot, cf. Fig.\,\ref{fig:1LRT2LRTcomparison}. As a result, a much larger fraction of the thermalization volume is measured in the temperature probe volume of 2LRT$_0$ measurements, leading to:
\begin{equation*}
2LRT_{0}:\,\,\,\,\,\,V_{abs}^{sim,exp}\,=\,V_{therm}^{sim}\,<\,V_{probe}^{sim,exp}\,<\,V_{therm}^{exp}.
\end{equation*}
As a result, we obtain a $\kappa^{2LRT_{0}}_{memb}$ value that is significantly lower than $\kappa^{1LRT}_{memb}$ (see Fig.\,\ref{fig:1LRT2LRTcomparison}), which already compares well to previously published results \cite{beechem_size_2016}. Nevertheless, a small fraction of heat carrying phonons still exists that exhibits $l^{MFP}\,>\,r_{probe}$, which in turn also artificially enhances $\kappa^{2LRT_{0}}_{memb}$. However, compared to the strong artificial increase of $\kappa^{1LRT}_{memb}$, this enhancement is minor as the majority of heat carrying phonons thermalizes in $V_{probe}^{sim,exp}$ at $T_{amb}\,=295\,\text{K}$, cf. Figs.\,\ref{fig:kxx_accum_mfp} and \ref{fig:1LRT2LRTcomparison}.

By further enlarging the temperature probe volume based on 2LRT measurements over $r_{range}\,\leq\,9\,\mu\text{m}$, all relevant $l^{MFP}$ values are comprised in the temperature probe volume:
\begin{equation*}
2LRT:\,\,\,\,\,\,V_{abs}^{sim,exp}\,=\,V_{therm}^{sim}\,<\,V_{therm}^{exp}\,\ll\,V_{probe}^{sim,exp}.
\end{equation*}
Now that the temperature probe volume encompasses the entire thermalization volume, the extraction of the thermal conductivity based on our Fourier model leads to our final thermal conductivity $\kappa^{\,2LRT}_{\,\,memb}\,=95^{+11}_{-7}\,$\Wu. The fraction of $P_{abs}$ that is measured in the temperature probe volume of all three experimental techniques is reported in Fig.\,\ref{fig:1LRT2LRTcomparison}, along with the main extensions $p_{abs}$, $r_{probe}$, and $r_{range}$ that predominantly limit the temperature probe volume during our Raman thermometry. These extensions then enable the comparison to our results of \textit{ab initio} calculations shown in Fig.\,\ref{fig:kxx_accum_mfp}.

\subsubsection*{\label{subsubsec:laser_choice}Choice of the heating laser\protect}

Let us note that we applied a cw 325\,nm heating laser for our 1LRT measurements and a cw 266\,nm heating laser for our 2LRT$_0$ and 2LRT measurements. The reason for this choice is twofold. First, based on our 266\,nm laser we were not able to acquire high quality resonant Raman spectra for 1LRT measurements. Despite the high ultraviolet sensitivity of our experimental setup, even integration times of 10\,min per resonant Raman spectrum did not yield a solid data basis. Thus, we transitioned to a 325\,nm cw laser for 1LRT measurements, utilizing a microscope objective optimized for this wavelength. However, for 2LRT$_0$ and 2LRT measurements, we still applied our 266\,nm cw laser due to the larger available range for $P_{abs}$, but also the dual-wavelength range design of our main microscope objective. This particular microscope objective is optimized around 266\,nm and the visible range around 500\,nm (high transmission and overlapping focal planes), but exhibits poor transmission at 325\,nm ($\leq\,15\%$). Thus, the light penetration depth for the heating laser changes from $p_{abs}\,=\,74\,\text{nm}$ \cite{kawashima_optical_1997} for 1LRT measurements to $p_{abs}\,=\,45\,\text{nm}$ \cite{kawashima_optical_1997} for 2LRT$_0$ and 2LRT measurements. However, this change in $p_{abs}$ will not affect the overall trend for all three $\kappa$ values in Fig.\,\ref{fig:1LRT2LRTcomparison}, on the contrary, the overestimation of $\kappa^{\,1LRT}_{\,\,memb}$ should even be more pronounced for a 266\,nm heating laser.

\section{\label{sec:DiscussionOutlook}Discussion and outlook \protect}

As a result of Fig.\,\ref{fig:1LRT2LRTcomparison}, 2LRT measurements appear as the only suited method to measure $\kappa$ in our photonic membrane made from III-nitride material. Still 1LRT measurements could also be applied to measure realistic values for $\kappa$. But for this purpose, the laser spot size and laser penetration depth would need to be tuned in a way such that their corresponding length scales ($r_{probe}\,\approx\,r_{abs}$ and $p_{probe}\,\approx\,p_{abs}$) are sufficient to probe the major fraction of heating carrying phonons that significantly contribute to $\kappa$, cf. Fig.\,\ref{fig:kxx_accum_mfp}. Let us note however that for direct bandgap semiconductors like GaN whose absorption coefficient exceeds $1\,\times\,10^{5}\,\text{cm}^{-1}$ above the bandedge \cite{Muth1997a}, this remains a challenging task. To achieve larger $p_{probe}$ values for 1LRT measurements in GaN, one would need to tune the laser wavelength around the onset of absorption in close energetic proximity to the bandedge \cite{callsen_excited_2018}. However, in this energetic regime the absorption features of band-to-band transitions and excitons still overlap (e.g., at $T_{amb}\,=\,295\,\text{K}$), rendering any precise determination of $P_{abs}$ a challenging task. This is especially true for the comparably large $T_{rise}$ values common for Raman thermometry \cite{beechem_invited_2015}. As soon as temperatures are so drastically increased, the energetically steep absorption onset of excitons and/or band-to-band transitions shifts towards lower energies \cite{kudlek_long-term_1990}, which in turn varies the absorption. 
In addition, depending on the local temperature, the balance between the different contributions to the absorption is altered. As a result of these thermal non-linearities, one lacks a constant value for the absorption coefficient at a fixed heating wavelength, which prevents any meaningful 1LRT measurements. In general, changes in the absorption coefficient are less pronounced deeper in the bands of a direct bandgap semiconductor compared to the region close to its bandedge. Nevertheless, for large values of $T_{rise}$, the absorption coefficient at a certain wavelength cannot always be considered as constant, even when the energetic spacing to the bandedge is increased. Clearly, such optical non-linearities remain an interesting task for future work. Potential optical non-linearities related to the absorption coefficient are another reason why we refrain from a more detailed analysis of the high temperature data shown in Figs.\,\ref{fig:1LRT} and \ref{fig:2LRT_wod} ($>\,150\,\text{K}$). Interestingly, the high absorption coefficient of GaN that complicates the Raman thermometry is directly connected to the reason why GaN has such a high relevance for photonic applications, which commonly rely on direct bandgap semiconductors for emitters.

For the thermal characterization via TDTR measurements, it is common practice to work with sufficiently large and well defined laser spot diameters \cite{li_gan_2020} on the order of several tens of micrometers. At the same time, the light penetration depths is less problematic compared to our Raman thermometry technique as usually metal transducers made from, e.g., gold, aluminum, and silver are employed. On the one hand, this makes TDTR measurements a quantitative, reliable, and consequently well-established thermal technique. On the other hand, the complications that arise for Raman thermometry also carry a strong potential for measuring the mean free path length of heat carrying phonons and their impact on $\kappa$. By controlling the heat and probe laser spot diameters during  transient thermoreflectance (TTR) measurements, Minnich \textit{et\,al.} have pioneered such phonon mean free path lengths measurements in silicon \cite{minnich_thermal_2011}. Interestingly, our results suggest that Raman thermometry could, similarly to TTR measurements, be suited for measuring phonon mean free path lengths, which is particularly promising due to the experimental simplicity of Raman spectroscopy compared to pump-probe techniques. Clearly, this potential for measuring heat carrying phonon mean free path lengths adds to the overall interesting aspects of 2LRT measurements that were shown in this work. Not only that a high spatial resolution can be achieved by our non-invasive experimental technique, but also the full angular dependence of $\kappa$ in highly non-symmetric semiconductors like $\beta$-Ga$_2$O$_3$ \cite{guo_anisotropic_2015,slomski_anisotropic_2017} could readily be explored. In addition, as in TDTR and micro-TDTR \cite{maire_heat_2017} measurements, Raman thermometry carries the potential for measurements at cryogenic temperatures. Still, the temperature limits of Raman thermometry for GaN and AlN need to be explored in full detail. Furthermore, our 2LRT measurements could even represent an interesting alternative to other non-invasive thermometry techniques like thermal transient grating (TTG) spectroscopy \cite{duncan_thermal_2020}. Thanks to the spatial resolution of 2LRT measurements one could potentially directly witness effects of non-Fourier-like thermal transport in real-space, such as a negative local effective thermal conductivity: close to the material's edge, the heat flux may have the same sign as the temperature gradient - a clear violation of Fourier's law ~\cite{hao_frequency-dependent_2009}. Therefore, our experimental setup may open up exciting opportunities to probe these effects, while remaining non-intrusive. 

\subsection{\label{sec:DiscussionOutlook_Minnich}Comparison to transient thermoreflectance \protect}

The direct comparison in between the pioneering work of Minnich \textit{et\,al.} \cite{minnich_thermal_2011} (TTR) and our Raman thermometry summarized in Fig.\,\ref{fig:1LRT2LRTcomparison} implies some additional insight into two different experimental perspectives on thermal transport as soon as the length scale of the heated volume in a bulk semiconductor approaches the mean free path lengths of the heat carrying phonons $l^{MFP}$.

First, we will start with a brief summary of the findings of Minnich \textit{et\,al.}. Below an ambient temperature of 200\,K they found that the effective thermal conductivity measured on high quality, bulk silicon declined with the diameter of their heater scaling from 60\,-\,15\,$\mu$m (1/e$^2$ definition of the diameter). For the case of these TTR measurements, the heater is given by an Al transducer that is heated by a focused laser. Exactly this Al transducer is then also used to measure the evolution of temperature via its relative reflectivity changes $\Delta\,R/R$ probed by a second laser. As this is a transient experiment, $\Delta\,R/R$ declines (specific to the choice of the metal transducer and probe laser wavelength) over time, which can be recorded in a, so-called, thermal transient. Subsequently, relevant material parameters like, e.g., $\kappa$ can be derived from these transients by comparison to a Fourier-based dynamic model based on the time-dependent heat equation \cite{schmidt_pulse_2008}. The interpretation of their findings is based on the assumption that the phonon bath can be divided into a purely diffusive and purely ballistic fraction, which represents a well-justified simplification. When lowering the diameter of their laser heat spot on the metal transducer (i.e, the heater) in comparison to a smaller and constant probe spot diameter (11\,$\mu\text{m}$), they measure a most pronounced decline of effective $\kappa$ values at, e.g., an ambient temperature of 60\,K. This decline is then attributed to an increasingly large fraction of heat carrying phonons that propagates ballistically. In the approximation of Minnich \textit{et\,al.} this sub-group of phonons undergoing ballistic transport does not contribute to thermal transport in contrast to the sub-group of diffusively propagating thermal phonons that they predominantly probe by their TTR experiment (regulated by the diameters of the heat and probe laser spots). As a result, the heat flux from the perspective of their transient temperature probe is lower than predicted by Fourier's law, which is then interpreted as a ballistic thermal resistance. Finally, the increase of this ballistic thermal resistance with declining diameter of their heater leads to the measurement of an effective reduction of $\kappa$ at $\leq\,200\,\text{K}$ in silicon.

As good experimental practice, during the reduction of the heater diameter the laser power density was kept constant on the surface of the metal transducer during the TTR experiments. Consequently, always an identical temperature rise should have been achieved in the Al transducer. However, due to the reduction of the diameter of the heat spot, this does not necessarily lead to always identical temperature rises in the silicon material underneath, when transitioning to an increasingly ballistic regime of thermal phonon transport. Here, the ratio in between the heat and probe laser focus spot diameters will impact the temporal evolution of the diminishing temperature rises (i.e., local cooling) that are indirectly measured via $\Delta\,R/R$ for ambient temperatures $\leq\,200\,\text{K}$. Thus, for the interpretation of the TTR measurements the problem already arises when translating $\Delta\,R/R$ values to temperature rises in the material underneath the metal transducer based on the oversimplified model given by the time-dependent heat equation. Based on this model, the thermal transients provide access to the thermal flux, while the temperature gradient around the heat spot in  silicon is overestimated, resulting in a decline of the effective $\kappa$ values with diminishing diameter of the heater. Thus, for TTR measurements the challenge rests in the direct determination of the temperature rise in the material as soon as the dimension of the heater approaches the $l^{MFP}$ values of a significant fraction of the heat carrying phonons.

Exactly at this point our Raman thermometry can provide an interesting additional perspective as, e.g., 1LRT measurements always directly probe absolute temperatures rises directly in the material of choice at the cost of all complications that come along with an optical heating of a sample, cf.\,Sec.\,\ref{sec:1LRT}. For our 1LRT measurements with the smallest achievable temperature probe volume, we directly measure a temperature rise that is overestimated by our numerical model based on the stationary heat equation and Gaussian surface heating, cf.\,S-Sec.\,IV. Exactly this small temperature rise is then interpreted as a sign of high thermal conductivity within the limitations of Fourier's law. Thus, despite of the fact that Raman thermometry can probe (here: steady state) temperature rises directly in the sample region of interest, it still suffers from the application of oversimplified modeling. Consequently, the TTR measurements from Minnich \textit{et\,al.} observe a reduction of the thermal flux with decreasing heat spot size, which is translated into a reduction of effective $\kappa$ values by a Fourier-based model. In contrast, we directly observe an absolute lowering of the temperature rise in the temperature probe volume of 1LRT measurements in comparison to our Fourier-based model, which is then misinterpreted as being caused by a large thermal flux due to a large $\kappa$ value. 

However, in the end the approach for TTR and Raman thermometry to measure realistic $\kappa$ values within the corset of any Fourier-based model is similar. By increasing the diameter of the heater during TTR measurements increasingly larger effective values for $\kappa$ are measured until a realistic value is reached by comprising the most significant fraction of $l^{MFP}$ of all heating carrying phonons in the experiment. For Raman thermometry the same happens when the temperature probe volume is increased (transition from 1RLT, over 2LRT$_0$, to 2LRT), just that increasingly smaller effective values for $\kappa$ are measured, until again a realistic value for $\kappa$ is obtained. Finally, this interesting comparison of TTR measurements and Raman thermometry directly illustrates the need for more appropriate modeling to take advantage of any direct temperature measurement enabled by Raman thermometry. Nevertheless, both experimental techniques provide promising experimental access to the $l^{MFP}$ distribution of all heat-carrying phonons.

\subsection{\label{sec:Carriers}Photo-generated carriers \protect}

In general, the main challenge with Raman thermometry is the complicated generation of heat via the absorption of laser light. Not only heat-carrying phonons are generated and contribute to $\kappa$, but also free carriers and excitons can be generated. Consequently, depending on the sample quality and $T_{amb}$, a complicated balance of different contributions remains to be disentangled. In this contribution, we simplified this problem, by characterizing a particularly suited photonic membrane by 2LRT measurements, which allowed us to focus on thermal transport dominated by heat-carrying phonons. All these potential additional contributions to thermal transport also render 2LRT measurements advantageous over 1LRT measurements, as the potentially complicated area where the laser light is absorbed can be excluded from the thermal characterization. Thus, the analysis of 2LRT data can begin, e.g., several micrometers away from the heating spot, which is in many direct bandgap semiconductors enough to exclude a dominant contribution of excitons and carrier transport to thermal transport as mean free path lengths are -\,in the best case scenario\,- limited by the radiative lifetime of these excitations \cite{weatherley_imaging_2021}. However, this simplification cannot be taken for granted in indirect bandgap semiconductors of sufficiently high crystalline quality, which can exhibit comparably longer mean free path lengths for excitons and carriers.

Furthermore, we can exclude that our trend for the $\kappa$ values extracted from 1LRT, 2LRT$_0$, and 2LRT measurements (see Fig.\,\ref{fig:1LRT2LRTcomparison}) is predominantly impacted by photo-generated carriers. For the 2LRT measurements, this argument trivially holds as the probed temperature region is sufficiently far away from the laser-induced heating spot in comparison to the mean carrier diffusion length $l_{\text{diff}}\leq\,60\,\text{nm}$ in our sample, cf. Sec.\,\ref{sec:Experiment} and S-Sec. II. See the top axis of Fig.\,\ref{fig:2LRTfit}(a) regarding the distance of the temperature probe region to the heating spot. In our 1LRT and 2LRT$_0$ measurements the situation is more complicated, yet we can estimate an upper bound for the photo-generated carrier concentration in our photonic membrane. At $T_{amb}\,=\,295\,\text{K}$ we measured an effective lifetime of $\tau_{QW}\,=\,60\,\pm\,5\,\text{ps}$ for the QW transition ($E_{1}\,=\,2.75\,\text{eV}$) in our sample based on time-resolved PL (TRPL) measurements \cite{Rousseau2018c}. Commonly, the carrier lifetime $\tau_{GaN}$ in the GaN matrix material that surrounds the QW is even lower as the QW traps the carriers. However, direct TRPL measurements on the bandedge emission of GaN ($E_{2}\,=\,3.41\,\text{eV}$) at $T_{amb}\,=\,295\,\text{K}$ were not feasible due to the low intensity of this band, which is caused by the intended carrier transfer to the QW. Thus, based on $\tau_{GaN}\,<\,\tau_{QW}$ and the corresponding excitation and temperature probe volumes, we can estimate an upper bound for the photo-generated carrier concentration that is achieved during our 1LRT and 2LRT$_0$ measurements. During our 1LRT measurements shown in Fig.\,\ref{fig:1LRT}, at the largest $T_{rise}\,=\,780\,\pm\,30\,\text{K}$ ($P_{abs}\,=\,13.5\,\text{mW})$, we reach the maximal number of photo-generated carriers with a concentration $\rho^{photo}_{1LRT}\,\approx\,1.7\,\times\,10^{19}\,\text{cm}^{-3}$ in the temperature probe volume. Clearly, if not photo-excited, such high carrier concentration can only be achieved by heavy doping of GaN \cite{Nenstiel2015a}. Let us note that this concentration represents an upper bound for $\rho^{photo}_{1LRT}$, because $\tau_{GaN}$ and $\tau_{QW}$ decline with increasing temperature due to the impact of non-radiative defects. At the same time with rising temperature the concentration of purely thermally excited free carriers also rises in GaN. The situation regarding the photo-excited carriers is similar for our 2LRT$_0$ measurements. However, due to the larger temperature probe volume we reach $\rho^{photo}_{2LRT_{0}}\,\approx\,3.2\,\times\,10^{18}\,\text{cm}^{-3}$ at $T_{rise}\,=\,730\,\pm\,130\,\text{K}$ ($P_{abs}\,=\,9.3\,\text{mW}$). Thus, any future detailed analysis of the full $T_{rise}$ trends that we show for 1LRT and 2LRT$_{0}$ measurements in Figs.\,\ref{fig:1LRT} and \ref{fig:2LRT_wod}, will require temperature-dependent TRPL measurements in order to be conclusive. However, as we focus on $T_{rise}\,\leq\,150\,\text{K}$ in this work to extract all $\kappa$ values, the photo-generated carrier concentrations decrease. For this lower value of $T_{rise}$, we obtain upper bounds for the photo-generated carrier concentrations of $\rho^{photo}_{1LRT}\,\approx\,6.4\,\times\,10^{18}\,\text{cm}^{-3}$ and $\rho^{photo}_{2LRT_{0}}\,\approx\,1.4\,\times\,10^{18}\,\text{cm}^{-3}$.

Overall, the phonon current and hence $\kappa$, can be degraded by the phonon-charge carrier interaction since this causes a transfer of momentum from the phonon system to the charge carrier system. However, the charge carriers themselves can also carry heat and contribute positively to the total $\kappa$. However, even in heavily doped semiconductors, the phonon contribution to $\kappa$ completely dominates the charge carriers' contribution and any electron drag enhancement of the phonon $\kappa$ has also been found to be negligible in GaAs\,\cite{protik_coupled_2020}, SiC\,\cite{protik_electron-phonon_2020}, and Si\,\cite{protik_elphbolt_2022, zhou_ab_2015}. For doped Si Liao \textit{et\,al.} analyzed the impact of free carriers and holes on $\kappa$ based on first principles \cite{liao_significant_2015}. For \textit{n}-type silicon they predict a moderate reduction of $\kappa$ in silicon of $2\,\text{to}\,7\%$ for free electron concentrations scaling from $1\,\times\,10^{18}\,\text{cm}^{-3}$ to $1\,\times\,10^{19}\,\text{cm}^{-3}$ due to the electron-phonon interaction. In GaN this electron-phonon interaction is enhanced due to the polar nature of acoustic phonons, which has already motivated several, mostly theoretical studies in this field \cite{yang_nontrivial_2016,tang_thermal_2020,muthukunnil_joseph_electron_2022,tang_phonon_2023}. Accordingly, a reduction of $\kappa$ with rising carrier concentration is commonly reported for homogeneously doped samples, which contradicts the experimental trend that we observe for our Raman thermometry measurements summarized in Fig.\,\ref{fig:1LRT2LRTcomparison}. Thus, as we observe a decrease in $\kappa$ with decreasing density of photo-generated carriers in the temperature probe volume, scaling from 1LRT, over 2LRT$_{0}$, to 2LRT measurements, we can exclude any prominent impact of photo-generated carriers on $\kappa$ in our photonic membrane. 

It nonetheless remains an interesting task to probe $\kappa$ over the full range of experimentally accessible $T_{rise}$ values in, e.g., GaN membranes of even higher quality in comparison to the sample that formed the basis for this work. Here, GaN membranes that originate from GaN grown on sapphire or even bulk GaN substrates could form a basis \cite{ciers_smooth_2021}. In such samples even higher values for $\tau_{GaN}$ and $\tau_{QW}$ can be expected, potentially leading to even higher achievable photo-generated carrier concentrations with a stronger impact on $\kappa$ and its temperature dependence. It also remains an interesting open question, how a local photo-excitation-induced carrier distribution can be compared to global carrier distributions induced by, e.g. doping and/or thermal excitation.

\section{\label{sec:Summary}Summary \protect}

In summary we show, how quantitative and non-invasive thermometry by optical means can be achieved by spatially resolved 2LRT measurements on a photonic semiconductor membrane made from III-nitrides. Thanks to the spatial resolution of the 2LRT technique, the analysis of temperature gradients in such a membrane can be limited to regions sufficiently far from the heating spot and the heat sink, which simplifies the data analysis. This simplification goes hand-in-hand with the possibility to disentangle different contributions to thermal transport given by, e.g., charge carriers and heat-carrying phonons by their deviating mean free paths. Interestingly, we demonstrated 2LRT measurements on a partially under-etched photonic membrane, which in very similar form has already formed the basis for nanolasers \cite{vico_trivino_high_2012,trivino_continuous_2015,rousseau_quantification_2017,jagsch_quantum_2018,Rousseau2018a,Rousseau2018c}. Consequently, our thermal analysis of the most fundamental building block of such nanolasers can form the basis for future thermal optimizations of these structures. For thermometry that applies a laser as the heat source, such light emitting structures inflict the particular challenge of a precise determination of the heating power because part of the absorbed laser power is converted to light in the In$_{x}$Ga$_{1-x}$N/GaN ($x=0.15$) QW structure that forms the core of our photonic membrane. Thus, based on our interlinked thermal and optical analysis in our fully customized optical setup, we showed that we can safely neglect the energy loss by the emission of photons. Furthermore, our experimental setup pioneers an approach for 2LRT measurements that only requires front-side access of the sample - a technical, but still relevant step regarding the applicability of 2LRT measurements to a large variety of samples. Commonly photonic membranes are never fully under-etched, which renders 2LRT measurements a preferential choice for their analysis, which is further strengthened by the fact that no metal transducer is required as for most alternative, reflectivity-based techniques.

Experimentally we introduced our 2LRT measurements in a step-by-step approach starting with 1LRT and 2LRT$_0$ measurements, exhibiting different heat and temperature probe volumes and a differing sensitivity to thermally induced stress. Based on \textit{ab initio} calculations of phonon transport we show that 1LRT measurements always overestimate $\kappa$ in our sample due to an insufficient size of the temperature probe volume. For these calculations, we took into account the main limiting factors of phonon transport in our photonic membrane, which comprises 3- and 4-phonon-scattering, phonon-isotope scattering, as well as phonon boundary scattering. As an outcome of this theory of phonon transport in a photonic membrane, we were also able to elucidate an intriguing scaling behavior for $\kappa$. From 1LRT, over 2LRT$0$ to 2LRT measurements the temperature probe volume increases, which is accompanied by a systematic decrease of $\kappa$. Based on the phonon mean free path lengths obtained from our calculations, we showed that only 2LRT measurements allow us to encompass the entire set of heat-carrying phonons that contributes to $\kappa$. This result is a central aspect of this work, as it opens the perspective for phonon mean free path spectroscopy based on a systematic variation of the temperature probe volume used for Raman thermometry. Precise control over the heated volume and the temperature probe volume is especially relevant for direct bandgap semiconductors often used for photonic applications due to the occurrence of absorption coefficients at the wavelength of the heat laser. Consequently, 1LRT measurements can only provide quantitative results, if the temperature probe volume defined by the absorption depth of the Raman laser and its focus diameter are adapted to the distribution of phonon mean free paths at the temperature of interest. Here, the quasi-ballistic transport of heat-carrying phonons at room temperature in, e.g., GaN poses a general challenge to Raman thermometry aiming for high spatial resolution, while maintaining the aim of a precise measure of $\kappa$. Therefore, 2LRT measurements provide a generally promising alternative, as the entire spatial temperature distribution across the sample can be probed. For future work, it is of utmost importance to directly model these experimental temperature distributions. Consequently, the comparison to calculations will no longer be limited to a comparison of $\kappa$ values, but extended to a direct comparison with non-Fourier-like temperature maps. This point is even further strengthened by the possibility to measure 2LRT down to cryogenic temperatures, which shall give direct experimental access to different phonon transport regimes by temperature imaging.

\hfill

\section*{Acknowledgment}
N.H.P. was supported by a Humboldt Research Fellowship from the Alexander von Humboldt Foundation, Bonn, Germany. M.E. and G.C. acknowledge funding from the Central Research Development Fund (CRDF) of the University of Bremen for the project "Joined optical and thermal designs for next generation nanophotonics". The research of M.E., W.S., K.D., D.V., and G.C. was funded by the Deutsche Forschungsgemeinschaft (DFG, German Research Foundation) – 511416444. G.R. acknowledges MIT-IBM Watson AI Laboratory (Challenge No. 2415). This work was supported by the Swiss National Science Foundation through Grant Nos. 200020\_162657 and 200021E\_175652.

\bibliography{1-testBib_ME}

\end{document}


\title[]{SUPPLEMENTAL MATERIAL: \\Optical and thermal characterization of a group-III nitride semiconductor membrane by microphotoluminescence spectroscopy and Raman thermometry}

\author{Mahmoud Elhajhasan}
\author{Wilken Seemann}
\author{Katharina Dudde}
\author{\\Daniel Vaske}
\author{Gordon Callsen*}
\email{gcallsen@uni-bremen.de}

\affiliation{Institut für Festk\"orperphysik, Universit\"at Bremen, Otto-Hahn-Allee 1, 28359 Bremen, Germany}

\author{Ian Rousseau}
\author{Thomas F. K. Weatherley}
\author{\\Jean-François Carlin}
\author{Raphaël Butt\'{e}}
\author{Nicolas Grandjean}

\affiliation{Institute of Physics, \'{E}cole Polytechnique F\'{e}d\'{e}rale de Lausanne (EPFL), CH-1015 Lausanne, Switzerland}

\author{Nakib H. Protik}

\affiliation{Institut f\"ur Physik und IRIS Adlershof, Humboldt-Universit\"at zu Berlin, 12489 Berlin, Germany}

\author{Giuseppe Romano}

\affiliation{MIT-IBM Watson AI Lab, IBM Research, Cambridge, MA, USA}

\date{\today}

\maketitle

\section{\label{S-Sec:Setup}Experimental details}

\subsection{\label{S-Sec:Setup}Customized experimental setup}

%
%
%
\begin{figure*}[h]
    \includegraphics[width=16 cm]{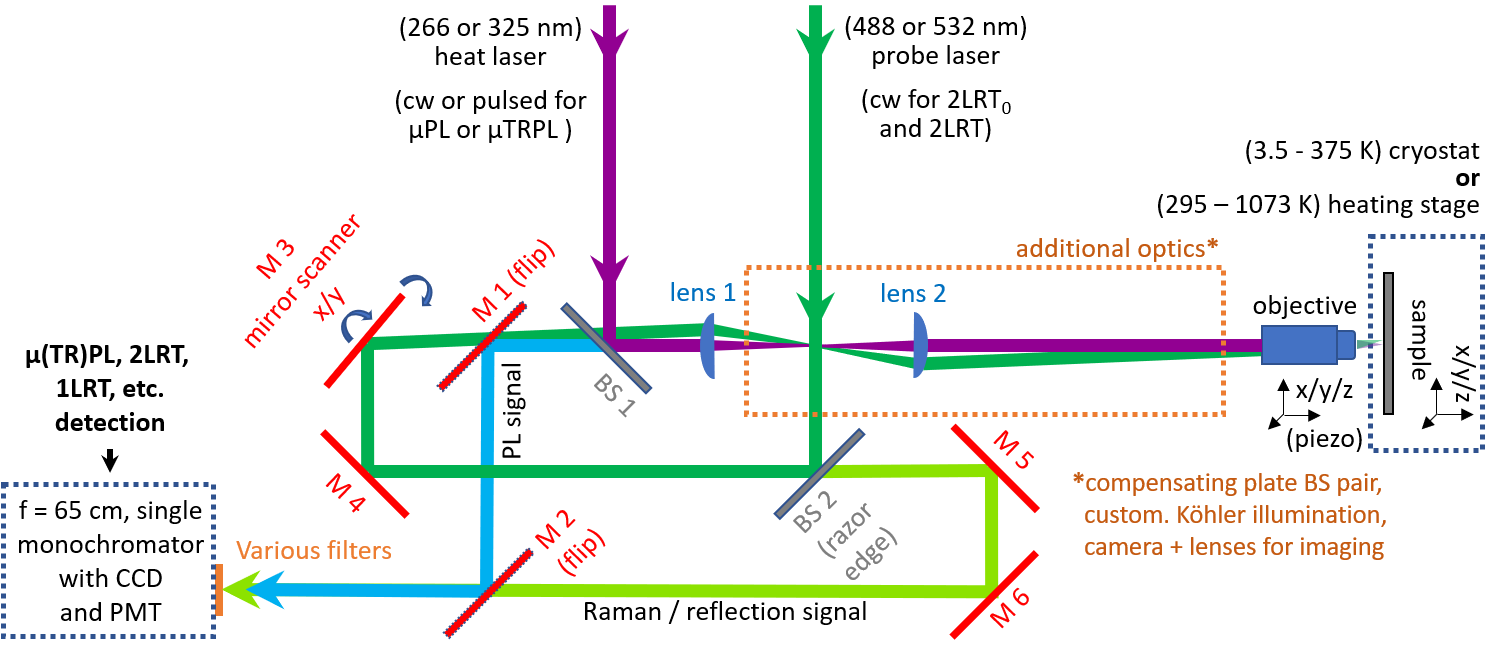}
    \caption{\textit{Sketch of the customized optical setup (not to scale) that can perform (time-resolved) $\mu$PL measurements in parallel to all three different types of Raman thermometry in use for the present work (1LRT, 2LRT$_0$\,, and 2LRT). In addition to the PL and Raman signal, the reflected light can also be recorded to support the determination of the absorbed laser power ($P_{abs}$). The beam path of the two different cw heating lasers with a wavelength of 325\,nm or 266\,nm is depicted in purple. A pulsed 266\,nm laser can also be used for time-resolved $\mu$PL measurements. The beam path followed by the $\mu$PL signal and/or the resonant Raman signal (1LRT) is depicted in blue. For probing the temperature or for conventional non-resonant Raman spectroscopy, a 488\,nm or 532\,nm laser is utilized as shown by the green beam path. The independent scanning of the laser focus points over the sample is realized either by a mirror scanner (M\,3), a three-axis piezo stage holding the microscope objective, or a low-temperature three-axis piezo stage inside a closed-cycle He-cryostat. The sample can either be placed in this He-cryostat or in a heating stage. Finally, the signal of interest (e.g., blue - PL signal, light green - non-resonant Raman signal) is guided towards the detection system. Please see the main text for further details.}}
    \label{S-Fig:full_setup}
\end{figure*}

The simplified sketch of our experimental setup from the main text (Fig.\,1) is drawn in more detail in S-Fig.\,\ref{S-Fig:full_setup}. The beam path for the ultraviolet (UV) continuous wave (cw) lasers that are in use for the micro-photoluminescence ($\mu$PL) measurements and the Raman thermometry is depicted in purple. First, the heating laser is reflected by the beamsplitter 1 (BS\,1), which is a 45$^{\circ}$ long-pass Raman beamsplitter from Semrock for either 266\,nm or 325\,nm. BS\,1 is followed by two scan lenses (lenses 1 and 2, distances not to scale), which compensate each other with respect to the UV laser in use. For $\mu$PL measurements we utilized either a 325-nm-laser from Kimmon (model: IK3501R-G) or a 266-nm-laser from Crylas (model: FQCW266-50). The 325\,nm laser was also used for the one-laser Raman thermometry (1LRT) on bulk GaN and the GaN membrane, cf. Fig.\,4 from the main text. Only with this UV laser we were able to record the required resonant-Raman spectra on GaN. For two-laser Raman thermometry with (2LRT) and without spatial displacement (2LRT$_0$) we switched to our 266\,nm laser due to its superior beam profile and improved laser pointing stability. Furthermore, our microscope objective in use (80$\times$ Mitutoyo Plan UV Infinity Corrected Objective) exhibits a dual-wavelength-design for around 266\,nm and 532\,nm. This particular design of our microscope objective guarantees a similar focal plane for these two wavelength ranges, while the corresponding transmission is also optimized. Thus, we selected probe lasers with a wavelength of 488\,nm (manufacturer: Coherent, model: 2\,W Genesis CX SLM) and 532\,nm (manufacturer: H\"ubner Photonics, model: 300\,mW Samba) for measuring non-resonant Raman spectra required for 2LRT$_0$ and 2LRT measurements. The particular dual wavelength design of our microscope objective then allowed us to have the heating laser (purple beam path in S-Fig.\,\ref{S-Fig:full_setup}) and the temperature probe laser of choice (green beam path in  S-Fig.\,\ref{S-Fig:full_setup}) in focus, while sufficient heating laser powers could straightforwardly be achieved on the sample. As any focusing process via a motorized micrometer screw ($<\,100\,\text{nm}$ step resolution) does not always necessarily yield the same laser spot diameters, we carefully measured the focus spot diameters of our heating and temperature probe lasers for every series of measurements we performed. Please see S-Sec.\,\ref{S-Sec:Spotsize} for further details regarding the laser focus spot size determination.

The temperature probe laser depicted in green in S-Fig.\,\ref{S-Fig:full_setup} is routed towards the entrance aperture of the microscope objective via a suited 45$^{\circ}$ long-pass Raman beamsplitter (BS\,2) from Semrock for the wavelength of choice, several mirrors, and a 2-axis mirror scanner (M\,3) from Thorlabs (model: GVS202). In combination with the two scan lenses (focal length: 15\,cm), this mirror scanner allows tilting the temperature probe laser beam in the entrance aperture of the microscope objective, which results in a movement of the corresponding laser spot in its image plane. We observed that scanning ranges up to 100\,$\times$\,100\,$\mu\text{m}^{2}$ did not show any scanning non-linearities nor laser spot variations beyond the error intervals given in S-Sec.\,\ref{S-Sec:Spotsize}. In addition to the scanning action of the temperature probe laser (step resolution $<\,100\,\text{nm}$), the heating laser can also be scanned over the sample surface by a 3-axis closed-loop piezo stage from "Physik Instrumente" (model: P-611.3S NanoCube XYZ), allowing for a scanning range of 100\,$\times$\,100\,$\mu$m$^{2}$ with a step resolution of $\ll\,50\,\text{nm}$. Furthermore, the absolute positioning reproducibility of both scanning system was always better than 10\% of the smallest achievable laser spot size diameter (FWHM). During the $\mu$PL spectroscopy and Raman thermometry the sample was kept in vacuum in a closed-cycle He-cryostat (cooling not in use in the present work) at a base pressure $<\,5\,\times$\,10$^{-6}$\,mbar. In this cryostat the sample was mounted on an open-loop, low temperature piezo stage from attocube (model: ANPX/Y/Z51) via a customized copper holder that was directly connected to the ground plate of the cryostat via copper filaments (known as a "thermal link"). In this work this low temperature piezo stage was only used for a coarse positioning of our sample with excellent absolute positioning stability ($\ll\,50\,\text{nm}$). The copper filaments allowed to transfer any heat that developed during the experiments to the ground plate of the cryostat, which acted as the heat sink. To favor heat transfer from the sample to the copper holder we used silver paint to bond our sample. In addition, our copper sample holder featured a temperature sensor directly positioned under the sample. Throughout all our 1LRT, 2LRT$_0$, and 2LRT measurements that relied on elevated powers of the heating laser, we also constantly monitored any temperature rise in our sample holder. All measurements shown in the main text showed temperature rises in the sample holder of $<\,1\,\text{K}$, which is below the best temperature resolution for Raman thermometry we achieved in this work. In addition, the sample can be mounted on a heating stage from Instec (model: HCS421VXY) in our setup to enable temperature calibration measurements, cf. Sec.\,\ref{S-Sec:Calibration}.

Upon excitation of the sample with the heating laser, the resulting PL and resonant Raman signals (shown in blue) are routed towards the detection system via the flip mirrors M1 and M2 shown in S-Fig.\,\ref{S-Fig:full_setup}. However, before entering the detection system, the light passes a long-pass filter to suppress the spurious laser light from the PL and resonant Raman (Stokes side) signals. The light is then guided into the detection system consisting of a monochromator with a focal length of 64\,cm (model: FHR640) and a nitrogen-cooled charge-coupled-device (CCD) from Horiba (model: Symphony II BIUV) via a lens that enables an aperture matched light in-coupling to maximize light throughput and spectral resolution. The monochromator is equipped with a set of gratings that either enable overview $\mu$PL spectra (150\,l/mm ruled grating, spectral window size around a central wavelength of $400\,\text{nm}$: $>\,300\,\text{nm}$, spectral resolution: $\leq\,0.5\,\text{nm}$) or high resolution Raman spectra (1800\,l/mm and 2400\,l/mm holographic gratings). For instance, for our 532\,nm temperature probe laser we achieved a spectral resolution in the range of the first order Raman modes of GaN (Stokes side of the Raman spectrum), i.e., better than $<\,0.03\,\text{nm}$, which corresponds to around $1\,\text{cm}^{-1}$. As this spectral resolution is around three times better than the narrowest linewidths we report for the $E_{2}^{high}$ Raman mode of GaN at 295\,K, we do not necessarily require any deconvolution techniques to quantitatively measure linewidths at 295\,K and above within the error intervals that we state \cite{olivero_empirical_1977}. 

Due to the particular design of our optical setup shown in S-Fig.\,\ref{S-Fig:full_setup}, the Raman signal can simultaneously be measured while the heating laser is engaged. The corresponding beam path is shown in S-Fig.\,\ref{S-Fig:full_setup} in light green. After leaving the microscope objective's aperture, the Raman signal passes the scan lenses 1 and 2, is transmitted through BS\,1, and then routed towards the detection system via mirrors M\,3\,-\,M\,6, while mirrors M\,1 and M\,2 are removed from the beam path. Naturally, our experimental setup shown in S-Fig.\,\ref{S-Fig:full_setup} also features many additional optics like beam expanders (control over the laser spot sizes), automated filter wheels (important for Raman thermometry), and polarizers (commonly required for Raman spectroscopy), which we consider as standard spectroscopic equipment. Furthermore, as illustrated by the orange dashed box in S-Fig.\,\ref{S-Fig:full_setup}, our experimental setup comprises a customized optical microscope, featuring a camera and a homemade K\"ohler illumination to pre-position the sample with respect to the heating laser and temperature probe laser based on a conventional bright-field microscope image.

As our customized optical setup routinely measures relative shifts of the Raman modes less than $<\,0.1\,\text{cm}^{-1}$, which is at least ten times smaller than our spectral resolution in the visible range, great care must be devoted to the temperature stability of the laboratory and the monochromator as described for a monochromator of shorter focal length in Ref.\,\cite{fukura_factors_2006}. Therefore, we always ensured the temperature stability of our monochromator and laboratory to within $\pm\,0.1^{\circ}\text{C}$ during high resolution Raman measurements. Similar care was devoted to the mechanical stability of the entire optical setup. Within typical recording times of 2LRT maps of around 8\,h, we measured lateral drifts between the heating laser spot and the temperature probe laser spot of less than $50\,\text{nm}$ (x/y-plane), which also matches the lateral shifts we observed between our nano-structured sample surface and both laser spots. At the same time, minor variations ($\leq\,100\,\text{nm}$) also occurred in the \textit{z}-direction, which were predominantly caused by mechanical building oscillations (the setup is positioned on a well-isolated optical table with active self-leveling isolators). However, such focus deviations only result in laser spot size variations below the corresponding errors that we state in S-Sec.\,\ref{S-Sec:Spotsize} due to the comparably low numerical aperture (NA\,=\,0.55) of our UV microscope objective. In general, our experimental setup shown in S-Fig.\,\ref{S-Fig:full_setup} represents a universally applicable design that is  particularly suited for measuring $\mu$PL in combination with Raman thermometry (1LRT, 2LRT$_0$, and 2LRT) for the case where the samples are not fully underetched as described in the main text. As a result, one only has optical access to one side of the sample, which complicates the determination of the absorbed laser power ($P_{abs}$) that leads to local sample heating. However, in return a much larger variety of sample structures can be analyzed based on our experimental design that combines different laser scanning techniques with high resolution spectroscopy.

\subsection{\label{S-Sec:Spotsize}Laser spot size determination}

%
%
%
\begin{figure*}[]
    \includegraphics[width=\textwidth]{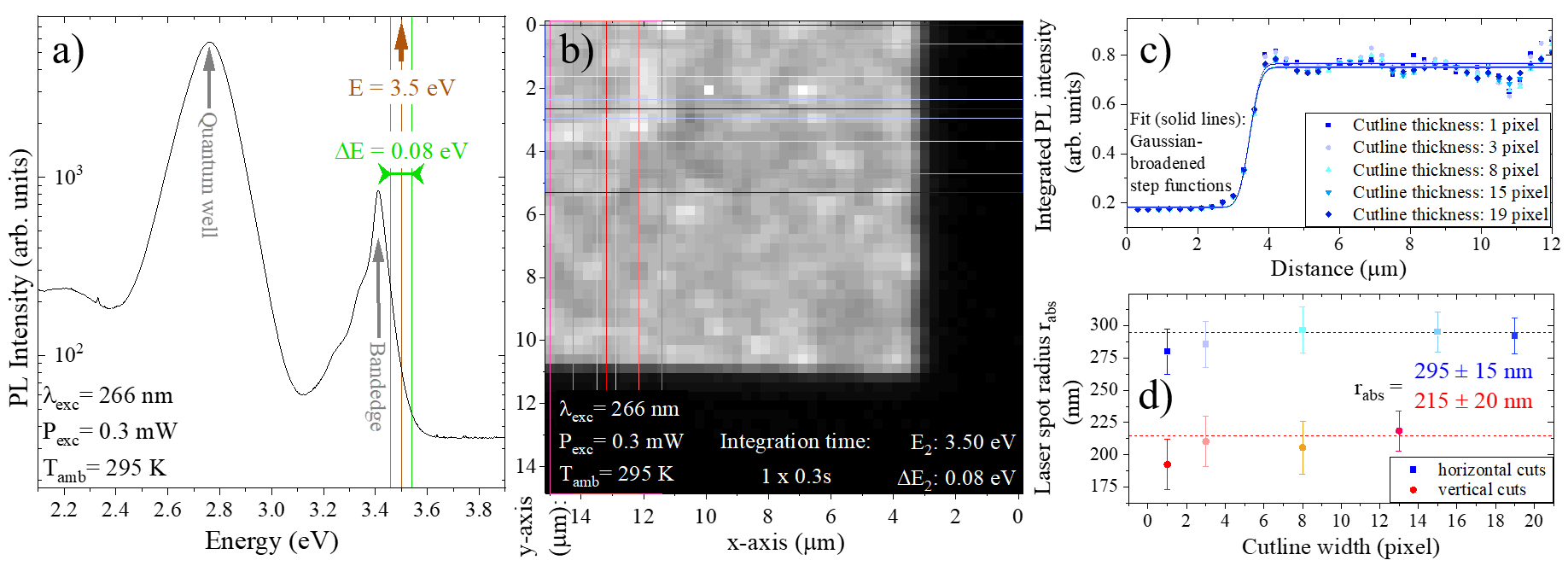}
    \caption{\textit{Laser spot size determination by $\mu$PL mapping. (a) Example of a $\mu$PL spectrum recorded in the freestanding corner region of the GaN pad at ambient temperature. Here, as for Fig.\,2 of the main text, the quantum well and GaN bandedge emissions are mainly visible. By defining a bandpass near $E\,=\,3.5\,\text{eV}$ (width: $\Delta\,E\,=\,0.08\,\text{eV}$), a spatially resolved intensity distribution can be plotted in a pixel mapscan as shown in (b). The spacing between the pixel equals 300\,nm. Cutlines through this pixel plot for different widths (given in pixel) can be extracted for horizontal (blue) and vertical (red) directions. Exemplary intensity distributions for such horizontal cuts (symbols) are shown in graph (c), which resemble Gaussian-broadened step functions that can be fitted (solid lines) by a suitable model. (d) As a result, we obtain the laser spot radius $r_{abs}$, which is accounting for the broadening of the step function. Here, the dependence of $r_{abs}$ on the direction and the binning width in pixel are also depicted. See the main text for any further details.}}
    \label{S-Fig:spotsize}
\end{figure*}

This section exemplifies the determination of the heating laser spot radius $r_{abs}$ for $\mu$PL mapscans recorded with our 266\,nm cw laser. The laser spot radius determination for the alternative heating laser (325\,nm) works in an identical manner. A similar determination of the laser focus spot sizes related to the temperature probe lasers (488\, nm or 532\,nm) builds on non-resonant Raman mapscans. Figure\,3 from the main text shows an example of such a Raman mapscan, which can be used to determine the corresponding temperature probe laser radius $r_{probe}$ in analogy to the method described hereafter. The only difference is given by the experimental implementation of the scanning of, e.g., the heating or temperature probe laser spot over the sample as described in S-Sec.\,\ref{S-Sec:Setup}. Similar results for the determination of $r_{abs}$ can also be obtained from resonant Raman mapscans, however, the much stronger $\mu$PL signal eases the determination of $r_{abs}$.

S-Figure \ref{S-Fig:spotsize}(a) shows the $\mu$PL signal of our sample, which is dominated by the signal of the quantum well (QW) and the bandedge emission of GaN at ambient temperature. S-Figure\,\ref{S-Fig:spotsize} can directly be compared to Fig.\,2(c) of the main text. For the following analysis, we chose a bandpass around $E\,=\,3.5\,\text{eV}$ (width: $\Delta\,E\,=\,0.08\,\text{eV}$) to generate a spatially resolved intensity plot as shown in S-Fig.\,\ref{S-Fig:spotsize}(b). The particular choice for the energy of this detection bandpass on the high energy side of the main emission peak related to the GaN bandedge emission (3.410\,eV) is of importance for the resulting $\mu$PL mapscan. At the given temperature the bandedge emission of GaN corresponds to an overlap of emission contributions related to the free A- and B-excitons as well as band to band transitions. Thus, in order to exclude any artificial spatial broadening in the $\mu$PL mapscans due to the propagation of light in the plane of the photonic membrane, it is important to position the detection bandpass sufficiently far above the onset of absorption in GaN at room temperature. As the energetic spacing between the excitonic absorption features and the absorption onset of the bandedge of GaN is given by the exciton binding energy (e.g., around 23\,meV for the A-exciton of GaN) \cite{callsen_excited_2018}, we chose to shift our detection bandpass by about 90\,meV to higher energies compared to the peak maximum of the GaN bandedge emission. Note that S-Fig.\,\ref{S-Fig:spotsize}(b) does drastically spatially broaden, if the bandpass is, e.g., shifted towards the QW emission shown in S-Fig.\,\ref{S-Fig:spotsize}(a). This can be understood as our photonic membrane made from GaN is designed to guide the light of the In$_{x}$Ga$_{1-x}$N ($x=0.15$) QW in its plane. Thus, one can directly observe how the light of the InGaN QW is outcoupled of the photonic membrane at its edges (not shown) when shifting the detection bandpass energy to the energetic range of the QW emission. A similar case occurs when shifting the bandpass to the low energy flank of the GaN bandedge emission. Here, however, only a minor additional spatial broadening of the image occurs as any in-plane transport of the corresponding light is still hindered by the absorption given by the QW, tail states in GaN, and deep levels in AlN.

The determination of the radius of the heating laser spot is based on either horizontal (blue) or vertical (red) cuts through the intensity mapscan from S-Fig.\,\ref{S-Fig:spotsize}(b). Cuts are illustrated for different pixel intervals, while the minimal pixel spacing amounts to 300\,nm. The resulting intensity distribution known as a cutline is exemplified in S-Fig.\,\ref{S-Fig:spotsize}(c) for horizontal cuts (blue) through the intensity distribution. For these horizontal cuts we summed and normalized the intensities for certain pixel intervals (1, 3, 8, 15, 19 pixels), aiming to improve the signal-to-noise ratio but also to gain insight into any rotation of our sample with respect to the scanning axes. Subsequently, we fitted each dataset (color-coded symbols) shown in S-Fig.\,\ref{S-Fig:spotsize}(c) by a Gaussian-broadened step function, which gives $r_{abs}$ as the origin of the broadening. The resulting trend for $r_{abs}$ depending on the pixel binning width is shown in S-Fig.\,\ref{S-Fig:spotsize}(d) for horizontal (blue hues) and vertical (red hues) cuts. We find that the resulting $r_{abs}$ values only show a minor dependence on the pixel binning widths, which illustrates that we successfully oriented the edges of our GaN membrane along the scan axes of our $\mu$PL mapping system. Nevertheless, by averaging the corresponding $r_{abs}$ values for the horizontal and vertical direction, we either obtain $r_{abs}\,=\,295\,\pm\,15\,\text{nm}$ (horizontal) or $r_{abs}\,=\,215\,\pm\,20\,\text{nm}$ (vertical). We associate this asymmetry to the ellipticity of the beam of our 266\,nm cw heating laser, which is caused by the two-step frequency doubling process in its cavity. Here, especially the last frequency doubling step from 532\,nm down to 266\,nm by a non-linear barium borate (BBO) crystal is known to cause such laser beam ellipticity, which then also appears in the associated focus spot. For the comparison between our data and the modeling, which is further detailed in S-Sec.\,\ref{S-Sec:Modeling}, we neglect this ellipticity in the focus spot of this heating laser by treating its heating spot as isotropic in-plane with $r_{abs}\,\approx\,250\,\text{nm}$. In particular for our 2LRT measurements, this small laser beam ellipticity is not of concern, as the temperature is probed by the probe laser (488\, or 532\,nm) at a distance of several micrometers away from the heating spot, cf.\,Fig.\,7 of the main text. 

Furthermore, it must be noted that our method to precisely measure $r_{abs}$ is based on the assumption that no additional mechanisms exist that might spatially broaden the intensity mapscan shown in S-Fig.\,\ref{S-Fig:spotsize}(b). Here we can exclude any direct impact of our sample as its edges are smooth down to the single-nm-range (see Fig.\,1 and Sec.\,I of the main text). Also we can exclude any projection effects from the silicon substrate as its surface is strongly roughened by the selective etching process described in the main text. As a result, no double excitation of our GaN membrane occurs based on spurious reflection of the heating laser from the silicon substrate. Furthermore, we also excluded mechanical vibrations or preferentially orientated sample drifts as additional broadening mechanisms. 

\section{\label{S-Sec:Cathodoluminescence}Cathodoluminescence mapping - carrier and exciton diffusion lengths}

%
%
%
\begin{figure*}[]
    \includegraphics[width=14 cm]{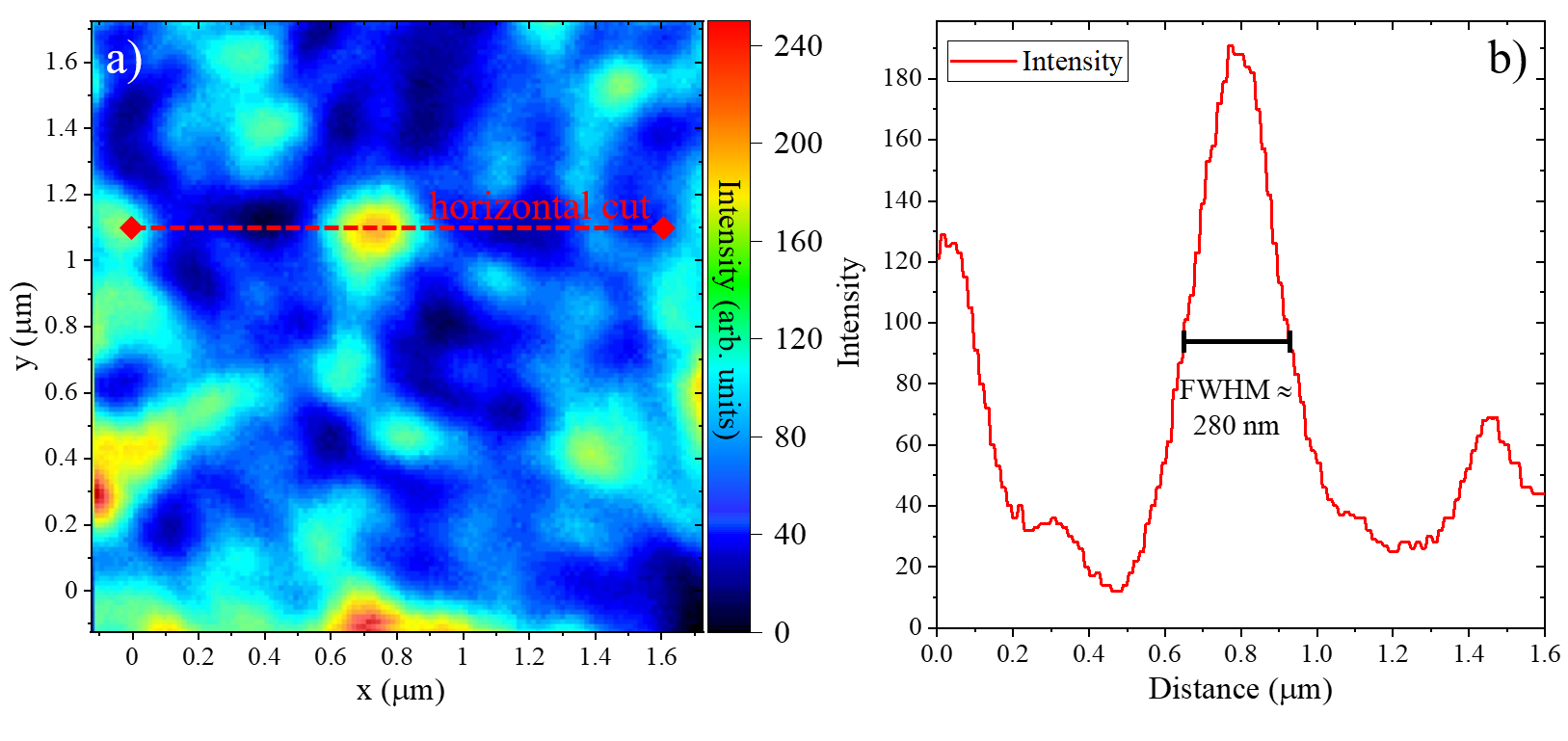}
    \caption{(a) Cathodoluminescence mapscan recorded with an acceleration voltage of 6\,kV and a probe beam current of 5\,nA in the corner region of the photonic membrane mainly made from GaN. This plot shows the intensity distribution for the bandedge emission of GaN for an energy of about 3.4\,eV with a bandpass of 50 meV at a temperature of 295\,K. Intensity fluctuations likely originate from carrier localization at defects. Plotting the corresponding intensity distribution along a horizontal cutline (red, dashed line) yields plot (b). Here, a selected and spatially well-isolated feature from the CL mapscan shows a spatial widths of $\approx\,280\,\text{nm}$ (FWHM). Consequently, an upper bound for the exciton and carrier diffusion length can be approximated at 295\,K. Please see the text for further details.}
    \label{S-Fig:CL}
\end{figure*}

Based on the $\mu$PL mapscan shown in Fig.\,2(b) of the main text, we already derived a first upper bound for the effective mean exciton and carrier diffusion length $l^{\mu\text{PL}}_{\text{diff}}\,\leq\,250\,\text{nm}$ in the upper GaN section of our photonic membrane. However, the precise determination of this value by a $\mu$PL mapscan was limited by the corresponding step size and can in general also be limited by the laser focus spot size that we obtain with our lasers emitting at either 266\,nm or 325\,nm. Thus, it is a common approach to estimate the exciton and carrier diffusion lengths by spatially resolved cathodoluminescence (CL) measurements \cite{liu_exciton_2016, hocker_determination_2016}, which can exhibit higher spatial resolution. At the given temperature and the non-resonant excitation conditions native to CL, it is not possible to further distinguish exciton from free carrier diffusion in the following. In S-Fig.\,\ref{S-Fig:CL}(a) we show a CL mapscan that was recorded with an acceleration voltage of 6\,kV and a probe beam current of 5\,nA in a Rosa 4634 system from Attolight at an ambient temperature of 295\,K. The location on our sample for these CL measurements again coincides with the corner region of our GaN membrane that is, e.g., depicted in S-Fig.\,\ref{S-Fig:spotsize}(b). The corresponding CL interaction volume in GaN at an acceleration voltage of 6\,kV has a lateral FWHM of $55\,\pm\,5\,\text{nm}$ as extracted from Monte-Carlo simulations with CASINO \raisebox{0.2cm}{{\tiny\textcircled{R}}} under consideration of an additional thermal broadening. As CL spectroscopy naturally generates hot carriers, the CL interaction volume is increased, which is taken into account by a corresponding convolution with a Gaussian as described in Refs.\,\cite{weatherley_imaging_2021,jahn_carrier_2022}. For the determination of this FWHM value we integrated the corresponding thermally broadened carrier generation rate over the cross-plane direction in our sample, as we are interested in the best achievable spatial resolution of our in-plane CL mapping. The CL mapscan from S-Fig.\,\ref{S-Fig:CL}(a) was recorded for the GaN bandedge emission of our sample around 3.4\,eV [see Fig.\,2(b) of the main text and S-Fig.\,\ref{S-Fig:spotsize}(b)] with a bandpass of 50\,meV over an area of $\approx\,1.8\,\times\,1.8\,\mu\text{m}^{2}$. The intensity fluctuations that were already visible in Fig.\,2(b) of the main text are now highly spatially resolved and originate from exciton and carrier localization at defects in GaN. The microscopic origin of these intensity fluctuations is unknown, but most likely point-defects related to deep donors in GaN \cite{callsen_optical_2012} originate the bright spots shown in S-Fig.\,\ref{S-Fig:CL}(a).

By evaluating the intensity distribution around such a point-defect it is possible to estimate an upper bound for the exciton and free carrier diffusion length in GaN, similar to previous observations made in the InGaN/GaN system, which mainly focus on dislocations \cite{liu_exciton_2016, hocker_determination_2016}. The intensity distribution that is associated to the horizontal cutline (red dashed line) shown in S-Fig.\,\ref{S-Fig:CL}(a) is plotted in S-Fig.\,\ref{S-Fig:CL}(b). Here the most prominent feature exhibits a FWHM of $\approx\,280\,\text{nm}$, which surpasses the CL interaction volume for the given acceleration voltage by $\approx\,225\,\text{nm}$. The additional broadening is originating from exciton and carrier diffusion in GaN at 295\,K. Thus, we can derive a better approximation for the upper bound of the mean exciton and carrier diffusion length of $l^{\text{CL}}_{\text{diff}}\,\lesssim\,115\,\text{nm}$ in the topmost GaN material of our photonic membrane at 295\,K. The motivation behind the analysis of the intensity distribution around this particular defect center is given by its well-isolated position. In this work we refrain from a more precise determination of $l_{\text{diff}}$ as an upper bound already suffices to conclude that all our 2LRT measurements are unaffected by the in-plane diffusion of photo-induced carriers generated in the heating spot. Figure\,7(a) of the main text showed that for 2LRT measurements one starts to evaluate the temperature distribution for distances exceeding $2.5\,\mu\text{m}$ from the heating spot. In addition, our 1LRT measurements and 2LRT$_0$ measurements cannot be strongly affected by photo-induced carriers as their concentration is low as long as the maximal temperature rise is limited to $T_{rise}\,\leq\,150\,\text{K}$ as described in Sec.\,VIII A of the main text.

In summary, a pronounced impact of photo-induced carriers on our Raman thermometry can be excluded as not only the concentrations of carriers is small, but even their mean free path lengths are low in GaN, which remains also true for all direct free excitons native to wurtzite GaN. Nevertheless, even for $l^{\text{CL}}_{\text{diff}}\,\lesssim\,115\,\text{nm}$ a certain fraction of photo-excited carriers can diffuse out of the laser-heated volume and temperature probe volume for 1LRT and 2LRT$_0$ measurements, rendering 2LRT measurements our preferred choice for quantitative, only optical thermometry on a photonic membrane made from a direct bandgap semiconductor like GaN.

\section{\label{S-Sec:Calibration}Temperature calibration}

%
%
%
\begin{figure*}[]
    \includegraphics[width=\textwidth]{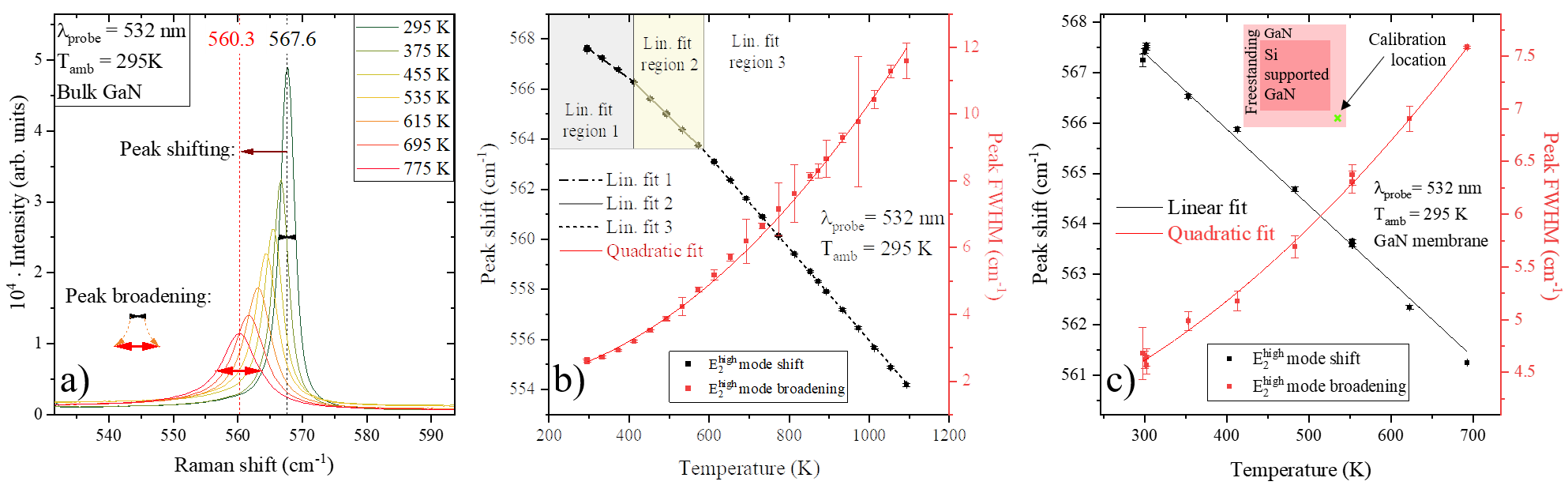}
    \caption{\textit{Temperature calibration on bulk GaN and the GaN pad. (a) Temperature dependence of the Raman spectra of bulk GaN in close energetic vicinity to the $E_{2}^{high}$ Raman mode. The Raman mode shifts towards lower relative wavenumbers and broadens with increasing sample temperature. (b) Summary of the bulk GaN Raman mode shift and the corresponding mode broadening with rising temperature (symbols). The solid lines depict fits to these two data sets. For the mode broadening a simple quadratic functions is sufficient for this fit, while the temperature evolution of the Raman mode shift can be best described by piecewise-defined linear functions (linear fit regions 1\,-\,3). (c) Similar data as shown in (b) but recorded on the GaN pad as illustrated in the inset. Here, the green cross marks the position of these temperature calibration measurements, which coincides with the positioning of the heating and probe laser of our 2LRT$_0$ measurements. Thus, this plot shows our local temperature calibration. The doubling of the datapoints at each temperature originates from the fact that we measured our Raman spectra during the heating and cooling of our heating stage to exclude any temperature hysteresis.}}
    \label{S-Fig:calibration}
\end{figure*}

The data that forms the basis of our temperature calibration is shown in S-Fig.\,\ref{S-Fig:calibration} for our bulk GaN substrate and GaN pad. S-Figure\,\ref{S-Fig:calibration}(a) shows a set of exemplary non-resonant Raman spectra, which were recorded for bulk GaN positioned in our heating stage, cf. S-Sec.\,\ref{S-Sec:Setup}. When increasing the global temperature of the sample that is positioned in vacuum, the $E_{2}^{high}$ Raman mode on the Stokes side of the Raman spectrum shifts towards lower relative wavenumbers, while it also increasingly broadens as visible in S-Fig.\,\ref{S-Fig:calibration}(a). Based on fitting our Raman spectra of this $E_{2}^{high}$ Raman mode by a suited model (Lorentz function with constant background), we extract the Raman mode positions and full width at half maximum (FWHM) values summarized in S-Fig.\,\ref{S-Fig:calibration}(b). This Raman mode shift and broadening with rising temperature originates from the volumetric expansion of the material and from multiple-phonon scattering events like 3-phonon scattering and with increasing temperature also 4-phonon scattering \cite{li_temperature_2000}. However, as we only aim to achieve a temperature calibration, we only need to fit the data from S-Fig.\,\ref{S-Fig:calibration}(b) with a function that can in the best case straightforwardly be inverted to enable the desired assignment from Raman mode shifts and broadenings to temperature. The mode broadening data from S-Fig.\,\ref{S-Fig:calibration}(b) can be fitted by a simple quadratic function, while the Raman mode shift is best approximated by a piecewise linear function (linear fit intervals 1\,-\,3) \cite{reparaz_novel_2014}. Clearly, all these fit functions can directly be inverted, allowing to extract the desired temperatures from Raman spectra that include the $E_{2}^{high}$ Raman mode. Here, the deviation between the fit functions and the data points limits the achievable temperature resolution to $\leq\,5\,\text{K}$ for the Raman mode shifts and $\leq\,20\,\text{K}$ for the corresponding broadenings.

For the GaN pad that is the center of interest for our Raman thermometry measurements, we decided to obtain a local temperature calibration. Therefore, we recorded the temperature calibration that is in use for all Raman measurements related to the GaN pad exactly in the corner region of the GaN membrane where we performed our 1LRT, 2LRT$_0$, and 2LRT measurements shown in the main text. The corresponding sample position of this local temperature calibration is marked by a green cross in the drawing shown in the inset of S-Fig.\,\ref{S-Fig:calibration}(c). We decided to perform such a local temperature calibration for our photonic membrane in order to lift any dependence between our temperature calibration and minor variations in strain and defect concentrations across the sample. Furthermore, this approach also aims to take the particular layer stack of our photonic membrane into account. Herein, different thermal expansion coefficients of, e.g., GaN and AlN can lead to an evolution of the $E_{2}^{high}$ Raman mode with rising temperature that deviates from the conventional case of bulk GaN material shown in S-Fig.\,\ref{S-Fig:calibration}(b).

However, disentangling the different origins for deviation in the temperature evolution of $E_{2}^{high}$ Raman mode recorded either on bulk GaN or on our photonic membrane comprising mostly GaN, but also AlN, and InGaN remains a task for future work. In this work, we only apply the data shown in S-Fig.\,\ref{S-Fig:calibration}(c) for our local temperature calibration. As shown herein, similar to the case of bulk GaN, the shift of the $E_{2}^{high}$ Raman mode can be approximated by a linear function in the temperature interval depicted in S-Fig.\,\ref{S-Fig:calibration}(c), while the corresponding Raman mode broadening can be fitted by a quadratic function. Here we chose to limit the maximum temperature to 700\,K to prevent any sample damage. By recording the corresponding Raman mode shifts and broadenings with increasing and decreasing temperatures, we excluded any temperature hysteresis in our temperature calibration. As a result, one can always observe at least two data points for the Raman mode shift and broadening at every temperature shown in S-Fig.\,\ref{S-Fig:calibration}(c). Again, the deviation between the fit functions and the data points limits the achievable temperature resolution to $\leq\,10\,\text{K}$ for the Raman mode shifts and $\leq\,50\,\text{K}$ for the corresponding broadenings. Compared to bulk GaN [S-Fig.\,\ref{S-Fig:calibration}(b)], the temperature resolution is moderately reduced for our GaN pad as the signal-to-noise ratio is decreased in the related Raman spectra recorded on a 250-nm-thick photonic membrane. Thus, the reduced scattering volume for our probe laser causes lower intensities of the $E_{2}^{high}$ Raman mode in our spectra. To compensate for thermally induced drifts of our sample, we recorded a mapscan of non-resonant Raman spectra for every temperature step shown in S-Fig.\,\ref{S-Fig:calibration}(c). We achieved a positioning accuracy for our calibrations greater than $\pm\,0.25\,\mu\text{m}$.

\section{\label{S-Sec:Modeling}Thermal modeling to extract the thermal conductivity}

To determine the thermal conductivity $\kappa$ from our measurements, we built a numerical model that aims to approximated our different thermometric approaches (1LRT, 2LRT$_0$, and 2LRT) by using the COMSOL Multiphysics\raisebox{0.2cm}{{\tiny\textcircled{R}}} software. The structure of this model is very similar from one experiment to the other, as it mostly differs through the the values of the light penetration depths ($p_{heat}$, $p_{probe}$) and the laser spot radii ($r_{heat}$, $r_{probe}$). Such laser spot radii are given by half the FWHM of the laser spot size for the corresponding laser wavelength, cf. S-Sec.\,\ref{S-Sec:Spotsize}. In addition, the evaluation of the temperature differs in the numerical model, depending on whether 1LRT, 2LRT$_0$, or 2LRT experiments are considered. In the following subsections, we will describe the geometry of these models (S-Sec.\,\ref{S-SSec:geometry}), the implementation of the heat transport (S-Sec.\,\ref{S-SSec:physImp}), the meshing of the structure (S-Sec.\,\ref{S-SSec:mesh}), the temperature evaluation (S-Sec.\,\ref{S-SSec:probe}) and the extraction of the thermal conductivity $\kappa$ (S-Sec.\,\ref{S-SSec:kappaExtr}). 
Furthermore, in S-Sec.\,\ref{S-SSec:BulkGaN} we report the 1LRT measurement on bulk GaN. 
The following S-Sec.\,A-D correspond to Step II of the thermal conductivity extraction described in the main text, and S-Sec.\,E corresponds to Step III.

    \subsection{\label{S-SSec:geometry} Geometry of the models}
\begin{figure*}[h]
        \includegraphics[width=\textwidth]{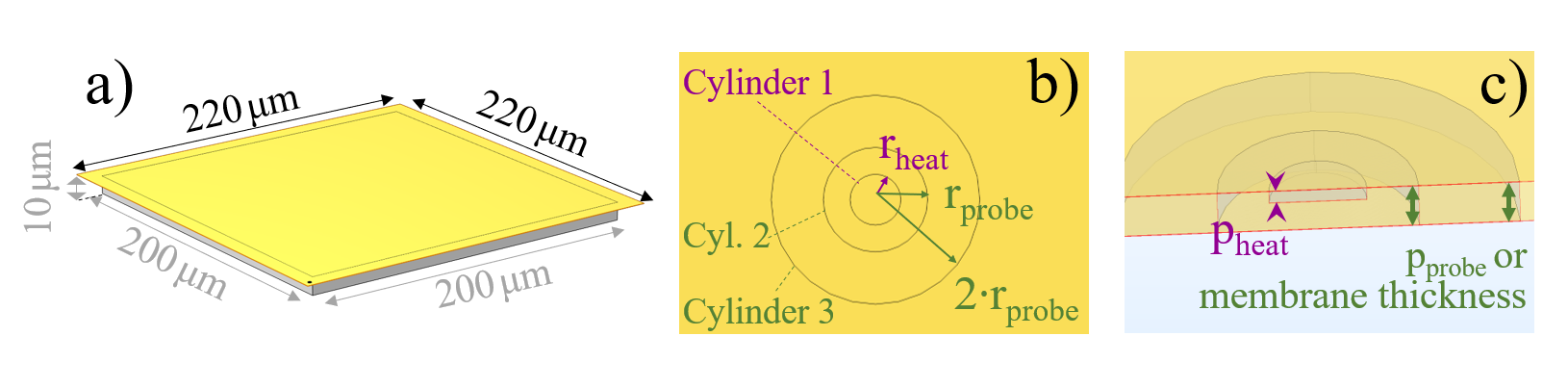}
        \caption{\textit{Illustration of the geometry of the 2LRT$_0$ numerical model. We show in yellow the GaN membrane and in gray its silicon support. (a) Depiction of the 220\,$\mu$m\,$\times$\,220\,$\mu$m$\,\times\,$250\,nm GaN pad over its 200\,$\mu$m\,$\times$\,200\,$\mu$m$\,\times\,$10\,$\mu$m silicon support. (b) Top view of the three cylinders described in S-Sec.\,\ref{S-SSec:geometry} indicating the various cylinder radii used in the modeling of the 2LRT$_0$ experiments. The center of the cylinders is located 4\,$\mu$m away from each side of the corner. (c) Clipping plane showing the profile of the membrane around the three cylinders, as used for modeling our 2LRT$_0$ measurements.
        }}
        \label{S-Fig:geom}
        \end{figure*}

    We model the geometry of our experiment by creating a 200\,$\mu$m\,$\times$\,200\,$\mu$m$\,\times\,$10\,$\mu$m silicon square box, over which we add the 220\,$\mu$m\,$\times$\,220\,$\mu$m$\,\times\,$250\,nm GaN pad [S-Fig.\,\ref{S-Fig:geom}(a)]. At one corner of this pad, centered 4\,$\mu$m away from each edge of the GaN pad, we create three concentric cylinders. The height of the first cylinder is equal to the penetration depth of the heating laser $p_{heat}$ and its radius equals the radius of the heating laser spot $r_{heat}$, while the second and the third cylinders have a height either equal to the penetration depth of the probe laser $p_{probe}$ or to the membrane thickness, if the light penetration depth is larger than the membrane thickness. The corresponding radii equal either one or two times the probe beam radius $r_{probe}$ for 2LRT$_0$ modeling, or to two and four times the radius of the heating laser spot for 1LRT modeling. These three cylinders are shown in S-Fig.\,\ref{S-Fig:geom}(b) and \ref{S-Fig:geom}(c). They impose a fine meshing in their vicinity, which is necessary to simulate a correct value of the temperature rise under the heating laser spot and to lighten the simulations by increasing the mesh density only around the heating spot. Additionally, the top surface of two innermost cylinders is used to introduce the heat source.
    
    The differences between the numerical models for the 1LRT, 2LRT$_0$ and 2LRT measurements are given by the different experimental parameters summarized in Tab. \ref{tab:summarize_param}.

    \begin{table} 
    \caption{\label{tab:summarize_param}Summary of our experimental parameters. The penetration depths originate from Ref.\,\cite{kawashima_optical_1997}.}
    \begin{ruledtabular}
    \begin{tabular}{llccccr}
     &  $\lambda_{heat}$   &   $\lambda_{probe}$    &   $r_{heat}$  &   $r_{probe}$ &   $p_{heat}$  &   $p_{probe}$ \\
    \hline
    1LRT    & 325\,nm & 325\,nm & 600\,nm & 600\,nm & 74\,nm & 74\,nm\\
    2LRT$_0$& 266\,nm & 488/532\,nm & 250\,nm & 650\,nm & 45\,nm & $\approx\,$500\,nm/60\,$\mu$m\\
    2LRT    & 266\,nm & 488/532\,nm & 250\,nm & 650\,nm & 45\,nm & $\approx\,$500\,nm/60\,$\mu$m\\
    
    \end{tabular}
    \end{ruledtabular}
    \end{table}

    \subsection{\label{S-SSec:physImp}Implementation of the thermal transport}
    The next step is the implementation of the thermal transport in our model. In COMSOL multiphysics\raisebox{0.2cm}{{\tiny\textcircled{R}}} this can be done by using the \textit{Heat Transfer in Solids} node of the \textit{Heat Transfer} module. The following equation is solved on a given mesh for a set of boundary conditions \cite{comsol_multiphysics_heat_2018}:
        \begin{align} \label{eq:heat1}
            \rho C_p \Vec{u} \cdot \nabla T + \nabla \cdot \Vec{q} &= Q + Q_{ted} \\
            \Vec{q} &= -\kappa \nabla T
        \end{align}
    with $\rho$ the density of the material (kg/m$^{3}$), $C_p$ the heat capacity at constant pressure (J/K), $\Vec{u}$ the velocity vector of translational motion (m/s) , $T$ the temperature (K), $\Vec{q}$ the local heat flux density (W/m$^{2}$), $Q$ the heat sources and sinks (J), and $Q_{ted}$ the thermoelastic dumping (J). We do not account for thermal expansion in the modeling and our sample is static, therefore $\Vec{u}\,=\,\Vec{0}$, and $Q_{ted}\,=\,0$. Combining and simplifying equations (1) and (2) leads to:
        
        \begin{equation} \label{eq:heat2}
            -\kappa \nabla^2 T = Q.
        \end{equation}

    We model the heat sink by setting the bottom of the silicon square depicted in S-Fig.\,\ref{S-Fig:geom}(a) to 295\,K, while all other boundaries, except the ones that introduce the heat source, as thermally insulating ($\Vec{n}\cdot \Vec{q} = 0$ with $\Vec{n}$ the unit vector perpendicular to a given surface). 
    Two additional loss mechanisms could in principle be included. First the power loss by black body radiation $P_{bbr}$ and second the intensity of the light emitted from the quantum well $P_{rad}$. However, both are negligible. From Stephan-Boltzmann's law, we estimate the power of black body radiation to be less than 240\,nW under the laser spot at our maximal measured power. Regarding the quantum well light emission power, time-resolved PL analysis yielded an internal quantum efficiency smaller than 0.1\,\% in our sample at 295\,K (see Sec.\,IV\,A\,3 of the main text).

    The laser power is introduced through the \textit{Deposited Beam Power} node of the \textit{Heat Transfer in Solids} node. For simplicity we choose a surfacic Gaussian heat source described by:
        
        \begin{equation}
            -\Vec{n} \cdot \Vec{q} = P_{abs} \cdot \frac{1}{2\pi \sigma^2}e^{-\frac{d^2}{2\sigma^2}} \cdot \frac{|\Vec{e}\cdot \Vec{n}|}{\norm{\Vec{e}}},
        \end{equation}
        
    with the power absorbed by the sample $P_{abs}$, the standard deviation $\sigma$, the beam orientation vector $\Vec{e}$, and the distance to the center of the beam spot $d$. In our case, the angle of incidence is considered perpendicular to the sample surface, yielding $\frac{|\Vec{e}\cdot \Vec{n}|}{\norm{\Vec{e}}} = \norm{\Vec{n}} = 1$. The definition of the beam waist built in COMSOL is related to the spot size we measured (Sec.\,\ref{S-Sec:Spotsize}) by:
        
        \begin{equation}
            \sigma = \frac{\text{FWHM}}{2\sqrt{2\ln(2)}}.
        \end{equation}
        
    This heat source is introduced on the top of the two innermost cylinders described in S-Sec.\,\ref{S-SSec:geometry} (see S-Fig.\,\ref{S-Fig:geom}(b)\,).
    A better model for the heat source would comprise the gradual attenuation of the laser beam in the material. Therefore we performed several numerical experiments to determine the impact of the penetration depth, which showed a maximal temperature deviation of less than 0.5\% between the temperature rises obtained via a surfacic heat source and such a volumetric heat source. While this remains true for materials with relatively low penetration depths, such as GaN or germanium, a better model would be needed as soon as the light penetration depth is is on the order of the laser spot radius. For instance, this would be the case for silicon when being heated by a 488\,nm laser.

    \subsection{\label{S-SSec:mesh} Meshing of our geometry}

    The meshing of the geometry is performed by COMSOL's algorithm. We mesh using a \textit{Free Tetrahedral Mesh}, with a \textit{normal} mesh element size. The cylinders defined in Sec.\,\ref{S-SSec:geometry} force the mesh size to shrink in their vicinity. We found that these three cylinders are sufficient for the temperature to converge. Reducing the mesh size in their volume or adding another cylinder did not significantly alter the final calculated temperatures. Between a mesh with a \textit{normal} meshing size and one with an \textit{extremely fine} mesh size, the difference of the measured temperatures in the heat spot, when $P_{abs}$ was set to the experimental maximum, amounted to $\sim$\,3\,\%.
    
    \subsection{\label{S-SSec:probe} Probing the temperature}
        
   In our experimental setup, we measure the average of the local temperatures weighted by the probe laser intensity profile \cite{beechem_invited_2015,ocenasek_raman_2015}:
   
       \begin{equation} \label{eq:probeT}
             T_{probe} = \frac{1}{p_{probe} \cdot w_e^2\cdot \pi} \int_V\,dr\,d\theta\,dz \,\cdot\, r\cdot T(r,\,\theta,\,z) \cdot  e^{-\frac{(r-r_0)^2}{w_e^2}} \cdot e^{-\frac{2z}{p_{probe}}}
        \end{equation}
        
    with $w_e$ the probe beam waist defined at the position at which the beam intensity falls to $\frac{1}{e}$ of its maximal value. This value is related to the radius of the focused spot and the standard deviation (Sec.\,\ref{S-SSec:physImp}) $\sigma$ by: $w_e = \sqrt{2}\sigma = \frac{r_{spot}}{\sqrt{\ln{2}}}$. $r_0$ is location of the probe beam, $p_{probe}$ the penetration depth of the probe laser, $T$ the local temperature, $(r, \theta, z)$ the cylindrical coordinates, and $V$ the volume of integration. The corresponding coordinates are set to vary as follows:
        
        \begin{itemize}
            \item[-] 0 to $2 \cdot r_{probe}$ along the radial coordinate $r$
            \item[-] 0 to 2$\pi$ along the angular coordinate $\theta$
            \item[-] 0 to $p_{probe}$ along the height coordinate $z$.
        \end{itemize}
        
    Regarding these integration limits, the radial limit is twice the radius of the probe spot, as any contribution to the signal outside of this limit is negligible (beam intensity falls to less than 0.05\% of its intensity at $r=r_0$). Regarding the depth limit, more than 95\% of the signal intensity originates from the sample region below this limit. At this point, because the 488\,nm laser has a penetration depth twice larger than the membrane thickness (Tab.\,\ref{tab:summarize_param}), the 2LRT$_0$ and 2LRT integration volumes are limited by the thickness of the membrane.

    Once the geometry (S-Sec.\,\ref{S-SSec:geometry}), the heat transport (S-Sec.\,\ref{S-SSec:physImp}), the meshing (S-Sec.\,\ref{S-SSec:mesh}) and the temperature probe (S-Sec.\,\ref{S-SSec:probe}) are implemented, we use the \textit{Parametric Sweep} node of the \textit{Study} node in COMSOL. Here we define the power absorbed $P_{abs}$ as a sweeping parameter and set it to the corresponding experimental values which lead to a measured $T_{rise}$ of less than 150\,K. $P_{abs}$ is the power that effectively enters the GaN membrane, meaning that reflection losses are accounted for. In the case of structures with a much higher internal quantum efficiency, $P_{abs}$ would be the power entering the membrane minus the power emitted as light, minus all other sources of power loss that lower the heating. In the same parameter sweep node in COMSOL, we also insert the variation over the thermal conductivity $\kappa$, but for the sake of clarity, this point is discussed in the next subsection.
        
    \subsection{\label{S-SSec:kappaExtr} Extraction of the thermal conductivity $\kappa$}
    
    The extraction of the thermal conductivity $\kappa$ from 1LRT and 2LRT$_0$ measurements differs from the corresponding analysis of 2LRT mapscans. Thus we split our explanations in the following subsections (1) and (2).

        \subsubsection{1LRT and 2LRT$_{0}$ measurements} \label{S-SSSec:kappaExtr_1LRT_2LRT0}
        
        For the 1LRT and 2LRT$_0$ measurements, we determine the thermal conductivity based on $T_{rise}$$(P_{abs})$ trends, as shown in Figs.\,4 and 5 of the main text. We limit the analysis to a range of powers for which $T_{rise}$ $\leq$ 150\,K holds. After the implementation of all experimental parameters (Secs.\,\ref{S-Sec:Modeling}\,A-D), the numerical model is swept over an adequate range of $\kappa$ values. For our photonic membrane, $\kappa$ was swept in steps of 5\Wu, from 20\Wu to 165\Wu. The stepping size was set to half the final error on the value of $\kappa$.
    
       Because the analysis is restrained to a regime where $T_{rise}\,\leq\,150\,$K, the thermal conductivity $\kappa$ is assumed independent of temperature, which is reflected in the linearity of the $T_{rise}$$(P_{abs})$ trends. By least square fitting to a linear function, we obtain the slopes of the experimental and numerical $T_{rise}$$(P_{abs})$ trends, then deduce $\kappa$ by matching the numerical slope to its closest experimental counterpart.
       To determine the uncertainty of $\kappa$, the upper bound is determined by matching the experimental slope minus its slope error to the numerical slopes, and the lower bound similarly with the slope error plus its slope error.

       \paragraph{Linearity of the T$_{rise}$(P$_{abs})$ trend:}
       In the case of semi-infinite samples (i.e., bulk samples), Lax \cite{lax_temperature_1977} derived a formula to compute the temperature rise under a Gaussian laser beam. The formula reads as follow:
        \begin{equation} \label{S-eq:Lax}
            T_{rise}(R, Z, W) = \frac{P_{abs}}{2 \sqrt{\pi} \kappa w_e} \cdot N(R,Z,W)
        \end{equation}
        where $R\,=\,u/w_e$ is the adimensional distance to the center of the focus of the heating laser spot, $u$ is the distance to the center of the heating laser, $Z\,=\,z/w_e$ is the adimensional distance to the sample surface, $W\,=\,\alpha w_e$ is a quantity that considers the attenuation of the beam, $\alpha$ is the absorption coefficient of the sample, $w_e$ the beam waist as defined in S-Sec.\,\ref{S-SSec:probe}, and $N(R,Z,W)$ is the normalized temperature rise. $N(R,Z,W)$ takes its maximum value when $R\,=\,Z\,=\,0$ (i.e., at the center of the laser beam, directly at the surface) and $W\rightarrow\infty$, when all the laser light is absorbed at the surface of the sample, with $\lim_{W\to\infty} N(0,0,W)=1$. 
        Although the sample under study is not a semi-infinite sample, we can reasonably expect $T_{probe} \, \sim \, \frac{P_{abs}}{\kappa}$. When $T_{probe} \, \leq \, 150\,K$, this is experimentally observed as shown in Figs.\,4(c) and \,5(c) of the main text.
        
       \paragraph{Asymmetry on the uncertainty of the thermal conductivity value:} 
        From S-Eq.\,\ref{S-eq:Lax}, we see that $\frac{T_{rise}}{P_{abs}} \sim \frac{1}{\kappa}$, therefore a symmetric error on the $T_{rise}(P_{abs})$ slope will lead to an asymmetric error on $\kappa$. Although S-Eq.\,\ref{S-eq:Lax} does not strictly hold for our GaN membrane, and the $\frac{T_{rise}}{P_{abs}} \sim \frac{1}{\kappa}$ scaling is not strictly followed, a larger value of $\kappa$ will result in a smaller slope of $\frac{T_{rise}}{P_{abs}}$, and inversely for a smaller value of $\kappa$, yielding asymmetric uncertainties.

       \subsubsection{2LRT measurements} \label{S-SSSec:kappaExtr_2LRT}
       To extract the thermal conductivity from a 2LRT mapscan, we numerically reproduce the experimental condition by setting $P_{abs}$ to the experimental value. The thermal conductivity $\kappa$ is swept over the same values as used in the analysis of 1LRT and 2LRT$_0$ measurements, 20\,\Wu to 165\,\Wu in steps of 5\,\Wu. 
       We extract two cut lines in respectively the \textit{a}- \& \textit{m}-directions in our photonic membrane, which will be compared to the exact same cut lines from the numerical model. This comparison is done on a reduced range of the cut line, sufficiently far away form the laser spot to avoid all complications related to the absorption of the laser and where the temperature rise does not exceed 150\,K. 
       By using a least square fitting method in the restricted range of our data points, we compute the slopes of our experimental and numerical cut lines. The thermal conductivity $\kappa$ is then derived by matching the experimental slopes to their closest numerical counterpart. 
       The previously described process can be applied to any of the crystal directions, for an adequate sample, highlighting one of the strengths of the 2LRT method.  
       Contrarily to the $\kappa$ extraction of the 1LRT and the 2LRT$_0$ measurements, for the 2LRT data analysis $\kappa$ was derived for a constant value of $P_{abs}$, and far away form the heating spot.
       To determine the uncertainty of our value of $\kappa$ derived from 2LRT measurements, we proceed similarly to what was done in S-Sec.\,\ref{S-SSSec:kappaExtr_1LRT_2LRT0} for the 1LRT and 2LRT$_0$ measurements.

        \paragraph{Linearity of the $T(\ln{(u)})$ curve:}
        This choice of range is motivated by the following form of the heat equation, derived for an infinite 2D membrane under the assumption of a temperature independent $\kappa$ and setting the origin of the coordinates to the center of the heat source \cite{reparaz_novel_2014}:
        \begin{equation} \label{S-eq:sebas}
            T(u) - T_0 = -\frac{P_{abs}}{2\pi d \kappa}\ln{(u/u_0)}.
        \end{equation}
        In Eq.\,\ref{S-eq:sebas} the variable $u$ is the distance to the center of the heating spot, $T_0$ is the temperature reached at the distance $u_0$ from the heating spot, and $d$ is the membrane thickness. This approach assumes that the heat source is point-like, which is an equivalent condition to setting our probe beam far enough from the heating laser spot. 
        Note that Eq.\,\ref{S-eq:sebas} cannot be used to directly derive the thermal conductivity $\kappa$, since our membrane features a particular geometry and a heat sink (the silicon substrate).

        \paragraph{Asymmetry on the uncertainty of the thermal conductivity value:}
        Similarly to the 1LRT and 2LRT$_0$ cases, but for the $T[\ln{(u)}]$ trend, we observe the following proportionality $\frac{T}{\ln{(u)}} \sim \frac{1}{\kappa}$, in the analyzed ranges. Again, although S-Eq.\,\ref{S-eq:sebas} does not strictly hold for our 2LRT experiment, $\frac{T(u)}{\ln{(u)}}$ scales inversely to $\kappa$, resulting in asymmetric uncertainties of $\kappa$.

\newpage
        
        \paragraph{Summary of the $\kappa$ extraction process:}
        For the 1LRT and the 2LRT$_0$ measurements on the one hand and the 2LRT measurements on the other hand, the derivation of $\kappa$ is analogous. We summarize it here:

            \begin{tasks}(2)
                \task[] \underline{1LRT} \& \underline{2LRT$_0$}
                \task[] \underline{2LRT}
                \task[-] get the slope of the experimental $T_{rise}(P_{abs})$ trend
                \task[-] get the slope of the experimental $T[\ln{(u)}]$ trend, for any of the desired crystal directions
                \task[-] simulate the experiment for a suitable range of $\kappa_{sim}$
                \task[-] simulate the experiment for a suitable range of $\kappa_{sim}$
                \task[-] get the slopes of the numerical $T_{rise}(P_{abs})$ trends, for each simulated $\kappa_{sim}$
                \task[-] get the slopes of the numerical $T[\ln{(u)}]$ trends, for each simulated $\kappa_{sim}$
                \task[-] get $\kappa$ by matching the experimental slope to the numerical ones
                \task[-] get $\kappa$ by matching the experimental slope to the numerical ones
                \task[-] get the uncertainty of the $\kappa$ value
                \task[-] get the uncertainty of the $\kappa$ value
            \end{tasks}

    \subsection{\label{S-SSec:BulkGaN}Thermal conductivity of bulk GaN}

    As described in Sec.\,IV A of the main text, we did not succeed in measuring the thermal conductivity of bulk GaN in our experimental setup. However, this observation represents in itself an interesting finding. As shown in Fig.\,4(c) of the main text, no heating can be observed on bulk GaN even at the maximum of the optically-induced heating power that we can reach with our 325-nm-laser. Nevertheless, when simulating this situation on bulk GaN for our 1LRT measurements based on the Fourier model described in S-Sec.\,\ref{S-Sec:Modeling}, even for the highest thermal conductivities ever reported for bulk GaN at room temperature $\kappa_{bulk}\,\approx\,300\,\text{W/mK}$ \cite{zheng_thermal_2019}, we should still have been able to observe a temperature rise ($T_{rise}$). Furthermore, we identified that neither the optical resolution of our experimental setup (see S-Sec.\,\ref{S-Sec:Setup}), nor the signal-to-noise ratios in our resonant Raman spectra of GaN constitute a limit for measuring $\kappa_{bulk}$ of high quality GaN. Thus, this apparent discrepancy between our simple thermal model and the experimental finding must again be explained by the increase in the thermalization volume $V_{therm}$ that is caused by phonons with a large mean free path in GaN, allowing them to escape from the laser absorption volume $V_{abs}$ (see Sec.\,VII\,B of the main paper). Thus, for 1LRT measurements with $V_{abs}\,=\,V_{probe}$ as a first approximation, the measured $T_{rise}$ values are lowered, leading to the extraction of artificially enhanced $\kappa_{bulk}$ values. This situation is similar to our findings from the main text summarized by Fig.\,14. Here, for 1LRT measurements we were still able to measure $T_{rise}$ with rising absorbed laser power $P_{abs}$ due to the lower thermal conductivity of our photonic membrane compared to bulk GaN. Nevertheless, any analysis of our 1LRT data aiming to extract $\kappa$ is somewhat problematic, as we rely on a model that only considers diffusive phonon transport, assuming a complete thermalization within $V_{abs}$. Therefore, it remains an important task to test our 2LRT measurements also on bulk GaN. However, we still do not know whether our heating laser is powerful enough to heat bulk GaN, such that we can probe the spatial evolution of $T_{rise}$ several micrometers away from the heating spot. For such measurements it would be promising to gain experimental access to GaN membranes made from high quality bulk GaN in order to suppress cross-plane thermal transport, aiming to enhance $T_{rise}$ due to purely in-plane thermal transport.


\bibliography{1-testBib_ME}